%% file: ms_o6major_x2.tex
\documentclass[twocolumn,twocolappendix]{aastex63}

\hypersetup{linkcolor=blue,citecolor=blue,filecolor=magenta,urlcolor=cyan}
\shorttitle{\oVI\ Kinematics of $z\approx0.2$ Star-forming Galaxies}
\shortauthors{Ho et al.}

\usepackage{natbib}
\usepackage{bm}
\usepackage{chngcntr}

\begin{document}

\input setdef.tex
\title{Kinematics of Circumgalactic \oVI\ Gas and Disk Rotation of $z\approx0.2$
Star-forming Galaxies}

%% you desire. In v6.3 you can include a footnote in the title.

\correspondingauthor{Stephanie Ho}
\email{shho@nmsu.edu}

\author[0000-0002-9607-7365]{Stephanie H.~Ho}

\affiliation{Department of Astronomy, New Mexico State University, Las Cruces, NM 88003, USA}

\author[0000-0001-9189-7818]{Crystal L.~Martin}
\affiliation{Department of Physics, University of California, Santa Barbara, CA 93106, USA}

\author[0000-0003-3926-1526]{Hasti Nateghi}
\affiliation{Centre for Astrophysics and Supercomputing, Swinburne University of Technology, Hawthorn, VIC 3122, Australia}

\author[0000-0003-1362-9302]{Glenn G.~Kacprzak}
\affiliation{Centre for Astrophysics and Supercomputing, Swinburne University of Technology, Hawthorn, VIC 3122, Australia}

\author[0000-0002-7541-9565]{Jonathan Stern}
\affiliation{School of Physics and Astronomy, Tel Aviv University, Tel Aviv 69978, Israel}

%%%%%%%%%%%%%%%%%%%%%%%%%%%%%%%%%%%%%%%%%%%%%%%%%

%%%%%%%%%%%%%%%%%%%%%%%%%%%%%%%%%%%%%%%%
% ABSTRACT
%%%%%%%%%%%%%%%%%%%%%%%%%%%%%%%%%%%%%%%%

\begin{abstract}
    
Quasar sightline observations
reveal that low-ionization-state gas corotates 
with the galaxy disk 
and often at sub-centrifugal velocities,
suggesting that the gas is spiraling towards
the galaxy disk.
However, 
while observations ubiquitously detect \oVI\ absorption
around 
low-redshift, $\sim$\lstar\ star-forming galaxies, 
the relationship between \oVI\ and the galaxy disk, 
especially the kinematics,
is not well-established.
This work focuses on the \oVI\ kinematics
and its comparison with that of the low ions 
and galactic disk rotation.
We present observations of 18 pairs of quasars
and $z\approx0.2$ star-forming galaxies.  
All quasar sightlines intersect 
the circumgalactic medium (CGM) within 45\deg\ 
from the galaxy major axes.
We show that while individual \oVI\ velocity components
do not correlate with disk rotation,
the bulk of \oVI\ gas in individual sightlines
rarely counter-rotates. 
We then match \oVI\ velocity components
with those of low ions by minimizing the difference of
their velocity centroids.
The \oVI\ velocity components
with successful low-ion matches 
are typically found at small sightline impact parameters
and are 
more likely to corotate with the disk.
We suggest that the low-ion-matched \oVI\ velocity components
trace the gas co-spatial with the low ions
near the extended disk plane in the inner CGM,
whereas those without low-ion matches
represent the gas at large 3D radii.
While the gas at large radii is theoretically expected
to kinematically correlate with the disk angular momentum,
this correlation is expected to be weaker 
due to the higher turbulent to mean rotation velocity ration 
at large radii, consistent with our results.

\end{abstract}


%%%%%%%%%%%%%%%%%%%%%%%%%%%%%%%%%%%%%%%%
% INTRODUCTION
%%%%%%%%%%%%%%%%%%%%%%%%%%%%%%%%%%%%%%%%
\section{Introduction}
\label{sec:intro}

The interplay between galaxies
and their surrounding gas 
shapes the evolution of galaxies.
Galaxies gather gas from their surroundings to form stars,
whereas winds driven by supernovae and active galactic nuclei (AGN)
remove gas from the galaxies
and slow star formation.  
The surrounding gas, known as the circumgalactic medium (CGM),
is a substantial reservoir of 
baryons and metals
and extends at least to the virial radius (\rvir)
of galaxies
(see \citealt{Tumlinson2017} and \citealt{FaucherOh2023} 
for reviews).
Because the low gas density 
makes direct imaging of the CGM challenging,
quasar spectroscopy serves as a powerful tool
for detecting the CGM in absorption
and characterizing CGM properties.

The ubiquitous detection of \oVIdb\ absorption 
around 
low-redshift, $\sim$\lstar\ star-forming galaxies 
in contrast to 
the low detection rate around 
quiescent galaxies have received significant
attention \citep{Tumlinson2011,Tchernyshyov2023}.
Dwarf galaxies also show the same contrast \citep{Dutta2024},
even though the \oVI\ detection rate and column density of 
star-forming dwarf galaxies
are lower than that of $\sim$\lstar\ galaxies
\citep{Johnson2015,Johnson2017,Tchernyshyov2022,Mishra2024}.
The origin of this ``\oVI\ dichotomy'' remains controversial.
AGN feedback possibly leads to a low CGM mass fraction,
resulting in the low \oVI\ detection rate of quiescent galaxies
\citep{Suresh2017,Nelson2018,Davies2020}.
Some work argued that 
the quiescent galaxies have higher average halo masses
than the star-forming galaxies in the observed galaxy samples,
and hence, the higher virial temperatures 
promote oxygen to higher ionization states.
In particular, the predicted CGM properties change significantly 
for galaxies with halo masses around 
$\sim10^{12}$\msun\ due to virialization
\citep{Oppenheimer2016,MathewsProchaska2017,Stern2018,Sultan2025}.
However, recent work with a mass-controlled galaxy sample
still found the ``\oVI\ dichotomy''. 
This implies a potential 
connection between \oVI\ and ongoing star formation,
for which 
the transformation of CGM happens concurrently
with the quenching of galaxies \citep{Tchernyshyov2023}.

Dedicated efforts with quasar absorption-line studies
have analyzed the circumgalactic \oVI\ properties 
of star-forming galaxies
and compared that with the cool gas ($\sim10^4$ K) 
traced by low-ionization-state ions (e.g., \mgII).
For $\sim$\lstar\ star-forming galaxies, 
the cool gas is typically detected only in the inner CGM
of $\sim100$ kpc, 
and the strength of the low-ionization-state absorption 
decreases sharply with increasing 
projected separation from the galaxies
\citep{Chen2010,Nielsen2013_ii,Huang2021}.
In contrast,
\oVI\ is detected in sightlines out to
at least two to three times
the galaxy virial radius,
and the \oVI\ column density only declines slowly
with the sightline impact parameters
\citep{Tumlinson2011,Tchernyshyov2022,Qu2024}.
Similarly for star-forming dwarf galaxies,
\oVI\ has a higher detection rate 
and a more extended distribution
compared to that of the low ions
\citep{Mishra2024,Dutta2024}.
But surprisingly,
both low ions and \oVI\
exhibit a bimodal distribution with azimuthal angle\footnote{
    The azimuthal angle $\alpha$ is defined as the angle between
    the galaxy major axis and the line drawn between
    the center of the galaxy and the quasar.
    }
\citep{Bouche2012,Kacprzak2012,Kacprzak2015,Schroetter2019}.
The bimodality in gas distribution is also supported by recent 
cloud-by-cloud multiphase, ionization modelings
that indicate a larger number of clouds 
in sightlines near the galaxy major and minor axes
\citep{Sameer2024}.

In addition to the spatial distribution, 
kinematics provides crucial information
for understanding the origin and properties of 
the CGM traced by \oVI\ and low-ionization-state ions.
While \oVI\ absorption has a larger velocity spread 
compared to that of the low ions,
they typically span similar velocity ranges
\citep{Werk2016,Nielsen2017}.
However, 
while the absorption of low ions becomes broader and stronger
towards the galaxy minor axes, 
a result attributed to gas kinematically disturbed by outflows
\citep{Bordoloi2011,Martin2019,Schroetter2019},
the \oVI\ velocity spread
does not vary with the sightline azimuthal angle 
and galaxy disk inclination
\citep{Nielsen2017}.
Furthermore, 
for sightlines near the galaxy major axes,
the cool gas (traced by \mgII) corotates with the galaxy disk
and often at sub-centrifugal velocities
\citep{Steidel2002,Kacprzak2010,Kacprzak2011ApJ,Ho2017,Zabl2019},
suggesting that the gas is infalling
\citep{HoMartin2020}.
In apparent conflict with these results, 
however, \oVI\ studies have not shown such a correlation
\citep{Kacprzak2019,Kacprzak2025}.

Given the spatially extended distribution of 
\oVI\ compared to the low ions,
it is perhaps unsurprising that
the \oVI\ kinematics 
do not seem to correlate 
with disk rotation
as the low-ionization-state counterparts.
In particular,
because the CGM angular momentum 
remains roughly constant with radius,
the outer CGM is expected to have 
a lower mean rotation velocity,
resulting in a weaker correlation with the disk rotation.
To date, however,
very few observational studies 
have focused on the kinematics comparison between
\oVI\ and the galaxy disk
\citep{Kacprzak2019,Nateghi2024multiphase,Kacprzak2025},
which is in contrast to the 
large number of works in previous decades %in the past decades
that observationally 
confirmed the corotation between the 
low-ionization-state gas and the disk
\citep{
Steidel2002,Kacprzak2010,Kacprzak2011ApJ,
Bouche2013,Bouche2016,Keeney2013,Chen2014,
DiamondStanic2016,
Ho2017,Martin2019,Zabl2019,HoMartin2020,
Lopez2020,Tejos2021}.

In this paper, 
we build a sample of star-forming galaxies
with quasar sightlines near 
the galaxy major axes
and exclude galaxies with face-on disks.
With quasar spectroscopy from the
\textit{Cosmic Origin Spectrograph (COS)} on the 
\textit{Hubble Space Telescope (HST)}
and galaxy rotation curve measurements,
we analyze and compare the kinematics of the CGM
traced by \oVI\ and  low ions (e.g., \siII, \siIII)
with the galaxy disk rotation.
We describe the sample selection, data reduction,
and measurements in Section~\ref{sec:data}.
In Section~\ref{sec:result},
we present the results of the kinematics comparison
among \oVI\ gas, low-ionization-state gas, and disk rotation.
We interpret our results and discuss their implications
in Section~\ref{sec:discussion}.
Finally, Section~\ref{sec:conclusion} summarizes the conclusions.
Throughout the paper, we adopt the cosmology 
from \citet{PlanckXIII2015}, 
with $h = 0.6774$, $\Omega_m = 0.3089$, 
$\Omega_\Lambda = 0.6911$, and $\Omega_b = 0.0486$.

%%%%%%%%%%%%%%%%%%%%%%%%%%%%%%%%%%%%%%%%
% DATA
%%%%%%%%%%%%%%%%%%%%%%%%%%%%%%%%%%%%%%%%
\section{Data and Observations}
\label{sec:data}

We present the analyses of 18 galaxy--quasar pairs 
to study the kinematics of \oVI\ and low-ionization-state absorbers
and their relationship with galaxy disk rotation.
We conducted quasar spectroscopy with \textit{HST/COS},
including new spectra from our \textit{HST} program
and archival observations.
We also acquired galaxy spectra from
multiple ground-based telescopes.
We describe our sample selection in Section~\ref{ssec:sample}.
Section~\ref{ssec:galspec} discusses 
the galaxy spectroscopic observations and 
rotation curve measurements, 
and Section~\ref{ssec:galprop} describes the galaxy properties.
In Section~\ref{ssec:hstcos},
we describe the \textit{HST/COS} observations
and measurements of absorption-line systems.

\subsection{Sample of Galaxy--Quasar Pairs}
\label{ssec:sample}

Our study focuses on
star-forming galaxies
with quasar sightlines 
near the major axes of the galaxies.
The galaxy--quasar pairs were selected from 
%published literature
published CGM studies
with known galaxy disk orientation;
the disk geometry was obtained either from  
published work or from 
the \texttt{PhotoObj} table of
the Sloan Digital Sky Survey
\citep[SDSS;][]{Blanton2017}.
Then, we restricted our sample to galaxies 
with inclined disks of inclination angles of $i \gtrsim 45$\deg, 
and each sightline lies within 45\deg\ from the galaxy major axis,
i.e., with azimuthal angle $\alpha \leq 45$\deg.
This selection with inclined disks and major-axis sightlines
excludes regions of the CGM proven to be disturbed by 
galactic outflows (Section~\ref{sec:intro}).  
Furthermore, we avoided sightlines near the galaxy minor axes
because of the small projected rotation speed 
from a rotating disk component, 
which would make it challenging to distinguish rotation
from other potential velocity components (e.g., conical outflows).  
The same reason also led us to exclude face-on disks,
in addition to their large uncertainties
in the disk position angles and sightline azimuthal angles.

With our sample selection criteria,
we selected our target galaxies
from 120 unique, $z\sim0.2$ galaxy--quasar pairs
from the COS-Halos survey \citep{Tumlinson2013,Werk2013},
the COS-Burst survey \citep{Heckman2017},
the Multiphase Galaxy Halos survey \citep{Kacprzak2019},
and the Quasars Probing Galaxies (QPG) program \citep{Martin2019},
among which 85\% of the galaxy candidates
were star-forming.
Except for the 50 star-forming galaxies in the QPG program,
the other three CGM studies were 
conducted with \textit{HST/COS}.
Because our work focuses on the kinematics of the \oVI\ gas
instead of the detection rate, 
we required \oVI\ being detected around 
the 52 star-forming galaxies
from those three CGM surveys,
reducing our \oVI-detected galaxy candidates to 32.  
Then, we imposed our disk inclination and azimuthal angle cuts,
and also visually inspected the images 
(from the survey papers and/or the SDSS $r$-band images)
to ensure the galaxies have reasonable disk geometry measurements.
We ended up with 11 galaxies from the three surveys,
four of which had no rotation curve data
at the time when the galaxies were selected (Section~\ref{sssec:mmt}).
As for the QPG program,
it focused on 
the \mgII\ kinematics around star-forming disk galaxies
with disk geometry and rotation curve measurements readily available
\citep{Ho2017,Martin2019}.
We selected galaxy--quasar pairs that satistied
our criteria on azimuthal angle and disk inclination,
regardless of whether or not \mgII\ was detected.
Then, we selected seven quasars bright enough in the 
far-UV (FUV) with $m_\mathrm{FUV} \lesssim19.5$
for new \textit{HST/COS} observations 
(Section~\ref{ssec:hstcos}).

\begin{figure}[htb]
    \centering
    \includegraphics[width=0.9\linewidth]{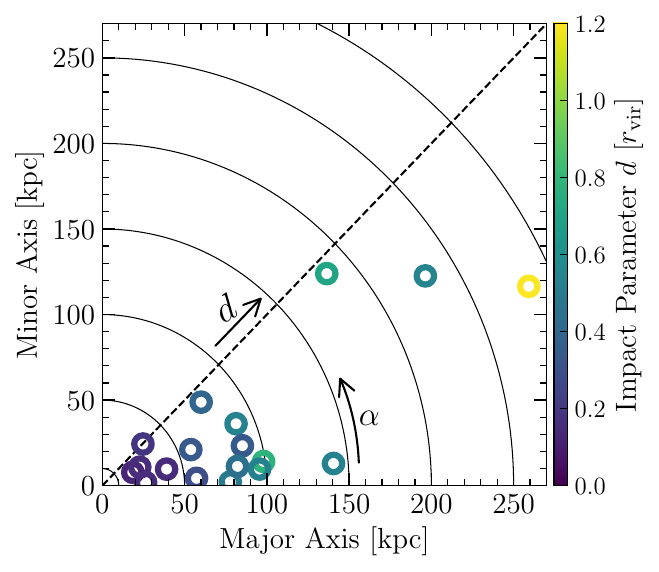}
    \caption{Ensemble of 18 sightlines.
                All selected galaxies have inclined disks with 
                inclination angles $i \gtrsim 45$\deg.
                Each sightline intersects the galaxy 
                near the major axis 
                with azimuthal angle $\alpha \leq 45$\deg.
                The impact parameter $d$ of the quasar sightlines 
                ranges between 20 and 284 kpc.
                The color of the markers represents
                the sightline impact parameter normalized by 
                the galaxy virial radius.
                Most sightlines intersect the inner CGM 
                within 100 kpc (i.e., within 0.5\rvir).
                }
    \label{fig:alphab_polar} 
\end{figure}

The final compiled sample 
consists of 18 galaxy--quasar pairs with
a median galaxy redshift of $z \approx 0.21$.
Tables~\ref{tb:obs_gal} and \ref{tb:galprops} specify
the coordinates and properties of galaxies, respectively,
and Table~\ref{tb:obs_qso} lists the coordinates of 
the paired quasars. 
Figure~\ref{fig:alphab_polar} shows the location 
of the quasar sightlines relative to 
the projected major and minor axes of the galaxies on the sky.  
The impact parameter $d$ of the quasar sightlines
span between 20 and 280 kpc.
The majority of the sightlines intersect the galaxies
within 100 kpc, 
which corresponds to 
less than half of the virial radius (\rvir)
as indicated by the colors of the markers 
(see Section~\ref{ssec:galprop}).

%%%%%%%%%%%%%%%%%%%%% FIG: alpha-b polar %%%%%%%%%%%%%%%%%%%%%%%%
%%%%%%%%%%%%%%%%%%%%%%%%%%%%%%%%%%%%%%%%%%%%%%%%%%%%%%%%%%%%%%%

%%%%%%%%%%%%%%%%%%%%%%%%
% Galaxy Spec
%%%%%%%%%%%%%%%%%%%%%%%%
\subsection{Galaxy Spectroscopy}
\label{ssec:galspec}

We used the galaxy spectra to measure
the systemic redshifts of the target galaxies
and their rotation curves
from the optical emission lines. 
We acquired new long-slit galaxy spectra
with the Blue and Red Channel Spectrographs 
at the MMT Observatory.
We also compiled previous observations 
\citep{Ho2017,Martin2019,Kacprzak2019}
from the Low Resolution Imaging Spectrometer
\citep[LRIS;][]{Oke1995,Rockosi2010} and
the Echellette Spectrograph and Imager 
\citep[ESI;][]{Sheinis2002}
at the W.~M.~Keck Observatory
and the Double Imaging Spectrograph (DIS) 
at the Apache Point Observatory (APO) 3.5 m telescope.\footnote{
  Instrument specifications can be found in the manual written 
  by Robert Lupton, which is available at \url{
  http://www.apo.nmsu.edu/35m_operations/35m_manual/Instruments/DIS/DIS_usage.html\#Lupton_Manual} 
    }

We summarize the galaxy observations in Table~\ref{tb:obs_gal}.
Section~\ref{sssec:mmt} describes the new MMT observations,
and Sections~\ref{sssec:keck} and \ref{sssec:apo}
briefly discuss the previous Keck and APO observations, 
respectively.
Section~\ref{sssec:rc} describes the measurements of 
the galaxy rotation curves from the emission lines 
(typically \halpha) detected in the galaxy spectra.

% =================================
% TABLE with Galaxy EMLine observations
\begin{center}
    \input{table_obsgal.tex}
\end{center}
% =================================

\subsubsection{MMT Blue and Red Channel Observations}
\label{sssec:mmt}

We conducted new galaxy spectroscopic observations 
using the MMT Blue and Red Channel Spectrographs.
On February 28, 2020,
we observed the galaxies with the Blue Channel 
and configured it with the 832 \linesmm\ grating blazed at 7790 \AA.
On January 9, 2021, we used the Red Channel  
with the 1200 \linesmm\ grating blazed at 7700 \AA.  
For both observations,
we applied the blue blocking filter R-63 
to block the higher order light. 
We used the default $1\times1$ detector binning,
which provided a pixel scale of 0\farcs3 for both channels
and a wavelength dispersion of 
0.72 \aapix\ and 0.80 \aapix\ 
for the Blue and Red Channel, respectively.
We observed the target galaxies by aligning the 
1\farcs0-wide long-slit with the galaxy major axes.
For both channels,
the 1\farcs0-wide slitlet provided 
a spectral resolution of 75--85 \kms\ 
full width at half maximum intensity (FWHM).

The data were reduced using IRAF \citep{Tody1986,Tody1993}
following standard procedures.  
For individual exposures, 
the zero-point bias level was corrected for,
and the fixed pattern noise was removed 
using exposures of the internal quartz lamp.
Cosmic rays were removed using L.A. Cosmic \citep{vanDokkum2001}.
Then,
the frames were combined using a sigma-clipping algorithm
to remove any remaining cosmic rays. 
A two-dimensional variance spectrum was computed 
when the exposures were combined.
Frames of arc lamp exposures were used to calibrate 
the wavelength of the spectra,
and the root-mean-square (rms) error 
of the dispersion solution was less than 15 \kms.
The wavelength solution was checked using
the calibrated night-sky emission-line spectrum
from the Ultraviolet and Visual Echelle Spectrograph 
\citep{Hanuschik2003}.
Zero-point corrections were applied when necessary,
which was caused by the difference in rotator angles
between the science exposures and the calibration frames.
This was followed by a heliocentric correction
to correct for the earth's seasonal motion.

From the two-dimensional galaxy spectra,
we measured the galaxy rotation curves from the 
emission lines (Section~\ref{sssec:rc}). 
We also extracted an one-dimensional spectrum 
for each galaxy
and determined the galaxy systemic redshift from
the redshifted wavelengths of the
\halpha, \nIIlam, and \sIIlam\ emission lines.

%%%%%%%%%%%%%%
% Keck Observations
%%%%%%%%%%%%%%
\subsubsection{Keck Observations}
\label{sssec:keck}

The Keck/LRIS observations were described in detail
in \citet{Ho2017} and \citet{Martin2019}.
The LRIS double spectrograph was configured 
using the D500 dichroic, 1200 \linesmm\ grism 
blazed at 3400\AA\ (LRISb) and
the 900 \linesmm\ grism blazed at 5500\AA\ (LRISr). 
The galaxy emission lines for rotation curve measurements
were captured by LRISr.
The detectors were binned by two 
in both spatial and dispersion directions,
which produced a spatial scale of $0\farcs27$
and a dispersion of 1.06 \aapix. 
The 1\farcs0-wide long-slit was either aligned with the 
position angle of the major axis of the target galaxies,
or masks of 1\farcs0-wide slitlets were designed to 
include nearby galaxies in the field.
The 1\farcs0-wide slitlets provided an 
average resolution of 75--105 \kms\ FWHM.
The data reduction procedures for the LRISr spectra 
were explained in \citet{Ho2017}
and were similar to those for reducing the MMT spectra
described in Section~\ref{sssec:mmt}.

The Keck/ESI observations were described in detail 
in \citet{Kacprzak2019}.
In brief, 
the galaxies were observed 
using the 1\farcs0-wide, 20\arcsec-long ESI slit, 
and the slitlet was aligned with the major axes of 
individual galaxies.
The detector was binned by two 
in both the spatial and spectral directions.
This produced a dispersion of 22 \kms\ pixel$^{-1}$
and pixel scales of 0\farcs27 to 0\farcs34
over the echelle orders that 
covered the galaxy optical emission lines.
The resulting spectral resolution from the 1\farcs0-wide slitlet
was 90 \kms\ FWHM.
Following the data reduction procedures
as outlined in \citet{Kacprzak2019},
we re-reduced the ESI spectra and  
re-examined the galaxy emission lines.
The new two-dimensional spectra were adopted to 
measure the galaxy rotation curves (Section~\ref{sssec:rc}).

%%%%%%%%%%%%%%
% APO/DIS
%%%%%%%%%%%%%%
\subsubsection{APO DIS Observations}
\label{sssec:apo}

For galaxies J142459+38211 and J155505+362848,
\citet{Ho2017} and \citet{Martin2019} 
conducted long-slit spectroscopy with APO/DIS 
to measure the galaxy rotation curves from the 
\halpha\ emission line.
The 1\farcs5-wide long-slit was aligned with 
the major axis of each galaxy.
The blue and red channels of DIS were 
configured with the 1200 \linesmm\
blazed at 4400\AA\ and 7300\AA, respectively.  
The \halpha\ emission was detected by the red spectra,
which had a spatial scale and dispersion 
of 0\farcs40 and 0.58 \aapix, respectively.
The DIS spectra were reduced as discussed in \citet{Ho2017}
and followed the same procedures as we reduced the MMT spectra.
The spectral resolution was measured using the arclamps' lines 
as 50 \kms\ FWHM.

\subsubsection{Galaxy Rotation Curves}
\label{sssec:rc}

We generate the galaxy position--velocity (PV) maps by
fitting a Gaussian profile to the galaxy emission lines
at different spatial locations on
the two-dimensional galaxy spectra.
Following \citet{Ho2017},
we convert the PV maps to rotation curve measurements
by deprojecting the position and velocity measured 
along the slit
onto circular orbits on the disk plane.
For a slitlet at an angle $\zeta$ from
the galaxy major axis,
we calculate the galactocentric radius $R$ 
from the projected distance $\rho$ along the slit
using
\begin{equation}
	R = \rho \sqrt{1 + \sin^{2}\zeta \tan^{2}i},
	\label{eq:rdepj}
\end{equation}
where $i$ represents the disk inclination angle.
We convert the line-of-sight velocity $v_\mathrm{los}(\rho)$
to rotation velocity $V_\mathrm{rot}(R)$ on the disk 
by
\begin{equation}
	V_\mathrm{rot}(R) = \frac{\sqrt{1 + \sin^{2}\zeta \tan^{2}i}}{\sin i \cos \zeta} v_\mathrm{los}(\rho).
	\label{eq:vdepj}
\end{equation}
In the middle column of Figure~\ref{fig:imrco6},
we show the line-of-sight velocity $v_\mathrm{los}$ and 
the rotation velocity $V_\mathrm{rot}$ 
of individual galaxies,
which are represented by the cyan solid squares and 
black open circles, respectively.

To model the intrinsic rotation curves of the galaxies,
we define a grid of rotation curves over a range of 
asymptotic rotation velocities $V_\mathrm{\infty}$ 
and turnover radii $R_\mathrm{RC}$ using an arctangent model,
\begin{equation}
    V_\mathrm{model}(R) = \frac{2}{\pi} V_\mathrm{\infty} \arctan{\frac{R}{R_\mathrm{RC}}}.
    \label{eq:arctan}
\end{equation}
For each galaxy, 
we project the models onto the slit, 
convolve these projected models 
with a Gaussian model for the seeing disk,
and compare them to the PV maps
measured from the galaxy emission lines.
We minimize the fit residual 
to estimate \vinf\  and $R_\mathrm{RC}$ 
for individual galaxies.  
In the middle column of Figure~\ref{fig:imrco6},
the seeing-convolved model for each galaxy
before and after the projection are shown 
using the black dashed  and cyan solid curves,
respectively.
Table~\ref{tb:galprops} lists 
the adopted \vinf\ and $R_\mathrm{RC}$ for individual galaxies.

%%%%%%%%%%%%%%%%%%%%% FIG: Image + RC + OVI %%%%%%%%%%%%%%%%%%%%%%%%
\begin{figure*}[htb]
    \centering
    \includegraphics[width=1.0\linewidth]{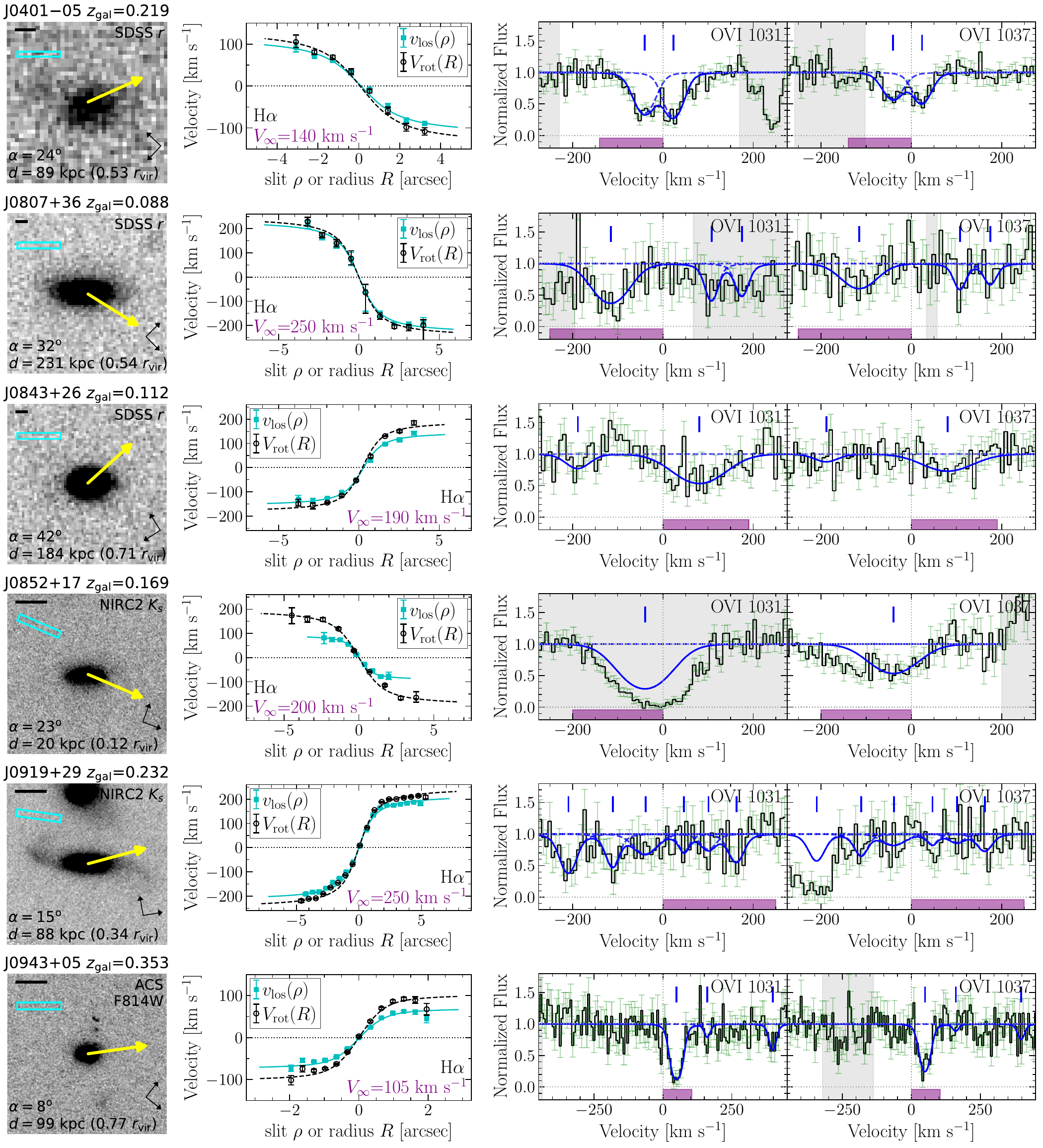}
    \caption{
        Images of sample galaxies, 
        their rotation kinematics, and 
        \oVI\ absorption measured in the quasar sightlines.
        \textit{Left:}
            Images of galaxies.
            The stamps are 10\arcsec$\times$10\arcsec\
            except for the SDSS $r$-band images;
            the black scale bar represents 2\arcsec.
            Images are oriented with the galaxy major axis 
            aligning with the image x-axis,
            and the quasar is at the $+$x direction.
            The yellow arrow points to the direction 
            of the quasar sightline.
            The cyan rectangle shows the 
            slit orientation of galaxy spectroscopic observation.
        \textit{Middle:}
            Position--velocity diagram 
            (cyan solid squares) 
            and derived rotation curve 
            (black open circles)
            of galaxy.
            The bottom labels state the emission line used 
            for deriving the rotation curve (black)
            and the galaxy asymptotic rotation speed \vinf\ (purple).
        \textit{Right:}
            \oVIdb\ absorption lines 
            relative to the galaxy systemic velocity.
            The blue dashed and solid curves 
            show the fitted Voigt profiles
            for individual velocity components and 
            the overall \oVI\ absorption, respectively.
            The blue ticks (top) mark the velocity centroids of 
            individual components.  
            The purple bar (bottom) covers the velocity range between
            the galaxy systemic velocity and  
            the asymptotic rotation velocity at 
            the quasar side of the galaxy major axis.
            Grey shaded regions are masked during Voigt profile fitting 
            due to contamination from intervening absorbers
            at other redshifts.
        }
    \label{fig:imrco6} 
\end{figure*}

\begin{figure*}%[h!]
    \centering
    \figurenum{2}
    \includegraphics[width=1.0\linewidth]{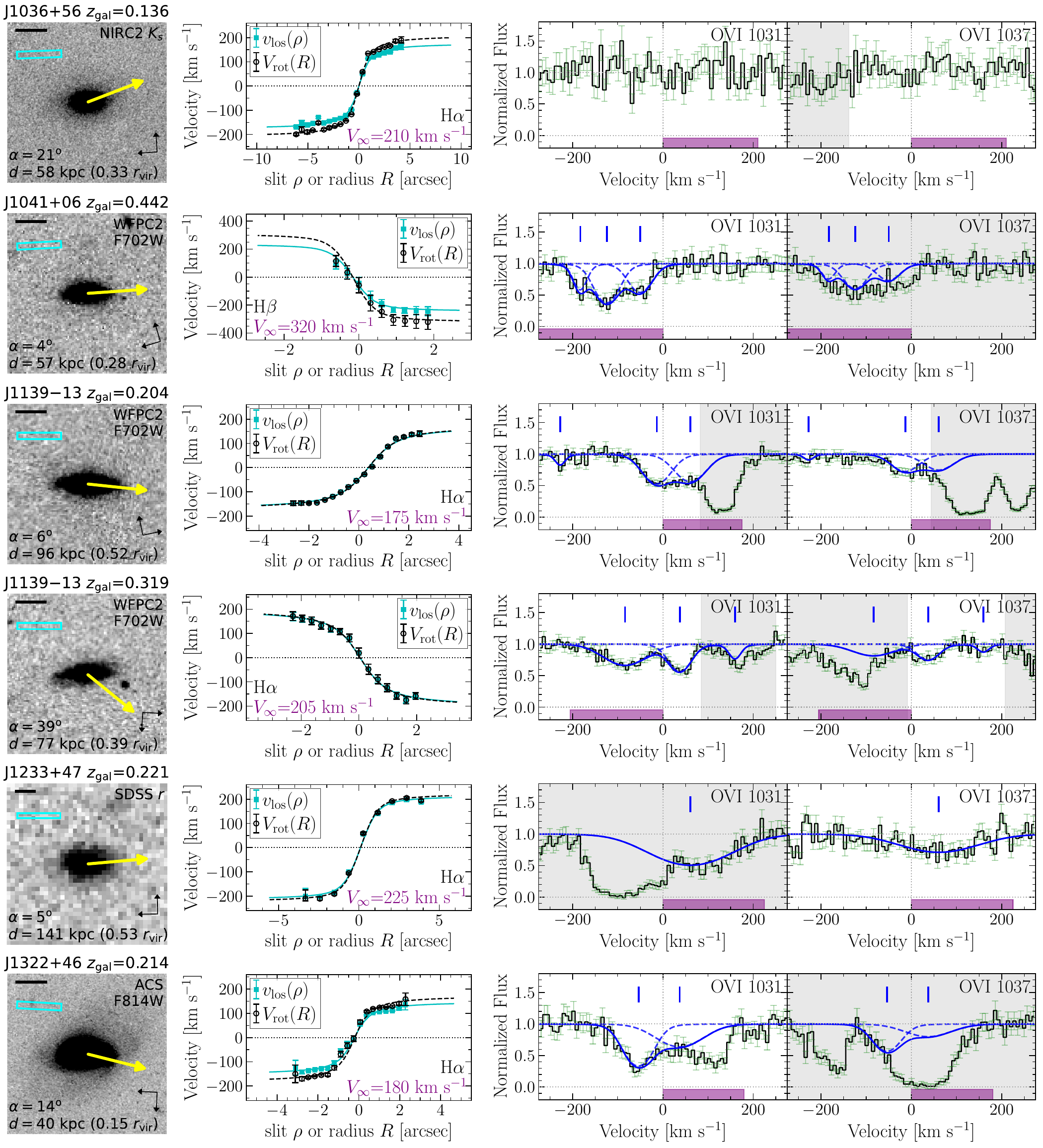}
    \caption{\textit{(Continued)}
            }
\end{figure*}
\begin{figure*}%[h!]
    \centering
    \figurenum{2}
    \includegraphics[width=1.0\linewidth]{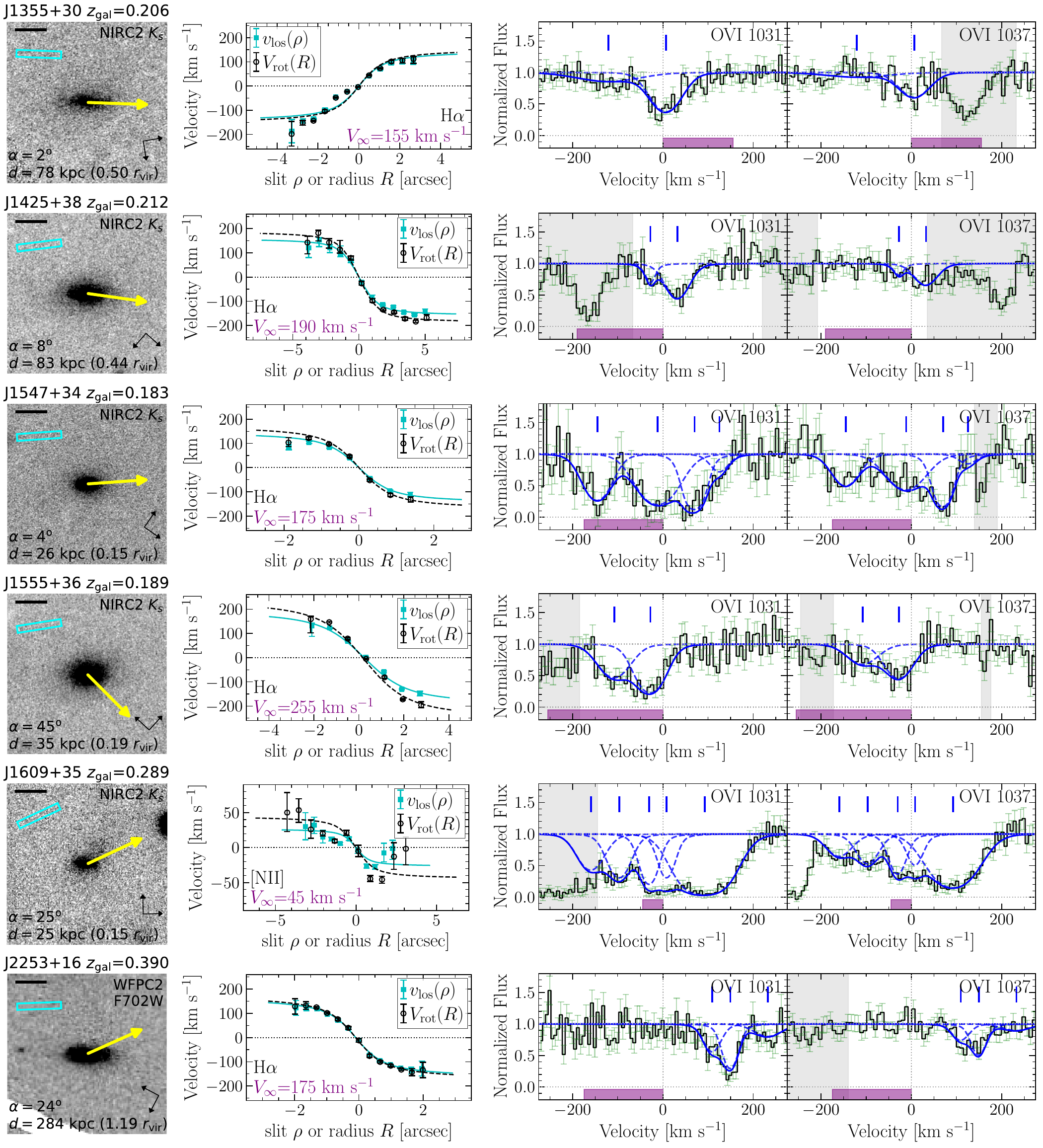}
    \caption{\textit{(Continued)}
            }
\end{figure*}
%%%%%%%%%%%%%%%%%%%%%%%%%%%%%%%%%%%%%%%%%%%%%%%%%%%%%%%%%%%%%%%

%%%%%%%%%%%%%%%%%%%%%%%%
% Galaxy Properties
%%%%%%%%%%%%%%%%%%%%%%%%
\subsection{Galaxy Properties}
\label{ssec:galprop}

We summarize the properties of the selected galaxies in 
Table~\ref{tb:galprops}. 
The systemic redshifts of the galaxies are 
spectroscopically determined from the optical emission lines
either from 
the new MMT spectroscopy 
or from previous work, 
and we specify the references in the Table.
Sections~\ref{sssec:galgeom} and \ref{sssec:galmasses}
describe the galaxy disk geometry and the stellar and halo masses,
respectively.
Then, we discuss the galaxy environments 
in Section~\ref{sssec:galenv}.

% =================================
% TABLE with Galaxy/Sightline properties
\begin{center}
    \input{table_galprop_new.tex}
\end{center}
% =================================

\subsubsection{Galaxy Disk Geometry}
\label{sssec:galgeom}

We adopted measurements of galaxy disk geometry
either from previous work
or from the SDSS DR17 catalog \citep{Blanton2017}.  
Table~\ref{tb:galprops} lists the adopted values and the references
for the disk inclination angles $i$ and 
the azimuthal angles $\alpha$ of the quasar sightlines 
with respect to the galaxy major axes.
Both \citet{Martin2019} and \citet{Kacprzak2019}
measured the galaxy disk geometry 
using high-resolution images:
the former obtained $K_s$-band images from
the Keck/NIRC2 camera with
Laser Guide Star Adaptive Optics system \citep{Wizinowich2006},
and the latter collected archival \textit{HST} images from 
the Advanced Camera for Surveys (ACS) and 
Wide Field and Planetary Camera 2 (WFPC2). 
For the remaining galaxies, 
we used the $r$-band photometry in the SDSS \texttt{PhotoObj} catalog
to derive the azimuthal angle
and the inclination angle from the axis ratio\footnote{
    We applied the Hubble formula \citep{Hubble1926} with $q_0 = 0.2$.
    }.
We show the images of individual galaxies
with the same instruments and bandpasses
used for measuring the parameters of the disk geometry
in Figure~\ref{fig:imrco6}.

\subsubsection{Stellar Masses, Halo Masses, and Virial Radii}
\label{sssec:galmasses}

The stellar masses \mstar\ of individual galaxies were obtained from
previous work and were rescaled 
to the Chabrier initial mass function \citep[IMF;][]{Chabrier2003}
following \citet{MadauDickinson2014}.
Stellar masses extracted from \citet{Martin2019} were 
in Chabrier IMF, 
whereas \citet{Werk2012} assumed the Salpeter IMF \citep{Salpeter1955},
and \citet{Heckman2017} adopted the Kroupa IMF \citep{Kroupa2001}
as in the MPA-JHU Value-Added Galaxy Catalog.  
Then, 
the halo mass \mvir\ of each galaxy was calculated
using the stellar mass--halo mass (SM-HM) relation derived from
abundance matching in \citet{Behroozi2010}.  
The virial radius \rvir\ was obtained following \citet{Ho2017} 
using the overdensity defined by \citet{BryanNorman1998}.

The above procedures were applied for 14 of the 18 galaxies,
for which the stellar masses 
were derived from multi-band galaxy photometry
using the spectral energy distribution fitting technique.
But for four galaxies from \citet{Kacprzak2019}, 
instead of first obtaining the stellar masses and then 
calculate the halo masses,
they derived the halo masses from  
the $r$-band Vega magnitudes of the galaxies
following the method in \citet{Churchill2013magiicat}
\citep[also see][]{Ng2019}.
We used their halo masses and calculated the stellar masses
of the galaxies using the SM-HM relation from \citet{Behroozi2010}.
Then, we derived the virial radii from the halo masses 
following the same procedures as we did for
other galaxies.

Table~\ref{tb:galprops} lists 
the stellar mass, halo mass, and virial radius of 
individual galaxies, 
and we label the references of the masses.
The median \logmstar\ and \logmvir\ are 
10.14 and 11.72, respectively.
The Table also shows
the sightline impact parameter 
normalized by the halo virial radius \brvir.
All except one sightline 
intersect the galaxies within
the virial radius (\brvir $< 1$),
and most of the sightlines lie within 0.5\rvir\
from the galaxies
(also see Figure~\ref{fig:alphab_polar}).

\subsubsection{Galaxy Environments and Potential Interactions}
\label{sssec:galenv}

The group environment of galaxies 
affect their CGM properties,
such as the covering fraction and velocity dispersion,
and the CGM becomes kinematically complex
\citep{Johnson2015,Pointon2017,Nielsen2018,Dutta2025}.
It also leads to uncertainties in identifying 
the host galaxies associated with the absorption 
detected in the sightlines.
Because we generally lack spectroscopic redshifts
for most galaxies in the our fields, 
we cannot classify the group environment
as detailed as in \citet{Johnson2013} and \citet{Chen2020}.
Nevertheless, 
using SDSS and SIMBAD \citep{SIMBAD2020},
we searched for galaxies near our targets
to confirm there were no other galaxies
with spectroscopic or photometric redshifts 
comparable to our target galaxies,
especially for galaxies with similar
$r$-band magnitudes and/or impact parameters 
as our targets.

However, the only exception is 
the target galaxy J091954+291345 at $z=0.23288$.
We found that 
two red galaxies identified by the SHELS galaxy redshift survey 
\citep{Geller2014}
have spectroscopic redshifts similar to our target.
At the northeast of our target galaxy,
a bright red galaxy 4\farcs4 away (17 kpc) has 
a redshift of 0.23208,
i.e., $-240$ \kms\ relative to our galaxy.
The other fainter, red galaxy has 
a redshift of 0.23179, i.e., $-327$ \kms,
at 8\farcs7 (33 kpc) south of our target.
Hence, these three galaxies form a group with 
velocity differences within a few hundred \kms.
While we present the spectroscopic measurements for this target galaxy
and the absorption lines measured in the quasar sightline,
we exclude this system from the CGM kinematics analyses
in later sections.

For galaxy J160951+353838, 
it shows a tail-like structure towards the northeast
(also see \citealt{Ho2017}).
This tail can be seen towards the upper right 
in the galaxy image stamp in Figure~\ref{fig:imrco6}.
The galaxy rotation curve also shows 
irregularity in the same direction.
The presence of the tail and the galaxy kinematics
together suggest interaction with another galaxy.
However, we do not find any bright galaxies with comparable 
photometric redshift that could have
interacted with our target galaxy within its virial radius.
While the presence of a less massive galaxy or satellite
potentially explains the irregular galaxy morphology and kinematics,
because our target is not in a close group as J091954+291345, 
we do not exclude this galaxy in the CGM kinematics analysis.

%%%%%%%%%%%%%%%%%%%%%%%%
% HST/COS
%%%%%%%%%%%%%%%%%%%%%%%%
\subsection{Quasar Spectroscopy with HST/COS}
\label{ssec:hstcos}

% =================================
% TABLE with HST/COS observations [Quasar Spectroscopy]
\begin{center}
    \input{table_obsqso.tex}
\end{center}
% =================================

The far-UV (FUV) spectroscopy of the background quasars of individual
galaxy--quasar pairs were either 
from our \textit{HST/COS} program (ID: 15866, PI: Ho) 
or from archival COS data.
These COS spectra were obtained
using the medium-resolution grating,
G130M and/or G160M, with a spectral resolution of $R\approx20,000$.  
Table~\ref{tb:obs_qso} lists the 
details of the quasar observations with COS.

For each quasar, 
the raw data from COS were processed by the 
\texttt{CALCOS} pipeline (v3.3.9), 
and the one-dimensional spectra (\texttt{x1d} files) were 
retrieved from MAST\footnote{\url{http://archive.stsci.edu}}.  
Custom software was used to 
align the one-dimensional spectra 
using the Milky Way absorption lines
and coadd the aligned spectra.
Because COS FUV spectra were oversampled, the coadded spectra 
were binned by three pixels to increase the signal-to-noise ratio.  
Finally, 
each coadded spectrum was continuum-normalized
by fitting the quasar continuum 
with an Akime Spline
at the absorption-free regions
using the 
\texttt{linetools}\footnote{\url{https://github.com/linetools/linetools}}
python package
\citep{linetoolsZenodo}.

%%%%%%%%%%%%%%%%%%%%%%%%
% Absorption Lines
%%%%%%%%%%%%%%%%%%%%%%%%
\subsection{Absorption-line Measurements: 
Voigt Profile Fitting and Column-density-weighted Velocity}
\label{ssec:abplines}

%%% Voigt Profile Fitting
We identified absorption-line systems within $\pm500$ \kms\
of the systemic velocity of the target galaxies.
We modeled individual velocity components with Voigt profiles
and adopted the atomic data from \citet{Morton2003}.  
The modeled Voigt profile characterized 
the velocity centroid $v$, 
the line width, i.e., the Doppler parameter $b$,
and the column density $N$
of individual velocity component.
Using the \texttt{VoigtFit} python package \citep{Krogager2018},
we convolved the models  
with the COS line-spread function
corresponding to the COS lifetime position
when the observation was conducted.
We compared the convolved line profiles to the data,
and different transitions of the same ionic species 
were fitted jointly.  
For each ionic species, 
we created different sets of models 
with different number of velocity components,
and the best-fit solution of each model set
was determined using a $\chi^2$ minimization algorithm.
Finally, we adopted the best-fit model that 
required the minimum number of velocity components
to produce a reasonable fit and with 
the reduced-$\chi^2$ value closed to one.

The right column of Figure~\ref{fig:imrco6}
shows the \oVI\ absorption lines
relative to the systemic redshifts
of individual target galaxies.
The fitted Voigt profiles of individual velocity components
and the overall \oVI\ absorption are shown
in blue dashed and solid curves, respectively,
and the blue ticks (top) mark the velocity centroids of 
individual components.  
Table~\ref{tb:voigt_o6} lists the 
parameters of the Voigt profiles
of the fitted \oVI\ absorption.\footnote{
    The Figures of the fitted line profiles of other ions
    and Voigt profile parameters
    are shown in Figure~\ref{fig:voigtfitset}
    and 
    Table~\ref{tb:voigt_all}, respectively, in the Appendix.
    }
Only the J103640$+$565125 sightline
did not detect \oVI\ around 
the target galaxy.  
In that case, we list the 
$2\sigma$ column density upper limit;
this was derived from 
the linear section of the curve-of-growth
and the upper limit 
of the rest-frame equivalent width of \oVI\ 1031,
measured using a velocity window 
that was three times the resolution element
centered at the galaxy systemic velocity.

It is worth noting that in three sightlines,
both \oVIdb\ transitions include
intervening absorption
at the same velocity range 
relative to the systemic velocity of our target galaxies.  
At +140 \kms\ from the $z = 0.204194$ galaxy 
in the J113910$-$135043 sightline,
the \oVIA\ absorption is overlapped with 
the \lyb\ absorption at $z=0.212$,
whereas \oVIB\ is blended with 
\sII\ $\lambda1250$ of the Milky Way (MW) absorption,
\lygamma\ at $z=0.317$, and 
\lydelta\ at $z=0.333$.
We masked \oVIB\ due to the complexity of 
absorption from multiple species at different redshifts.  
As for \oVIA, 
we first attempted to perform Voigt profile fitting
together with the \hI\ absorption at $z=0.212$, 
but that resulted in huge uncertainties 
in the Voigt profile parameters.  
So instead, we fitted the \hI\ lines at $z=0.212$ alone 
(from \lya\ to \lydelta),
through which we determined that 
the $z=0.212$ \lyb\ absorption fully accounted for 
the +140 \kms\ absorption near the \oVIA\ line 
of our target galaxy.
We then fixed the Voigt profile parameters 
of the $z=0.212$ \lyb\ absorption line
while fitting the \oVIdb\ doublet for our target galaxy.
For the J132222+464546 sightline, 
the absorption at +100 \kms\ of \oVIdb\
is blended with the
\sII\ $\lambda$1253 and \siII\ $\lambda$1260 absorption 
of the MW, respectively.   
We masked the absorption of \oVIB\ and MW \siII\ $\lambda$1260
due to the saturation of the MW \siII\ absorption lines.
Then, we fitted the \oVIA\ 
and the MW \sII\ $\lambda\lambda$1253, 1259 absorption together,
for which the MW \sII\ $\lambda$1253 fully accounted for 
the extra absorption at +100 \kms\ from our galaxy.
As for the J160951+353843 sightline, 
while the \oVI\ doublet showed absorption at $-$250 \kms, 
that part of the absorption in the \oVIB\ panel 
was identified as the \cII\ $\lambda$1036 absorption 
of our target galaxy.
Nevertheless, for all three cases, 
if part of the intervening absorption 
actually includes \oVI\ contribution from our target galaxies, 
then the bulk of \oVI\ absorption would become more Doppler shifted 
towards the same direction as the disk rotation,
which thereby strengthened our conclusion that
the bulk of \oVI\ gas rarely counter-rotates
(Section~\ref{sec:result}).

% =================================
% Table for Voigt Profile Fitting [OVI only]
\begin{center}
    \input{table_voigt_o6_new.tex}
\end{center}
% =================================

%%% COLDENSWT VELOCITY
For each sightline, 
we calculate the column-density-weighted velocity $\langle v \rangle_w$
of the overall absorption.
Using the results from Voigt profile fitting, 
we perturb the fitted velocity centroid and 
column density of each velocity component by 
adding a random value drawn from
a Gaussian distribution with standard deviation
equals to
their 1-$\sigma$ error bars,
and we repeat this process 1000 times.  
During the $i$-th iteration,
we determine the weighted velocity $\langle v \rangle_{w,i}$ by\\
\begin{equation}
    \langle v \rangle_{w,i} = \frac{\sum_j N_j v_j}{\sum_j N_j},
    \label{eq:coldenswtvel}
\end{equation}
where $j$ indicates the $j$-th velocity component,
and $v_j$ and $N_j$ represent 
the perturbed velocity centroid and column density, respectively.
From the resultant distribution of the weighted velocity,
we adopt the median as the column-density-weighted velocity 
and the 68\% confidence interval as the uncertainties.
We show the \oVI\ column-density-weighted velocity \velwovi\
for each sightline 
in the rightmost column of Table~\ref{tb:voigt_o6}.

%%%%%%%%%%%%%%%%%%%%%%%%%%%%%%%%%%%%%%%%
% Results: O VI Measurements + RC comparison
%%%%%%%%%%%%%%%%%%%%%%%%%%%%%%%%%%%%%%%%
\section{CGM Kinematics and Galaxy Disk Rotation}
\label{sec:result}

With our sample of quasar sightlines
near the major axes of galaxies,
we examine the relationship between 
CGM kinematics and galaxy disk rotation.
Section~\ref{ssec:ovikin} and Section~\ref{ssec:lowionkin}
focus on the \oVI\ CGM and 
the low-ionization-state gas, respectively.
In Section~\ref{ssec:lowhighkin},
we then compare 
the kinematics of the low ions and \oVI\ detected
in individual sightlines.
In our analyses, 
we exclude galaxy J091954+291345 
due to its close group environment (Section~\ref{sssec:galenv})
and focus on the remaining 17 galaxy--quasar pairs.

%%%%%%%%%%%%%%%%%%%%%%%
% OVI Kinematics
%%%%%%%%%%%%%%%%%%%%%%%
\subsection{Kinematics of \oVI\ Gas}
\label{ssec:ovikin}

\begin{figure*}[thb]
    \centering
    \includegraphics[width=1.0\linewidth]{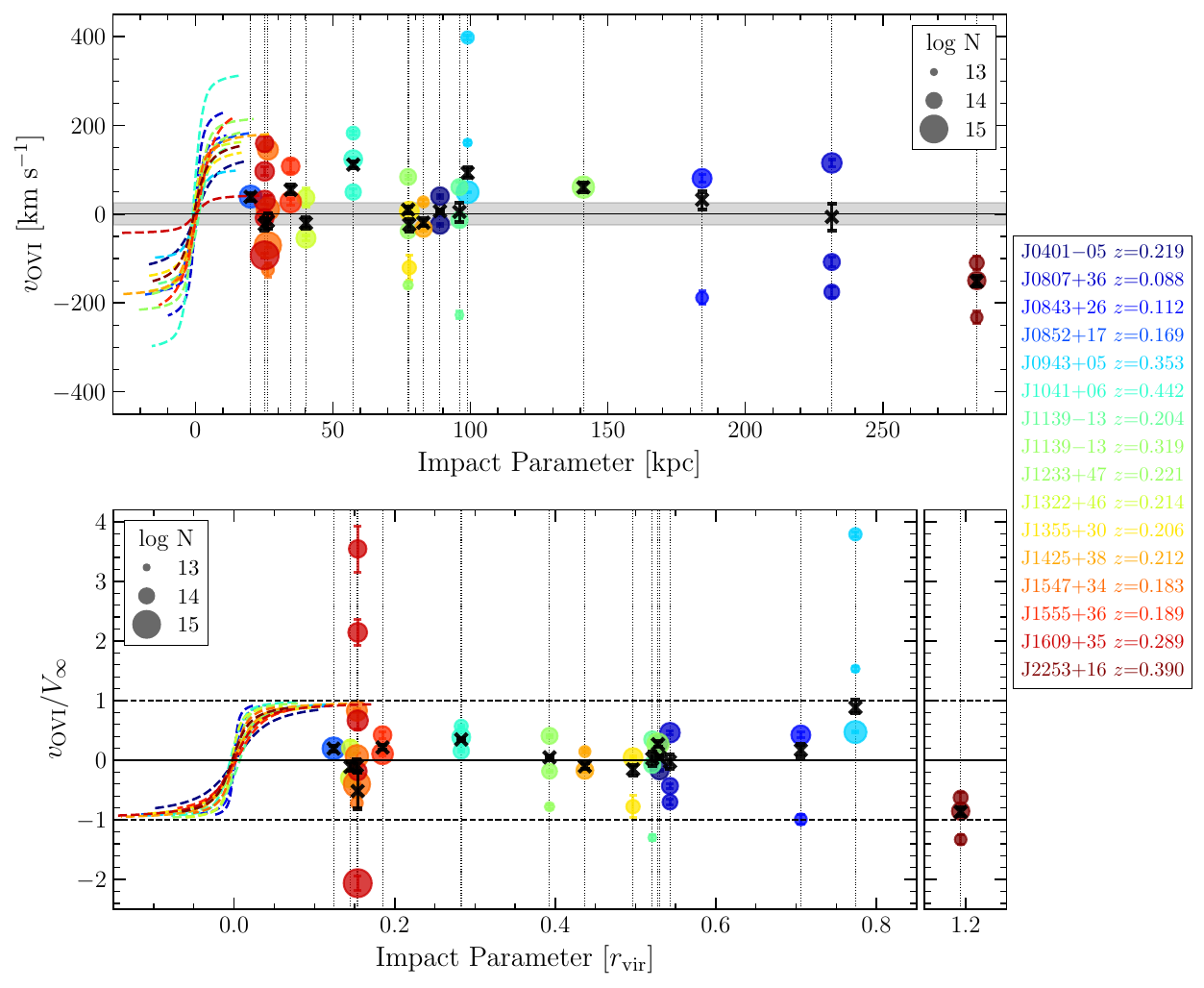}
    \caption{Comparison of \oVI\ gas kinematics to galaxy disk rotation.  
                The curves represent the rotation curves 
                of individual galaxies.
                Each circle indicates the velocity centroid of 
                individual \oVI\ velocity component, 
                and the marker size 
                scales with the column density.
                For the purpose of uniformity in this figure,
                galaxy rotation curves are defined 
                with a positive velocity
                at the quasar side of the galaxy major axis.  
                In the top panel, the vertical and horizontal axes show
                the line-of-sight \oVI\ velocity $v_\mathrm{O\ VI}$ and
                quasar sightline impact parameter,
                and the bottom panel normalized  
                the axes 
                by the galaxy asymptotic rotation speed \vinf\ and 
                virial radius \rvir, respectively.
                The grey shaded region 
                indicates the velocity range 
                where the gas is 
                indistinguishable from being at the 
                systemic velocity of the galaxy ($\pm 25$ \kms).
                Each black cross shows
                the column-density-weighted \oVI\ velocity \velwovi\
                of each sightline,
                and the associated error is typically 
                comparable to
                the size of the cross.
                The only net counter-rotating \oVI\ absorber
                (i.e., \velwovi\ and disk rotation have
                opposite Doppler sign 
                and \absvelwovi $> 25$ \kms)
                is detected at the sightline 
                with the highest impact parameter
                of 284 kpc (1.19\rvir).
                }
    \label{fig:o6vb}
\end{figure*}

Among the 17 galaxy--quasar pairs,
all sightlines except J103640+565125 have detected \oVI\
around the host galaxies.
Figure~\ref{fig:o6vb} compares the kinematics of the \oVI\ gas
and that of the galaxy disk rotation.
The circles and the dashed curves  
represent the velocity centroids of
individual \oVI\ velocity components
and the galaxy rotation curves, respectively.
With similar numbers of \oVI\ velocity components that
corotate (21) and counter-rotate (15) 
with the disk (Table~\ref{tb:o6class}),
it seems to suggest 
a lack of correlation between disk rotation
and the kinematics of individual \oVI\ velocity components.
However, except for the sightline 
at the highest impact parameter of 284 kpc, 
the column-density-weighted velocities (black crosses)
suggest that the bulk of \oVI\ gas either corotates with the disk
or is consistent with being at the systemic velocity of 
the host galaxies
(within the grey band).\footnote{
    Velocities smaller than the measurement uncertainties
    of the galaxy redshift of 25 \kms\
    are considered as being consistent with the
    galaxy systemic velocity.}
In other words, we rarely found sightlines with 
net counter-rotating \oVI\ absorbers.

In the bottom panel of Figure~\ref{fig:o6vb}, 
we normalized
the velocities and impact parameters
by the asymptotic rotation speeds and virial radii 
of the galaxies, respectively.
The velocity centroids of most \oVI\ components (34 out of 41)
are smaller than the asymptotic rotation speed,
i.e., $|v_\mathrm{O\ VI}/V_\mathrm{\infty}| < 1$.
Sightlines with \oVI\ components exceeding \vinf\
typically have large impact parameters of $\geq$0.5\rvir.
The only exception is 
the sightline around host galaxy J160951+353838,
but this galaxy has a tail-like structure 
and an abnormally low asymptotic rotation speed
of 45 \kms\ (Section~\ref{sssec:galenv}).
This explains the high $|v_\mathrm{O\ VI}/V_\mathrm{\infty}|$ values
even though the measured \oVI\ velocities in \kms\ are comparable 
to those in other sightlines.

%%%%%%%%%%%%%%%%%% FIG: OVI Doppler vs. b %%%%%%%%%%%%%%%%%%%%%
%%%%%%%%%%%%%%%%%%%%%%%%%%%%%%%%%%%%%%%%%%%%%%%%%%%%%%%%%%%%%%%

% =================================
% TABLE: Numbers of OVI corot/sys/counter (+matching low ions)
\begin{center}
    \input{table_o6classification.tex}
\end{center}
% =================================

%%%%%%%%%%%%%%%%%%%%%%%
% LIS Kinematics
%%%%%%%%%%%%%%%%%%%%%%%
\subsection{Corotation of the Low-ionization-state Gas
with the Disk}
\label{ssec:lowionkin}

Low-ionization-state ions
trace cooler gas compared to \oVI\ 
due to their lower ionization potentials (IP),
e.g., \siII\ and \siIII\ with IP of 8.2 eV and 16.3 eV,
respectively, compared to 113.9 eV for producing \oVI.
For our sample, 
only 13 sightlines have
COS spectra that cover  
the redshifted wavelengths
of \siII\footnote{
    We focus on the \siIIdb\ and \siIIlam\ transitions,
    because other \siII\ transition lines
    within the spectral coverage 
    are either too weak to be detected or
    blended with other absorption lines.
    }
and \siIII\ of the host galaxies,
including the J103640+565125 sightline without \oVI\ detection.
Among these 13 sightlines,
\siIII\ exhibits a higher detection rate compared to \siII, 
with 7 and 11 sightlines
showing \siII\ and \siIII\ detection, respectively.

\begin{figure*}[htb]
    \centering
    \includegraphics[width=1.0\linewidth]{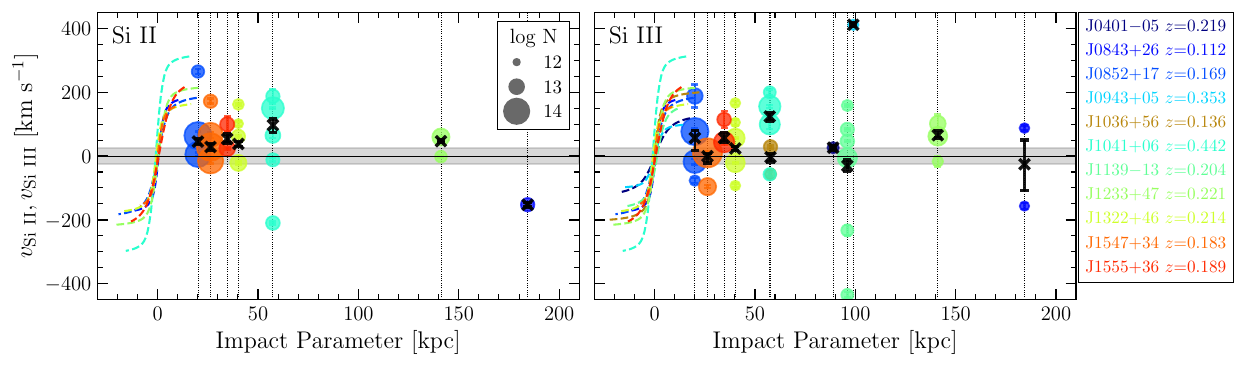}
    \caption{Comparison of \siII\ and \siIII\ 
                gas kinematics to galaxy disk rotation.
                This figure is analogous to the top panel 
                of Figure~\ref{fig:o6vb},
                but each circle in the left and right panels
                represents the velocity centroid of
                individual  \siII\ and \siIII\ velocity component,
                respectively.
                Different colors correspond to individual sightlines
                the same way as in Figure~\ref{fig:o6vb}.
                J1036+56 is added; 
                it is not shown in Figure~\ref{fig:o6vb} 
                due to \oVI\ non-detection.
                Among the 13 sightlines with spectral coverage
                of \siII\ and \siIII,
                7 and 11 sightlines have detected 
                \siII\ and \siIII, respectively.
                Net counter-rotating \siII\ and \siIII\ absorbers
                are rare.
                As for individual velocity components,
                while very few of them counter-rotate with the disk,
                \siIII, which has a higher ionization than \siII, 
                shows a higher counter-rotation rate. 
                }
    \label{fig:si23vb}
\end{figure*}

Figure~\ref{fig:si23vb} compares the kinematics
of the \siII\ (left) and \siIII\ (right) gas
to disk rotation.
Except for the sightline 
at 184 kpc from galaxy J084356+261855,
six out of the seven \siII-detected sightlines
show net corotating \siII\ absorption,
and only two (out of 21) \siII\ velocity components
counter-rotate with the disk.
Net counter-rotating \siIII\ absorbers are also rare,
but the fraction of \siIII\ velocity components 
that counter-rotate with the disk 
becomes higher
compared to that of \siII.

%%%%%%%%%%%%%%%%%% FIG: SiII/SiIII Doppler vs. b %%%%%%%%%%%%%%%%%%%%%
%%%%%%%%%%%%%%%%%%%%%%%%%%%%%%%%%%%%%%%%%%%%%%%%%%%%%%%%%%%%%%%

%%%%%%%%%%%%%%%%%%%%%%%
% Compare/Match with low-ions
%%%%%%%%%%%%%%%%%%%%%%%
\subsection{Comparison Between \oVI\ and Low-ionization-state Gas Kinematics}
\label{ssec:lowhighkin}

\begin{figure}[htb]
    \centering
    \includegraphics[width=0.9\linewidth]{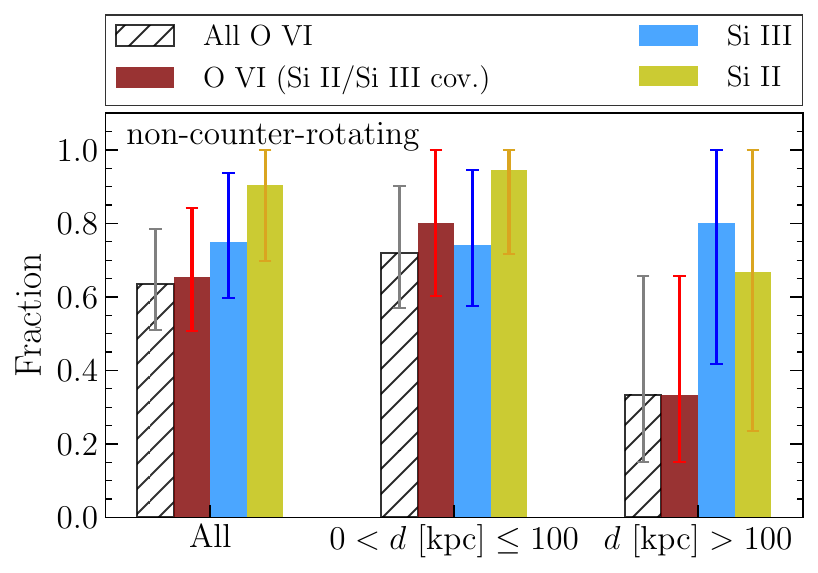}
    \caption{Fraction of non-counter-rotating
                velocity components of \siII, \siIII, and \oVI\
                compared to the galaxy disk rotation.
                Different colors represent different ions;
                ``All \oVI'' represents all \oVI\ velocity components,
                whereas ``\oVI\ (\siII/\siIII\ cov.)'' 
                only shows the \oVI\ components in the 13 sightlines 
                with \siII\ and \siIII\ spectral coverage.
                Regardless of the sightline impact parameter (left), 
                \siII\ shows the highest fraction of 
                non-counter-rotating velocity components. 
                By dividing the sightlines into 
                impact parameter bins of below (middle) 
                and above 100 kpc (right),
                \oVI\ shows a higher non-counter-rotating fraction 
                in sightlines with smaller impact parameters
                compared to sightlines further away.
                }
    \label{fig:kinhist}
\end{figure}

Sections~\ref{ssec:ovikin} and \ref{ssec:lowionkin}
have shown that
the low-ionization-state gas traced by \siII\ and \siIII\
rarely counter-rotate with the disk compared 
to \oVI,
suggesting that \oVI\ gas
exhibits different kinematics compared to 
the lower ionization counterparts.
In Figure~\ref{fig:kinhist},
we show
the fraction of non-counter-rotating velocity components
of \siII, \siIII, and \oVI,
i.e., the gas either corotates with the disk
or consistent with being at the systemic velocity 
of the host galaxy.\footnote{
    Both black hatched and maroon filled histograms represent \oVI,
    but the former includes all sightlines, 
    whereas the latter only includes the 13 sightlines
    with \siII\ and \siIII\ spectral coverage.
    }
The error bars 
represent the Poisson confidence limits calculated from 
the number of non-counter-rotating velocity components,
and then divided by the total number of velocity components
of the same ion.
Considering all sightlines 
regardless of their impact parameters (``All''), 
\siII\ has the highest fraction of 
non-counter-rotating velocity components of 90\%.
The fraction drops with increasing ionization potential,
for which only 75\% (60\%) of the \siIII\ (\oVI) velocity components
are not counter-rotating.
If we only focus on the sightlines 
with impact parameters below 100 kpc, 
the fraction of non-counter-rotating velocity components
for all three ions
become comparable  (between 75\% and 95\%).  
However, in sightlines beyond 100 kpc from the host galaxies,
\oVI\ shows a lower non-counter-rotating fraction
compared to \siII\ and \siIII\ in the same impact parameter range
and also compared to the \oVI\ detected at small impact parameters.
Therefore, 
in contrast to low-ionization-state gas
that rarely counter-rotates with the disk, 
whether individual velocity components of the \oVI\ gas
counter-rotate possibly
depends on the sightline impact parameter, 
with sightlines at larger impact parameters
more frequently detect counter-rotating \oVI\ gas.

%%%%%%%%%%%%%%%%%% FIG: HIST: Non-counter-rotate %%%%%%%%%%%%%%%%%%%%%
%%%%%%%%%%%%%%%%%%%%%%%%%%%%%%%%%%%%%%%%%%%%%%%%%%%%%%%%%%%%%%%

% %%%%%%%%%%%%%%%%%%%%% FIG: OVI-SiIII-SiII Voigt %%%%%%%%%%%%%%%%%%%%%%%%
\begin{figure*}[htb]
    \centering
    \includegraphics[width=1.0\linewidth]{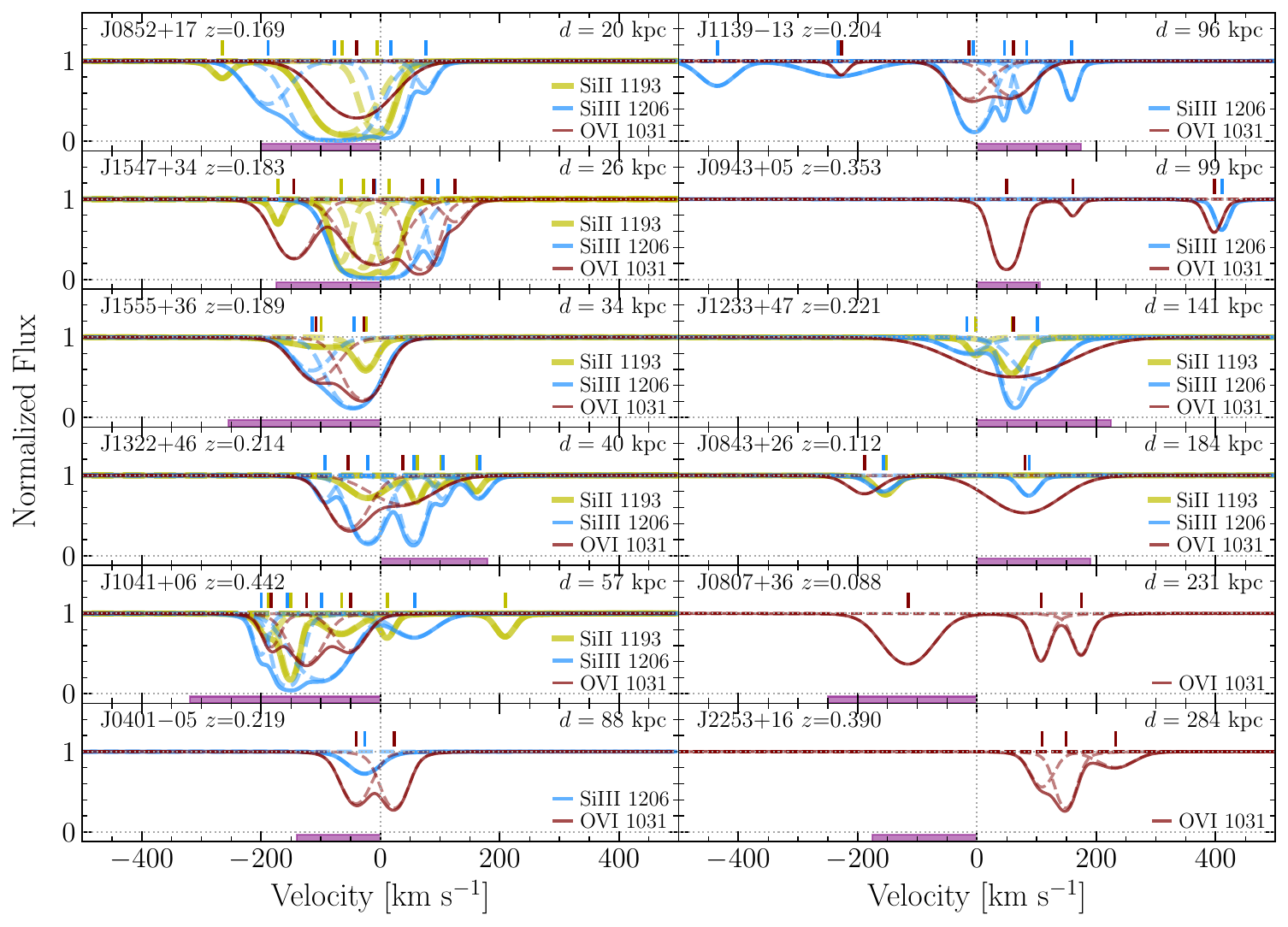}
    \caption{Comparison of Voigt profile fits for
                all 12 sightlines with \oVI\ (thin maroon) detected 
                and simultaneously with spectral coverage for
                \siII\ (thickest yellow) and \siIII\ (thick cyan).
                Panels are arranged in increasing 
                sightline impact parameter
                (upper right label).
                Dashed and solid lines 
                represent individual velocity components
                and the overall fitted Voigt profile, respectively.
                Similar to Figure~\ref{fig:imrco6}, 
                ticks at the top mark 
                the velocity centroids of individual components.
                The purple bar at the bottom covers 
                the velocity range between
                the galaxy systemic velocity and  
                the asymptotic rotation velocity at 
                the quasar side of the galaxy major axis.
                Despite the complex line profile structures,
                the overall absorption of different ions
                covers similar velocity ranges.
                }
    \label{fig:o6_si3_si2}
\end{figure*}
% %%%%%%%%%%%%%%%%%%%%%%%%%%%%%%%%%%%%%%%%%%%%%%%%%%%%%%%%%%%%%%%

Figure~\ref{fig:o6_si3_si2} compares
the Voigt profile fits of \oVI\ to those of \siII\ and \siIII\ 
measured in individual sightlines.\footnote{
    For the J154741$+$343357 sightline,
    the \siIII\ absorption-line profile is
    incomplete, because it falls at the reddest end 
    of the spectrum. 
    The J103640+565125 sightline is not shown due to 
    \oVI\ non-detection.
    }
We arrange the panels in increasing impact parameter
of the quasar sightlines.
First,
\siII\ and \siIII\ are more frequently detected
in sightlines at low impact parameters, 
and the two sightlines 
beyond 200 kpc from the host galaxies 
neither detect \siII\ nor \siIII.
Secondly, the line profiles of different ions
have complex component structures,
especially sightlines closer to the galaxies
tend to have more velocity components.
While some \oVI\ components have velocities comparable 
to their low-ionization-state counterparts, 
some do not affiliate with the low ions.
But regardless, the overall absorption
of different ions typically occupy similar velocity ranges.

% %%%%%%%%% FIG: OVI Doppler vs. b (highlight SiII, SiIII) %%%%%%%%%%%%
\begin{figure*}[htb]
    \centering
    \includegraphics[width=1.0\linewidth]{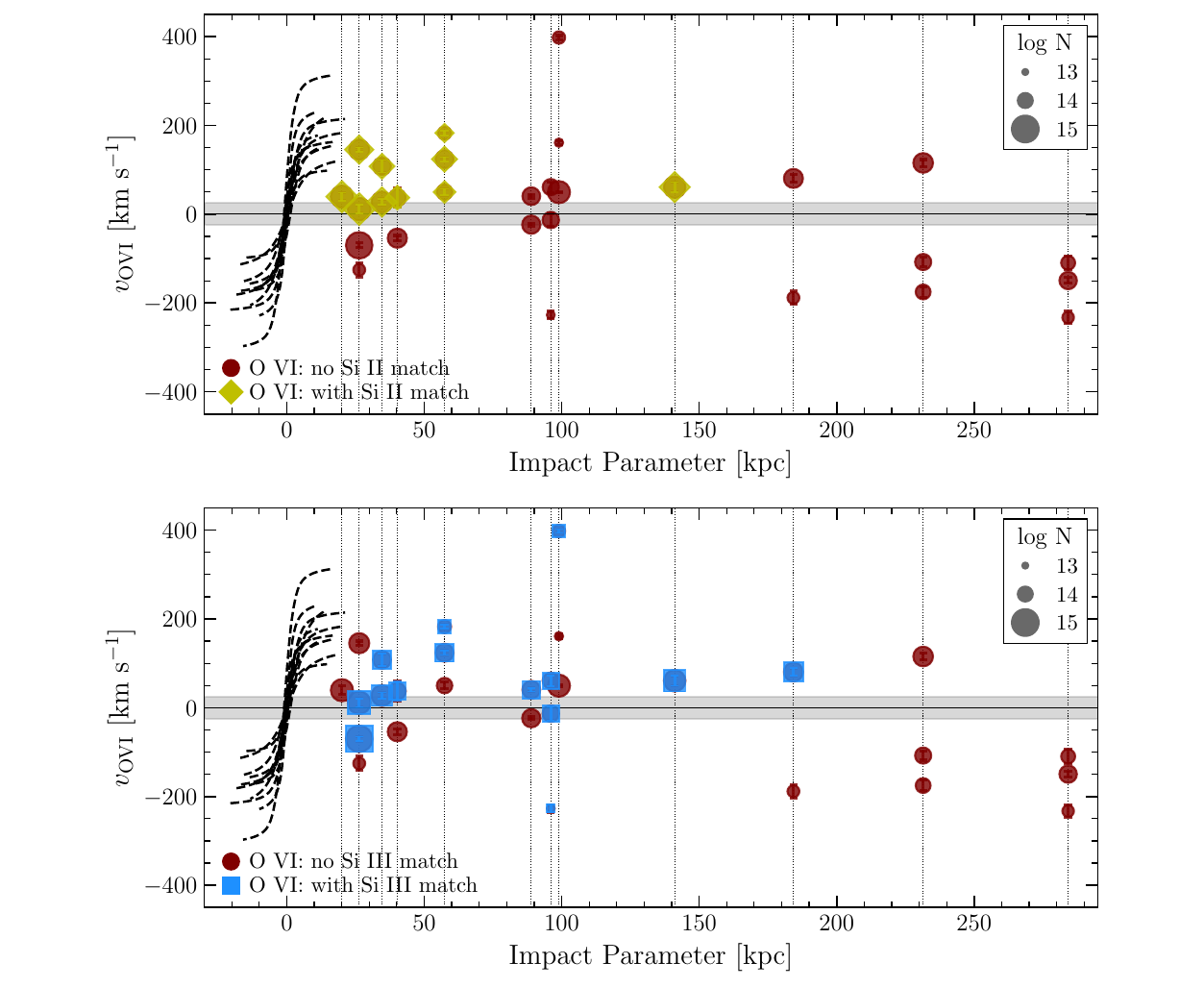}
    \caption{Comparison of \oVI\ gas kinematics 
                with \siII- and \siIII-matched velocity components
                to galaxy disk rotation. 
                Similar to the top panel of Figure~\ref{fig:o6vb}, 
                but this Figure only shows the sightlines with 
                \siII\ and \siIII\ spectral coverage.
                In the top (bottom) panel, 
                the \oVI\ velocity components successfully matched 
                with \siII\ (\siIII) are highlighted in
                yellow diamonds (cyan squares).
                Maroon circles represent the 
                unmatched \oVI\ velocity components.
                Most of the \siII- or \siIII-matched 
                \oVI\ velocity components 
                corotate with the galaxy disk,
                and none of the \siII-matched components
                counter-rotates.  
                }
    \label{fig:o6vb_o6si2}
\end{figure*}
% %%%%%%%%%%%%%%%%%%%%%%%%%%%%%%%%%%%%%%%%%%%%%%%%%%%%%%%%%%%%%%%

Motivated by Figure~\ref{fig:o6_si3_si2} 
that shows that only some \oVI\  components
have velocities comparable to those of the low ions,
we examine how the kinematics association 
between \oVI\ and low ions (or the lackthereof) 
relates to the comparison of \oVI\ kinematics 
and disk rotation.
In each sightline, 
we attempt to match individual \oVI\ velocity components
with those of \siII\ and \siIII\
by minimizing the difference of their velocity centroids.
We consider an \oVI\ velocity component 
to be successfully matched with a \siII\ or \siIII\ component
if the velocity difference is under 30 \kms,
a threshold that includes the uncertainties 
in defining the galaxy systemic velocity, 
the wavelength calibration of the COS spectra\footnote{
    According to the ``Cycle 26 COS/FUV 
    Spectroscopic Sensitivity Monitor (ISR 2020-06)'', 
    the COS wavelength zeropoint uncertainties are 3 pixels, 
    i.e., $\sim10$ \kms.
    },
and the typical uncertainty of the velocity centroid
from Voigt profile fitting.\footnote{
    While increasing (decreasing) the 30 \kms\ threshold value
    will increase (reduce) the number of low-ion matched \oVI\ 
    velocity components, 
    it does not affect the main conclusions
    from the kinematics comparison
    between \oVI\ components with and without low-ion matches.
    }
Each \siII\ or \siIII\ velocity component can only be matched
with one \oVI\ component (and vice versa).
Hence, if more than one \siII\ or \siIII\ component satisfy 
the matching criteria with \oVI, 
then we select the one with the smallest velocity difference.

Analogous to Figure~\ref{fig:o6vb},
Figure~\ref{fig:o6vb_o6si2} compares the \oVI\ kinematics 
with galaxy disk rotation
by showing the 12 sightlines with \oVI\ detection 
and spectral coverage of \siII\ and \siIII.
Each marker represents the velocity centroid
of an \oVI\ component, 
and the yellow diamonds (top) and cyan squares (bottom)
highlight the \siII- and \siIII-matched
\oVI\ components, respectively.
Both panels clearly show that
\siII- or \siIII-matched \oVI\ velocity components
rarely counter-rotate with the galaxy disk
(also see Table~\ref{tb:o6class}).
In fact, 
none of the \siII-matched \oVI\ velocity components 
counter-rotates (0/10),
in contrast to the 10/19 (53\%) no-\siII-matched \oVI\ components 
that counter-rotate with the disk.  
While counter-rotating \siIII-matched \oVI\ components exist 
(2/14; 14\%),
that occurs less frequently compared to the 
\oVI\ components without a \siIII\ matched
(8/15; 53\%).
The fact that 
half of the unmatched \oVI\ components 
counter-rotate
also implies a true distinction in the \oVI\ kinematics
between velocity components with and without low-ion matches.
Consider a hypothetical scenario that 
\oVI\ is randomly moving and has no correlation 
with the kinematics of the low ions at all. 
Then we should find
corotating and counter-rotating \oVI\ components 
each at 50\% of the time.
When we match \oVI\ components with \siII\ and \siIII\
by minimizing their velocity differences,
we will find several \oVI\ components
aligning with the low ions by chance. 
Because low ions tend to corotate with the disk 
(Section~\ref{ssec:lowionkin}),
the \oVI\ components with low-ion matches
will also tend to corotate, 
whereas the remaining unmatched \oVI\ components 
will generally counter-rotate.
However, our measurements show that 
only about 50\% of the unmatched \oVI\ components 
counter-rotate.
Therefore, this invalidates the hypothetical scenario,
indicating that 
the successful matches between \oVI\ and low ions 
did not occur by chance.  
In other words, 
while the low-ion matched \oVI\ components
rarely counter-rotate,
the unmatched \oVI\ components
do not preferentially corotate nor counter-rotate
with the disk.

Not only the \siII- and \siIII-matched \oVI\ components 
rarely counter-rotate,
but they 
are also more likely to 
corotate with the disk
compared to the unmatched \oVI\ velocity components.
Specifically, 9/10 (90\%) of the \siII-matched \oVI\ components
corotate with the disk 
in contrast to only 7/19 (37\%) of the \oVI\ components
without a \siII\ match.
The contrast becomes less prominent for \siIII. 
But still, a higher fraction (10/14; 71\%)
of the \siIII-matched \oVI\ components
corotate with the disk
compared to that of the
no-\siIII-matched \oVI\ components (6/15; 40\%).
Our results show that the \oVI\ velocity components
with and without low-ion matches
have different kinematics properties
and possibly trace gas that resides at 
different 3D distances from the host galaxies.
Because \oVI\ velocity components with matched low-ion
are more likely to corotate with the disk,
a plausible scenario is that the gas
resides on an extended disk plane.
We further discuss the implications of our results in 
Section~\ref{sec:discussion}.

%%%%%%%%%%%%%%%%%%%%%%%%%%%%%%%%%%%%%%%%
% Discussion
%%%%%%%%%%%%%%%%%%%%%%%%%%%%%%%%%%%%%%%%5
\section{Discussion}
\label{sec:discussion}

In Section~\ref{sec:result},
we have presented the kinematics of the \oVI\ gas
and compared that with galaxy disk rotation
and the kinematics of low-ionization-state gas
traced by \siII\ and \siIII.
While individual velocity components of
the \oVI\ gas do not seem to 
correlate with the galaxy disk rotation,
the column-density-weighted velocity of \oVI\ 
suggests that the bulk of \oVI\ gas rarely counter-rotate
with the disk
(Figure~\ref{fig:o6vb}).
We have also matched velocity components of \oVI\ 
with \siII\ and \siIII\
by minimizing the differences 
of their velocity centroids.
This analysis has shown that 
the \siII- or \siIII-matched \oVI\ velocity components
more frequently corotate with the disk
compared to velocity components 
that cannot be matched with the low ions
(Figure~\ref{fig:o6vb_o6si2}).
In this Section,
we discuss the implications and interpretation
of these results.

%%%%%%%%%%%%%%%%%%%%%%%%
% Location of Detected Gas
%%%%%%%%%%%%%%%%%%%%%%%%
\subsection{Location of the Multiphase Gas
Traced by \oVI, \siII, and \siIII}
\label{ssec:location}

Among the 12 sightlines with 
\oVI\ detection and spectral coverage
of \siII\ and \siIII, 
\oVI\ is detected in sightlines 
out to larger impact parameters compared to 
the low ions 
(Figures~\ref{fig:o6vb}, \ref{fig:si23vb}, and \ref{fig:o6_si3_si2}).
This result is commonly found in previous work.
While low ions
are only detected in the inner CGM out to $\sim100$ kpc
around star-forming galaxies,
\oVI\ is detected out to impact parameters 
of several hundred kpcs 
(see Section~\ref{sec:intro}).
This implies that \oVI\ is more spatially extended
compared to the low-ionization-state gas
that resides in the inner CGM.

In general, 
quasar absorption-line analysis only produces
line-of-sight measurements
and do not reveal the location of the absorbing gas
in 3D space.
However, our kinematics comparison between \oVI\ and low ions
provides the clues for 
distinguishing the location of the detected gas.
Figure~\ref{fig:o6vb_o6si2} shows
that \siII- and \siIII-matched 
\oVI\ velocity components more frequently corotate
with the disk compared to those components without low-ion matches,
and 
most of the matched components
are found in sightlines with impact parameters below 100 kpc.
This may not seem surprising,
because low ions are  
often detected in the inner CGM
and  corotate with the disk,
suggesting that the low ions trace
the gas residing near the extended disk plane of the galaxy
\citep[e.g.,][]{Kacprzak2011ApJ,Ho2017}.
But with the subset of \oVI\ components 
having velocities that matched with the low ions 
and also corotating with the disk,
this suggests that 
these low-ion-matched \oVI\ components
trace the gas that is co-spatial with the low ions
\citep[also see][]{Werk2016},
i.e., near the extended disk plane in the inner CGM.
On the other hand, 
the unmatched \oVI\ velocity components 
correspond to gas at the outer CGM.  
In fact, \citet{Werk2016} concluded
that the unmatched \oVI\ components trace
the volume-filling hot corona at the halo virial temperature,
and our results do not contradict with that description.

The interpretation of the location of the absorbing gas
explains two results.
First, it explains why 
the \siII- and \siIII-matched \oVI\ velocity components
are more likely to be corotating with the disk compared
to those \oVI\ components without a low-ion match.
It also explains
the complicated line structures detected in sightlines
with small impact parameters (Figure~\ref{fig:o6_si3_si2}),
because these sightlines
intersect both the inner and outer CGM
with different  \oVI\ kinematics.  
This is similar to the result from
multi-sightline probes of the Andromeda Galaxy
that demonstrated
the decreasing kinematic complexity of the CGM
with increasing radii
\citep{Lehner2025}.
Therefore,
our results imply that
while the outer \oVI\ CGM tends to move randomly,
the inner \oVI\ CGM carries angular momentum
and corotates with the galaxy disk
similar to the low-ionization-state gas.

\begin{figure}[bht]
    \centering
    \includegraphics[width=1\linewidth]{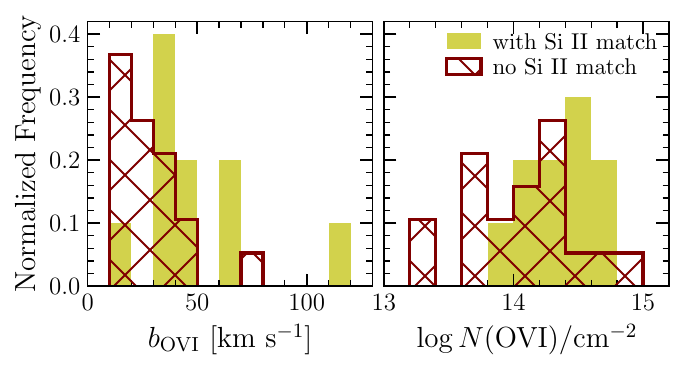}
    \caption{
            Distribution of 
            Doppler parameter (left) and column density (right)
            of \oVI\ velocity components
            with (solid yellow) and without \siII\ matches (hatched maroon).
            Performing the KS test 
            on the Doppler parameter rejects
            the null hypothesis at $\approx$96\%
            that the two \oVI\ populations
            are drawn from the same distribution,
            but the null hypothesis cannot be ruled out
            for the column density measurements.
            }
    \label{fig:o6si2_voigtparam}
\end{figure}

Furthermore, 
the \oVI\ velocity components with \siII\ matches
tend to have larger Doppler parameters 
compared to those without \siII\ matches.  
Figure~\ref{fig:o6si2_voigtparam} shows the distributions
of the Doppler parameter and column density
for the two \oVI\ populations.
Performing the Kolmogorov-Smirnov (KS) test on the Doppler parameter
rules out the null hypothesis at $\approx$96\% that
the \oVI\ components with and without \siII\ matches
are drawn from the same population distribution.  
However, 
the null hypothesis cannot be rejected for 
the column density populations.
On the other hand, when we repeat the same analysis
for the \oVI\ components with and without \siIII\ matches,
the null hypothsis cannot be ruled out 
for both measured quantities.
In fact,
since \siII-matched \oVI\ components
are more likely to be corotating with the disk
compared to those without \siII\ matches, 
the larger Doppler parameter of the former
plausibly suggests that
the rotation motion (in addition to turbulence)
has broadened the absorption lines.
Because the net CGM angular momentum 
remains roughly constant with radius in 3D space,
the rotation velocity is expected to decrease towards the outer CGM.
With \siIII\ generally being more spatially extended compared
to the lower-ionization \siII\ counterpart,
we expect the \oVI\ components with \siIII\ matches to
trace the gas out to larger 3D radii
and have lower rotation velocities on average
compared to those with \siII\ matches.
As a result, 
the rotation motion has produced
a measurable increase to the absorption-line width 
of the \oVI\ velocity components with \siII\ matches
but not those with \siIII\ matches,
which also explains 
why the latter have comparable Doppler parameters
as the no-\siIII-matched \oVI\ components.

% %%%%%%%%% FIG: OVI (match/unmatch with SiII) HIST %%%%%%%%%%%%
% %%%%%%%%%%%%%%%%%%%%%%%%%%%%%%%%%%%%%%%%%%%%%%%%%%%%%%%%%%%%%%%

%%%%%%%%%%%%%%%%%%%%%%%%
% Angular Momentum
%%%%%%%%%%%%%%%%%%%%%%%%
\subsection{Angular Momentum of the CGM and Radial Infall}
\label{ssec:am}

% %%%%%%%%% FIG: AM (RV constant) HIST %%%%%%%%%%%%
\begin{figure}[htb]
    \centering
    \includegraphics[width=1\linewidth]{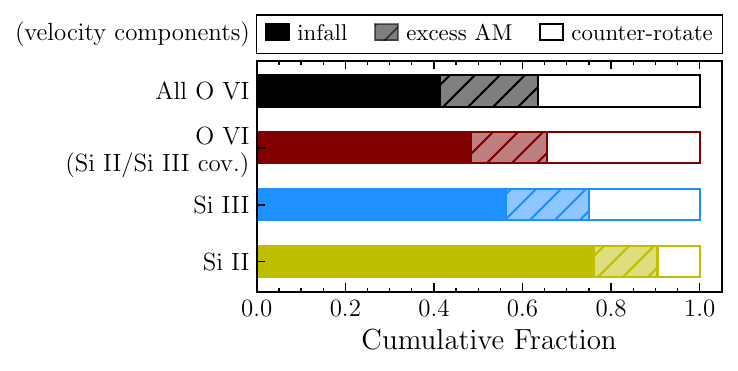}
    \caption{
            Comparison of angular momentum
            of \siII, \siIII, and \oVI\ velocity components
            with extended galaxy disk.
            Considering the hypothesis that
            non-counter-rotating velocity components
            trace gas on the extended disk plane,
            we compare the circular velocities of 
            individual velocity components
            to that required for the gas
            to stay on circular orbits.
            Fraction of velocity components
            with lower (higher) velocities 
            than needed are labeled as ``infall'' (``excess AM'')
            and are shown by the filled 
            (light-colored hatched) histograms.
            Unfilled histograms represent counter-rotating 
            velocity components.
            The fraction of infalling velocity components
            decreases with increasing ionization potential.
            This Figure follows the same color scheme as
            Figure~\ref{fig:kinhist}.
            }
    \label{fig:aminfall}
\end{figure}
% %%%%%%%%%%%%%%%%%%%%%%%%%%%%%%%%%%%%%%%%%%%%%%%%%%%%%%%%%%%%%%%

While the absorption-line measurements do not reveal 
where the gas is located along the line of sight, 
we have argued that some of the \oVI\ velocity components 
are tracing gas near the extended disk plane of galaxies
(Section~\ref{ssec:location}).  
But does the corotating gas have sufficient angular momentum
to stay on circular orbits? 
If not, the gas could be infalling to the galaxy,
a plausible scenario that explains
the corotating low-ionization-state gas 
at sub-centrifugal velocities
\citep[e.g.,][]{Ho2017,HoMartin2020}.

We explore this radial inflow hypothesis
by assuming 
all the non-counter-rotating gas being on the extended disk plane.
Then, we calculate the implied galactocentric radii $R_\mathrm{ion}$
and the circular velocities $V_\mathrm{c,ion}$
for individual \siII, \siIII, and \oVI\ velocity components.
We compare these circular velocities to 
that required for the gas to stay on circular orbits,
assuming the galaxy disks 
maintain the flat, arctangent rotation curve
out to the galactocentric radii.  
The histograms in
Figure~\ref{fig:aminfall} show the fractions of  
\siII, \siIII, and \oVI\ velocity components 
with circular velocities too low
to be on circular orbits,
i.e., the gas lacks sufficient angular momentum
and is infalling (filled histograms)\footnote{
    Velocity components that are 
    consistent with being 
    at the galaxy systemic velocities
    are considered as 
    lacking sufficient angular momentum
    to stay on circular orbits.
    }.    
The fraction of infalling velocity components
clearly increases with decreasing ionization potential,
for which only about 45\% for \oVI\ velocity components
are tracing infalling gas,
and this fraction increases to 80\% for \siII.
On the other hand, 
about 15\% of the velocity components of each ion
have excess angular momentum to 
be on the circular orbits (hatched).
This gas is not expected to be accreting onto 
the disk.

With the infall interpretation, 
we can also determine the radii where the infalling gas joins
the disk.
Assuming the specific angular momentum of 
the gas remains unchanged,
i.e., $R_\mathrm{ion}V_\mathrm{c,ion}$ stays constant,
we calculate the radius where 
the gas has the same specific angular momentum 
as the extended disk
(also known as the circularization radius).
For \siII, \siIII, and \oVI\ gas,
the calculation produces typical radii of under 50 kpc,
which is comparable to the radii of extraplanar \hI\ gas
of nearby disk galaxies
from 21-cm emission observations
\citep[e.g.,][]{Marasco2019}.

The caveat of the above calculations is that
while we assume the angular momentum is conserved
for the gas clouds 
associated with individual velocity components,
this may not be case according to cosmological simulations.
Simulations have shown that 
the CGM has a broad angular momentum distribution
\citep{Stewart2013,DeFelippis2020,Hafen2022}.
This is due to the large turbulent velocity of the CGM, 
for which 
the turbulent motion adds a random angular momentum vector
to the gas clouds
\citep{Fielding2017,Lochhaas2020,Pandya2023,Kakoly2025}.
However, this turbulent angular momentum component tends to
cancel out when the gas accretes 
onto the galaxy disk \citep{Hafen2022}.
In other words,
the gas clouds often interact and exchange angular momentum
with each other,
implying that the angular momentum 
of a particular gas cloud
is unlikely to be conserved during its infall to the disk.

% %%%%%%%%% FIG: AM (RV constant) HIST %%%%%%%%%%%%
\begin{figure}[htb]
    \centering
    \includegraphics[width=1\linewidth]{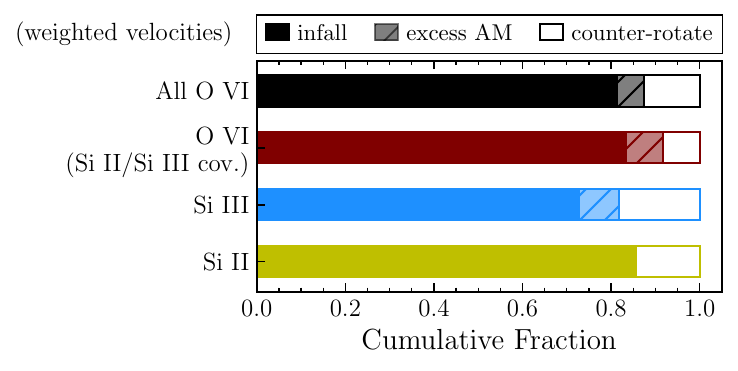}
    \caption{Comparison of angular momentum
            of the extended galaxy disk with 
            the average \siII, \siIII, and \oVI\ gas
            in individual sightlines.
            Analogous to Figure~\ref{fig:aminfall},
            but instead of considering individual velocity components,
            this Figure shows the result using 
            the column-density-weighted velocities
            of the gas detected in individual sightlines.
            Different ions show similar fractions
            of net counter-rotating gas and infalling gas.
            In particular, 
            \oVI\ has a lower counter-rotation fraction
            and a higher infall fraction compared to 
            that from the calculation of 
            using individual velocity components 
            in Figure~\ref{fig:aminfall}.
            }
    \label{fig:aminfall_coldensv}
\end{figure}
% %%%%%%%%%%%%%%%%%%%%%%%%%%%%%%%%%%%%%%%%%%%%%%%%%%%%%%%%%%%%%%%

Therefore, we repeat the analysis
by comparing 
the circular velocities required for the gas 
to stay on circular orbits
to the column-density-weighted velocities
of different ions in individual sightlines.
This enables us to consider 
the average angular momentum
of the gas intersected along the sightlines
instead of the gas clouds associated
with individual velocity components.
The result is shown in Figure~\ref{fig:aminfall_coldensv}.
First of all, 
the fraction of counter-rotating \oVI\ gas is only about 10\%,
which is four times lower than 
that calculated using 
individual velocity components ($\approx40$\%).
This reiterates the result from Figure~\ref{fig:o6vb}
that net counter-rotating \oVI\ absorbers are rare.
Also, the infalling gas fractions of \siII, \siIII, and \oVI\ gas
all reach $\approx80$\%, 
which are higher than the calculations
with individual velocity components, 
especially for \oVI\ (45\%).
Because \oVI\ traces gas out to larger 3D radii 
compared to the low ions,
and the expected CGM rotation velocity 
decreases with increasing radius
due to the net CGM angular momentum being roughly radius-independent 
(also see Section~\ref{ssec:location}),
the \oVI\  corotation and/or infall signatures 
could be easily masked by the turbulent motion.
Consequently,
the corotation and/or infall of \oVI\ gas is more likely
to be observed 
only when we consider the average angular momentum
using the column-density-weighted velocities in individual sightlines 
instead of individual velocity components.

%%%%%%%%%%%%%%%%%%%%%%%%
% Compare to previous work
%%%%%%%%%%%%%%%%%%%%%%%%
\subsection{Comparison with previous work on \oVI\ kinematics 
and disk rotation}
\label{ssec:comparek19}

Unlike the numerous literatures that
analyzed the relationship between low-ionization-state \mgII\ gas
and disk rotation,
very few work focused on
this comparison with \oVI\
(see Section~\ref{sec:intro}).
Using 10 major-axis sightlines
with azimuthal angle below 25\deg,
\citet{Kacprzak2019}
showed that 
both the individual velocity components 
and weighted velocities per sightline
of the \oVI\ gas
does not correlate with the disk rotation kinematics
(their Figure~4).
They also demonstrated that 
a simple monolithic rotating halo model 
failed to explain the full velocity spread of
the \oVI\ absorption.
In contrast,
when they ran AMR cosmological simulations,
they clearly found infalling \oVI\ filaments 
that rotate in the same direction as the galaxy disks.
The inflowing \oVI\ gas decelerates when it approaches the galaxies
and becomes rotationally-dominated.
However, the inflowing/rotating \oVI\ signatures 
were washed out by other velocity structures in the CGM
when they computed integrated velocities 
along lines of sight through 
the entire simulated galaxy halos.
Therefore, inspired by their simulation analysis,
they concluded that
while corotating or accreting \oVI\ gas could still exist
in their observed sightlines,
the signature was either weak or was masked by 
other kinematics ongoing within the CGM.

More recently, 
\citet{Kacprzak2025} conducted the COS-EDGES survey
and analyzed the multiphase CGM 
of nine edge-on disk galaxies ($i \geq 60$\deg)
probed by major-axis sightlines 
at impact parameters below $\approx$0.3\rvir. 
By comparing the stacked absorption-line profiles
of different ions (e.g., \mgI, \mgII, and \oVI)
in the small ($\leq 0.2$\rvir)
and large ($> 0.2$\rvir) impact parameter bins,
they showed that while the Doppler sign of 
the weighted velocities  
are consistent with disk rotation, 
the velocity magnitude
decreases with increasing impact parameters.
More importantly,
the amount of absorption that corotates with the disk
declines with both increasing ionization potential
and sightline impact parameters;
in contrast to over 90\% of the \mgII\ absorption 
that corotates with the disk,
only 80\% (60\%) of the \oVI\ absorption 
corotates at small (large) impact parameters.
Hence, they concluded that the CGM has
a radially-dependent kinematic structure.
Within 0.2\rvir,
the CGM hosts cool and dynamically broad gas 
that tightly coupled with disk rotation,
whereas beyond 0.2\rvir,
especially for \oVI,
the CGM becomes less kinematically correlated
with disk rotation
due to the presence of corotating, lagging, 
and volume-filling components.

In general, our analysis agrees with both
\citet{Kacprzak2019} and \citet{Kacprzak2025}
that the corotation signature 
of \oVI\ gas is challenging to observe
due to kinematics components other than rotation.
In particular, 
while individual \oVI\ velocity components
do not seem to correlate with disk rotation,
the column-density-weighted velocities suggest
that the bulk of \oVI\ gas rarely counter-rotates (Figure~\ref{fig:o6vb}).
We have attributed this ``discrepancy''
to the turbulent motion of the CGM,
and averaging is required to reveal
the net \oVI\ CGM angular momentum while 
searching for the corotation signature
(Section~\ref{ssec:am}).
Furthermore,
we have also compared the \oVI\ kinematics
to that of the low ions.
Not only do low ions more often corotate with the disk
compared to \oVI\ (Figure~\ref{fig:si23vb}),
but the \oVI\ velocity components with low-ion matches
are more likely to corotate
compared to those without low-ion matches (Figure~\ref{fig:o6_si3_si2}).
These results suggest that low ions 
are better indicators
for identifying the corotating gas, 
which also matches the conclusion 
from \citet{Nateghi2024multiphase}
and \citet{Kacprzak2025}
that the lower the ionization potential,
the more likely the ion is kinematically consistent with 
the rotation of the galaxy disks 
and supports the accretion scenario.

%%%%%%%%%%%%%%%%%%%%%%%%%%%%%%%%%%%%%%%%
% Conclusion
%%%%%%%%%%%%%%%%%%%%%%%%%%%%%%%%%%%%%%%%
\section{Conclusion}
\label{sec:conclusion}

We presented a compilation of new and archival 
\textit{HST/COS} observations of quasars 
behind $z\approx0.2$, star-forming galaxies.
All 18 quasar sightlines probe the CGM within 45\deg\
of the major axes of the host galaxies
with inclined disks ($i \gtrsim 45$\deg),
doubling the existing number of major-axis sightlines
in the literature
for studying the relationship between
the \oVI\ and disk rotation kinematics.

We compared the kinematics of the \oVI\ gas
with that of the low-ionization-state gas
(traced by \siII\ and \siIII) and disk rotation.
We showed that while individual velocity \oVI\ components
do not correlate with disk rotation,
the bulk of \oVI\ gas,
as measured by the column-density-weighted \oVI\ velocity
in each sightline,
rarely counter-rotates with the galaxy disk 
(Figure~\ref{fig:o6vb}).
In general, \siII\ and \siIII\ gas are less likely to 
counter-rotate with the disk compared to \oVI\
(Figures~\ref{fig:si23vb} and \ref{fig:kinhist}).
But more importantly, 
we matched velocity components
between \oVI\ and low ions (\siII\ and \siIII)
by minimizing the difference of their velocity centroids.
We showed that the \oVI\ velocity components
with matches with low ions 
were mainly found in sightlines at small impact parameters
of below 100 kpc,
whereas those \oVI\ velocity components
without low-ion matches were on average at
larger impact parameters.
These low-ion-matched \oVI\ velocity components
are also more frequently 
corotate with the disk compared to those
\oVI\ velocity components without a low-ion match
(Figure~\ref{fig:o6vb_o6si2}).

Previous literature has shown that 
low-ionization-state gas 
is commonly detected in the inner CGM,
and major-axis sightlines often detect
such gas to corotate with the galaxy disk.
Although absorption-line measurements 
do not reveal where the gas lies
along the line of sight,
our results of the low-ion matched \oVI\ velocity components
sharing these properties
suggest that
these \oVI\ components
trace the gas that is co-spatial with
that of the low ions,
i.e., on the extended disk plane in the inner CGM.
In contrast, the \oVI\ velocity components
without a low-ion match
traces gas at large 3D radius.
Furthermore, 
if some of the \oVI\ gas resides on the extended disk plane,
we asked the question that whether the gas 
has sufficient angular momentum to stay on circular orbits.
If not, this could suggest the gas is infalling.
We showed that about 45\% of the \oVI\ velocity components 
lack sufficient angular momentum
to maintain a circular orbit.
This fraction increases with decreasing ionization potential,
with 80\% of \siII\ velocity components tracing
radially infalling gas
(Figure~\ref{fig:aminfall}).
Interestingly,
repeating this calculation 
using the column-density-weighted velocities
increases the inflow fractions of \oVI\ and \siIII\
also to 80\%.
We interpret this result as a consequence of 
the turbulent motion adding random angular momentum vectors
to the gas clouds associated with individual velocity components.
Therefore,
the corotation and inflow signatures 
are more likely to be observationally detected 
only when we consider 
the average angular momentum of the gas
using the column-density-weighted velocities.
Our future work will include photoionization modeling
and geometric models 
for explaining the circumgalactic gas flows
of \oVI\ and low-ionization-state gas
and to understand the CGM angular momentum 
and gas accretion onto galaxies.

%%%%%%%%%%%%%%%%%%%%%%%%%
% Acknowledgement (+ facilities)
%%%%%%%%%%%%%%%%%%%%%%%%%
\vspace{2em}

We thank the referee for thoughtful comments 
and suggestions that improved the manuscript.
This work was supported by  
NASA through HST-GO-15866.002-A 
from the Space Telescope Science Institute.
We gratefully acknowledge the support for
SHH from the National Science Foundation under AST-2202328.
JS was supported by the Israel Science Foundation (grant No.~2584/21).
Some of the data presented in this article were obtained 
from the Mikulski Archive for Space Telescopes (MAST) at 
the Space Telescope Science Institute. 
The specific observations analyzed can be accessed via 
\dataset[doi:10.17909/2b4f-h722]{https://doi.org/doi:10.17909/2b4f-h722}.
Some of the ground-based data collections 
were supported by the National Science Foundation under AST-1817125.
Some observations reported here were obtained at the MMT Observatory, 
a joint facility of the University of Arizona 
and the Smithsonian Institution.
Some of the data presented herein were obtained at the W.~M.~Keck 
Observatory, which is operated as a scientific partnership among 
the California Institute of Technology, the University of California 
and the National Aeronautics and Space Administration. 
The Observatory was made possible by the generous financial support 
of the W. M. Keck Foundation.
Some observations were supported by 
Swinburne Keck programmes 2014B W018E, 2015 W187E, 
and 2016A W032E, W056E.
Some presented data are based on observations obtained with 
the Apache Point Observatory 3.5 m telescope, 
which is owned and operated by 
the Astrophysical Research Consortium.

\facilities{HST (COS, ACS, WFPC2), 
MMT (Blue Channel spectrograph, Red Channel Spectrograph), 
Keck:I (LRIS), Keck:II (ESI, NIRC2), ARC}

%%%%%%%%%%% REFERENCES %%%%%%%%%%%
\bibliography{master_papers}

%%%%%%%%%%%%%%%%%%%%%%%%%%%%%%%%%
% Appendix: Individual 
%%%%%%%%%%%%%%%%%%%%%%%%%%%%%%%%%
\appendix
\section{Absorption System Associated with Individual Galaxies}

This Appendix includes the Voigt profile fits of absorption 
associated with the target galaxies in individual sightlines.  
Figure~\ref{fig:voigtfitset} 
(including the complete figure set in the online journal)
shows the Voigt profile fits,
and Table~\ref{tb:voigt_all} lists the parameters of the fits.
Because not all ions were used in the analysis of this paper,
we only present the measurements in the Appendix.

% =================================
% LONG TABLE for Voigt Profile Fitting results
\begin{center}
    \input{table_voigt_all.tex}
\end{center}
% =================================

\newpage

\begin{figure*}[htb]
    \centering
    \includegraphics[width=1\linewidth]{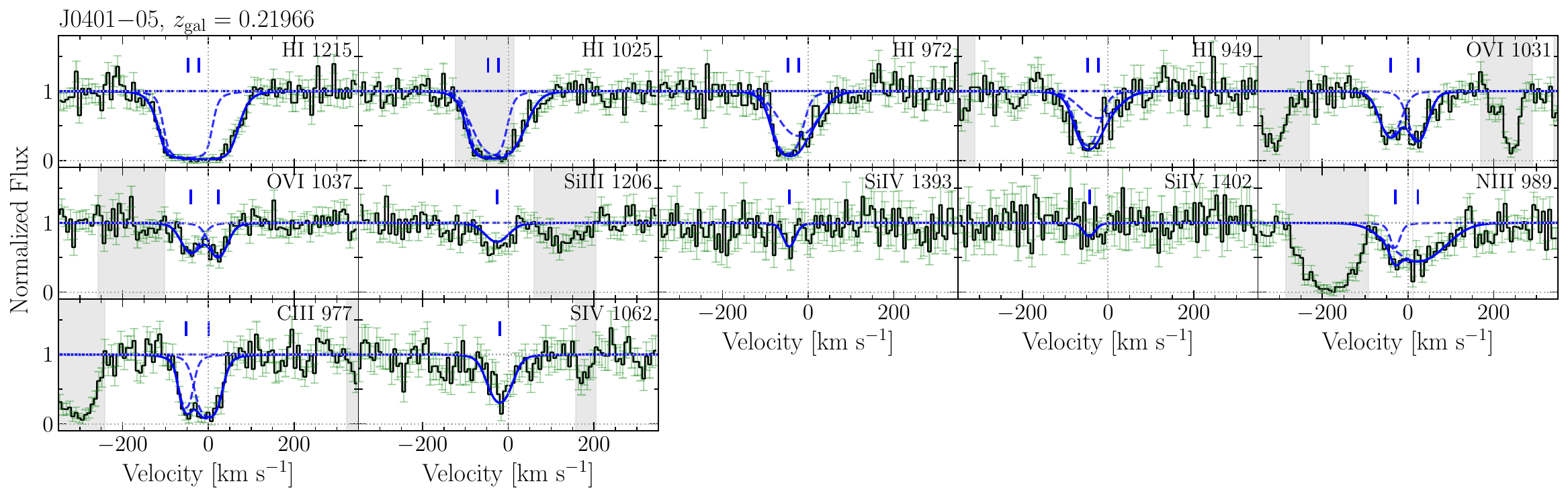}
    \includegraphics[width=1.0\linewidth]{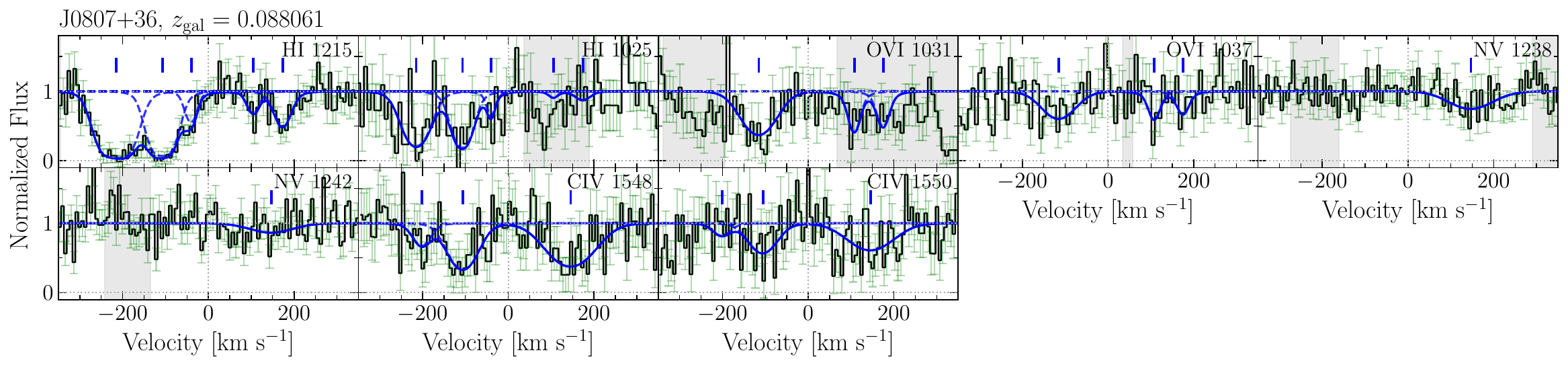}
    \includegraphics[width=1.0\linewidth]{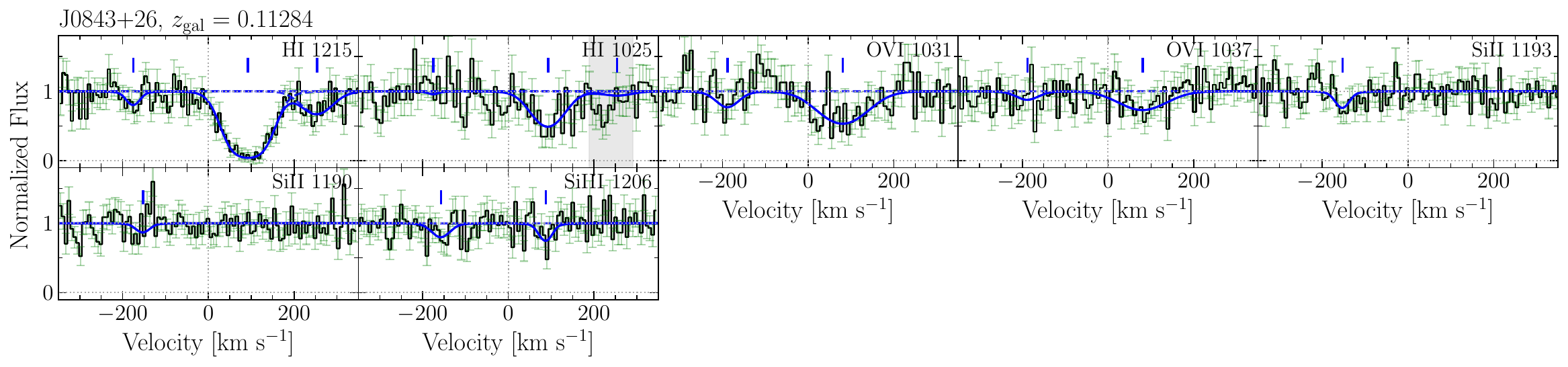}
    \includegraphics[width=1.0\linewidth]{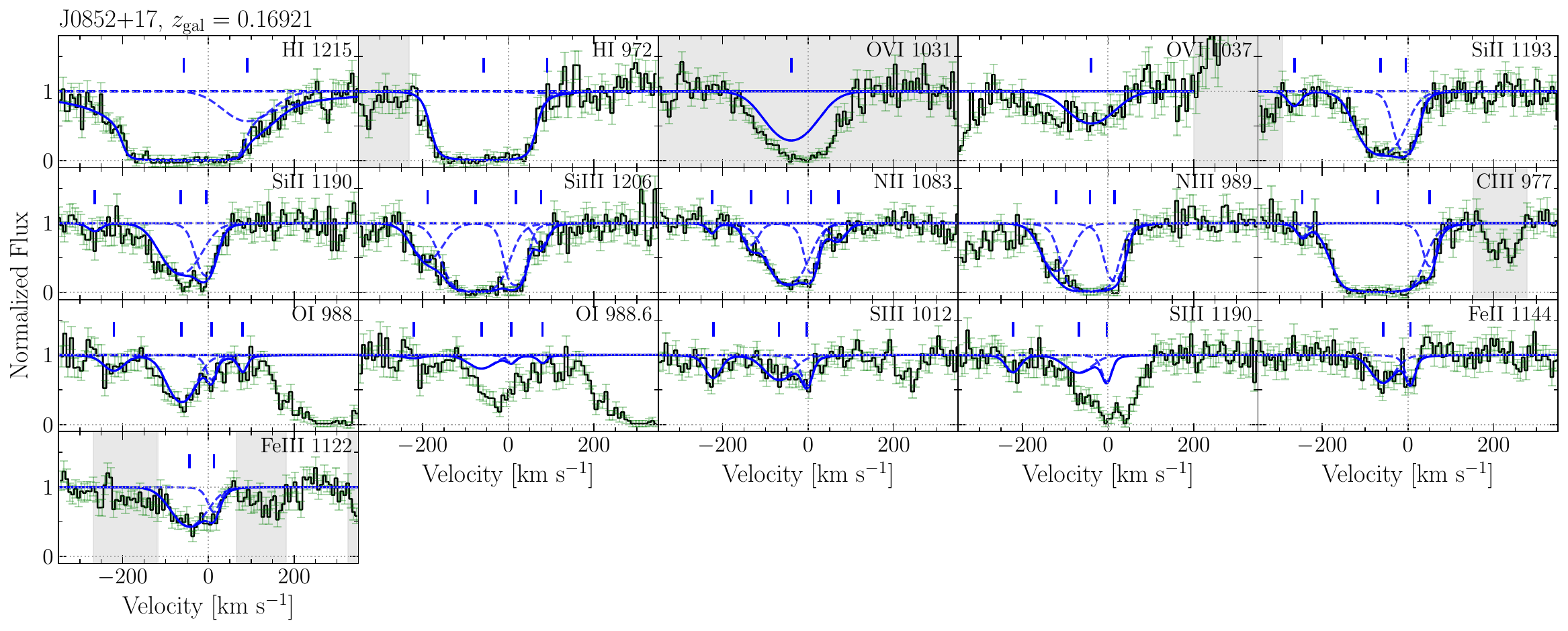}
    \caption{Voigt profile fits of absorption lines
                relative to the systemic velocity of the host galaxy
                in individual sightlines.
                The ionic species are labeled on the upper right.
                The blue dashed and solid curves 
                show the fitted Voigt profiles
                for individual velocity components and 
                the overall absorption of the ion,
                respectively.
                The blue ticks at the top 
                mark the velocity centroids of 
                individual components.  
                Grey shaded regions are masked 
                during Voigt profile fitting 
                due to contamination from intervening absorbers
                at other redshifts.
                }
    \label{fig:voigtfitset} 
\end{figure*}

\begin{figure*}%[h!]
    \centering
    \figurenum{11}
    \includegraphics[width=1.0\linewidth]{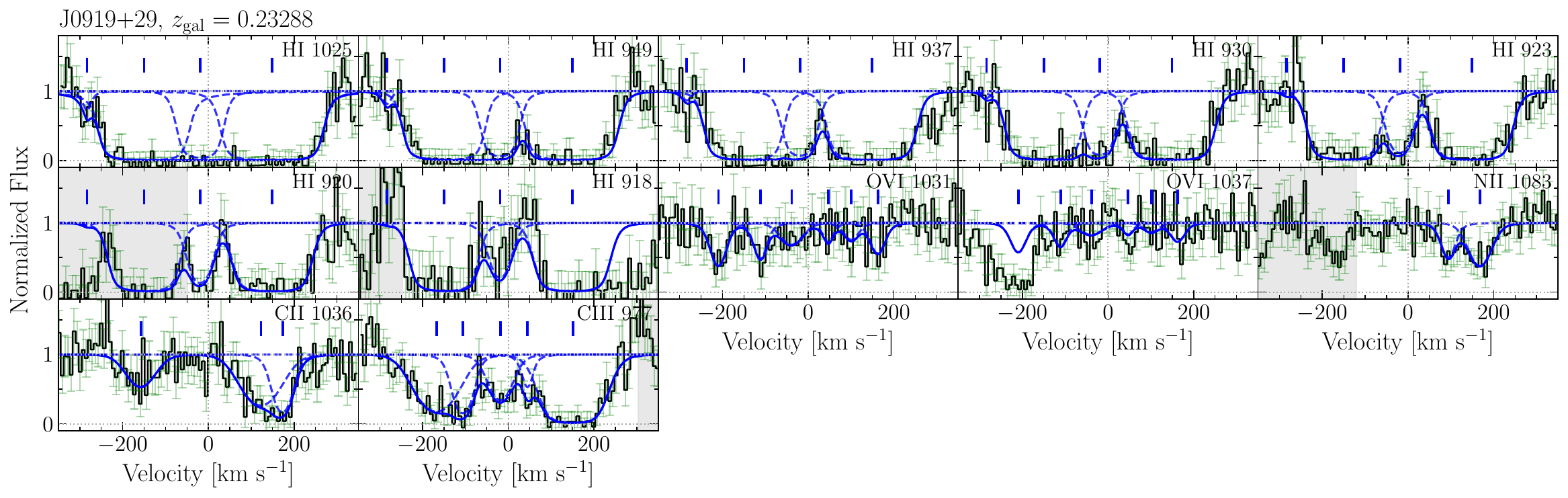}
    \includegraphics[width=1.0\linewidth]{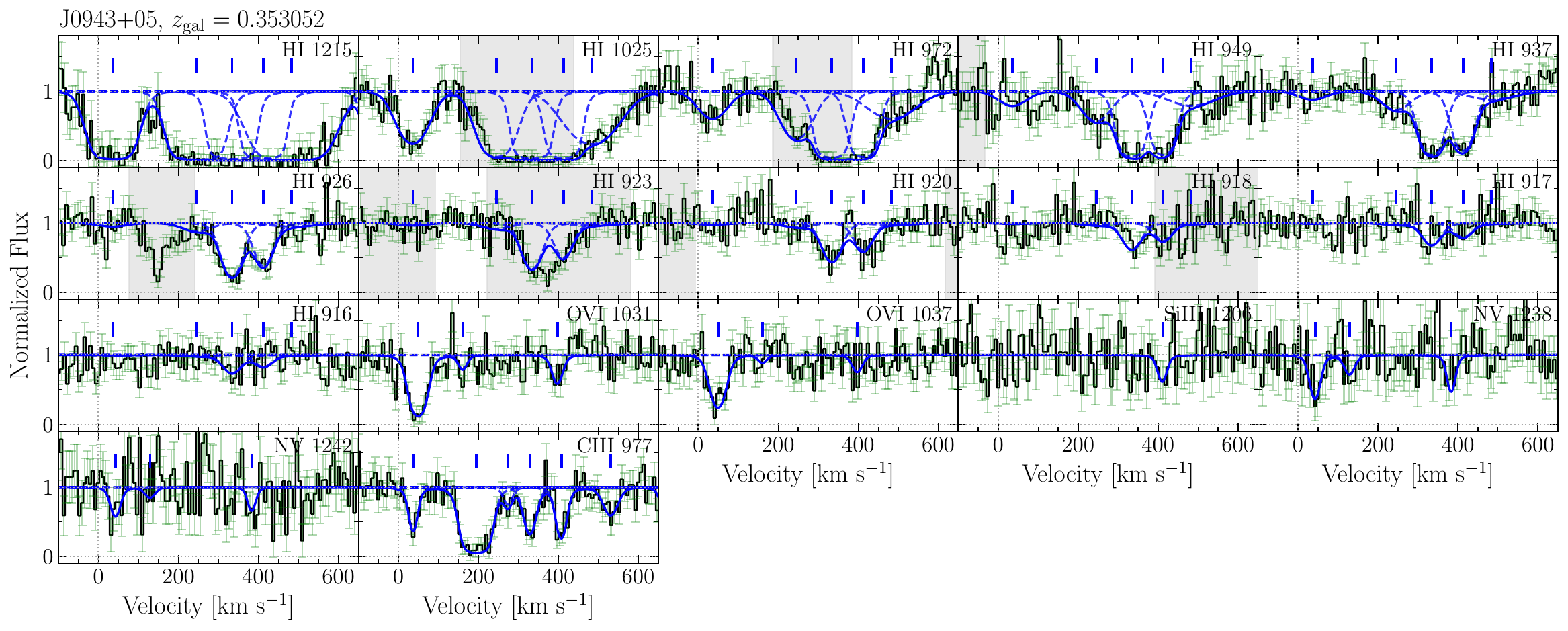}
    \includegraphics[width=1.0\linewidth]{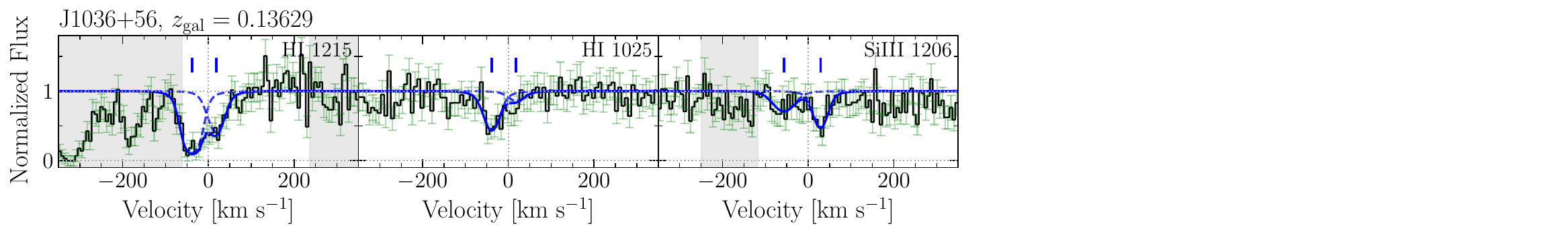}
    \includegraphics[width=1.0\linewidth]{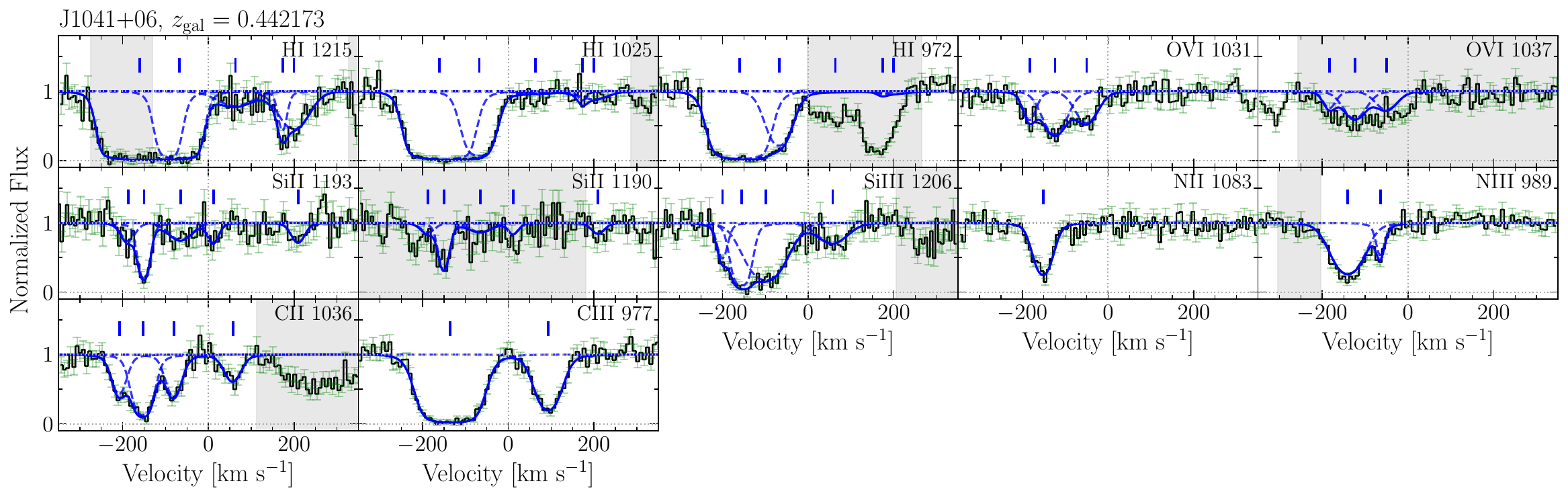}
    \caption{\textit{(Continued)}
            }
\end{figure*}

\begin{figure*}%[h!]
    \centering
    \figurenum{11}
    \includegraphics[width=1.0\linewidth]{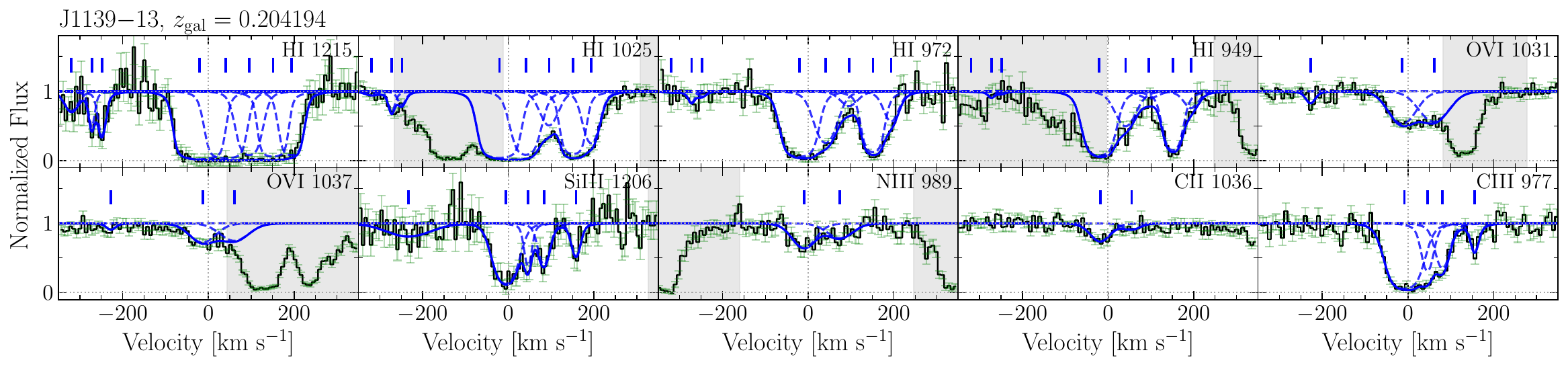}
    \includegraphics[width=1.0\linewidth]{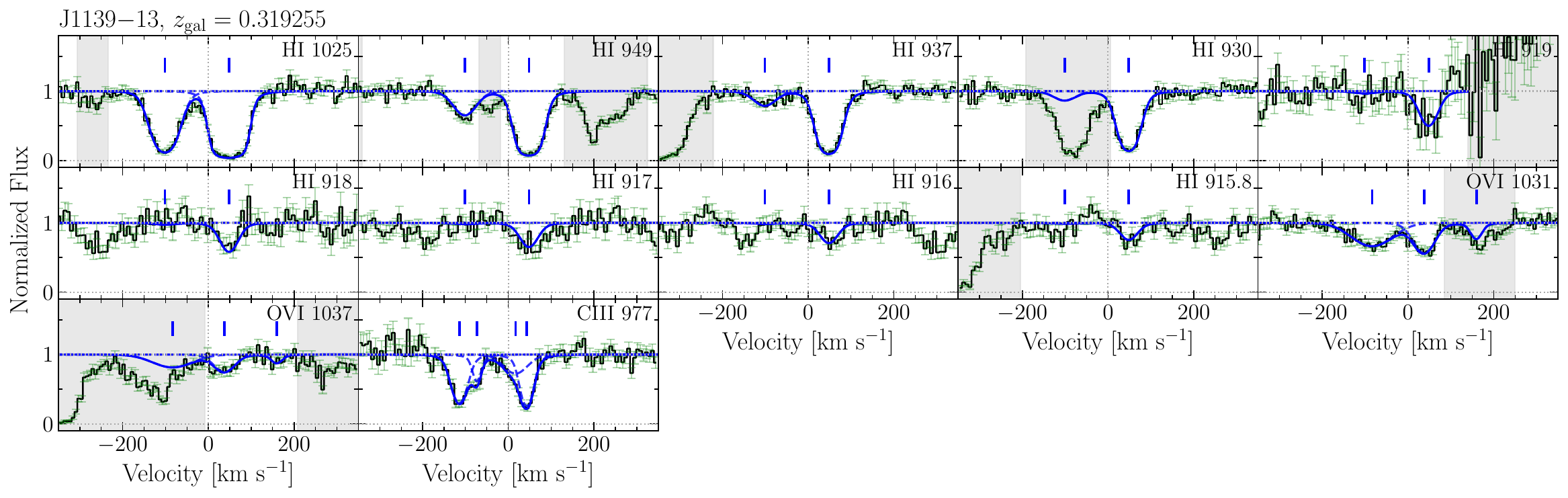}
    \includegraphics[width=1.0\linewidth]{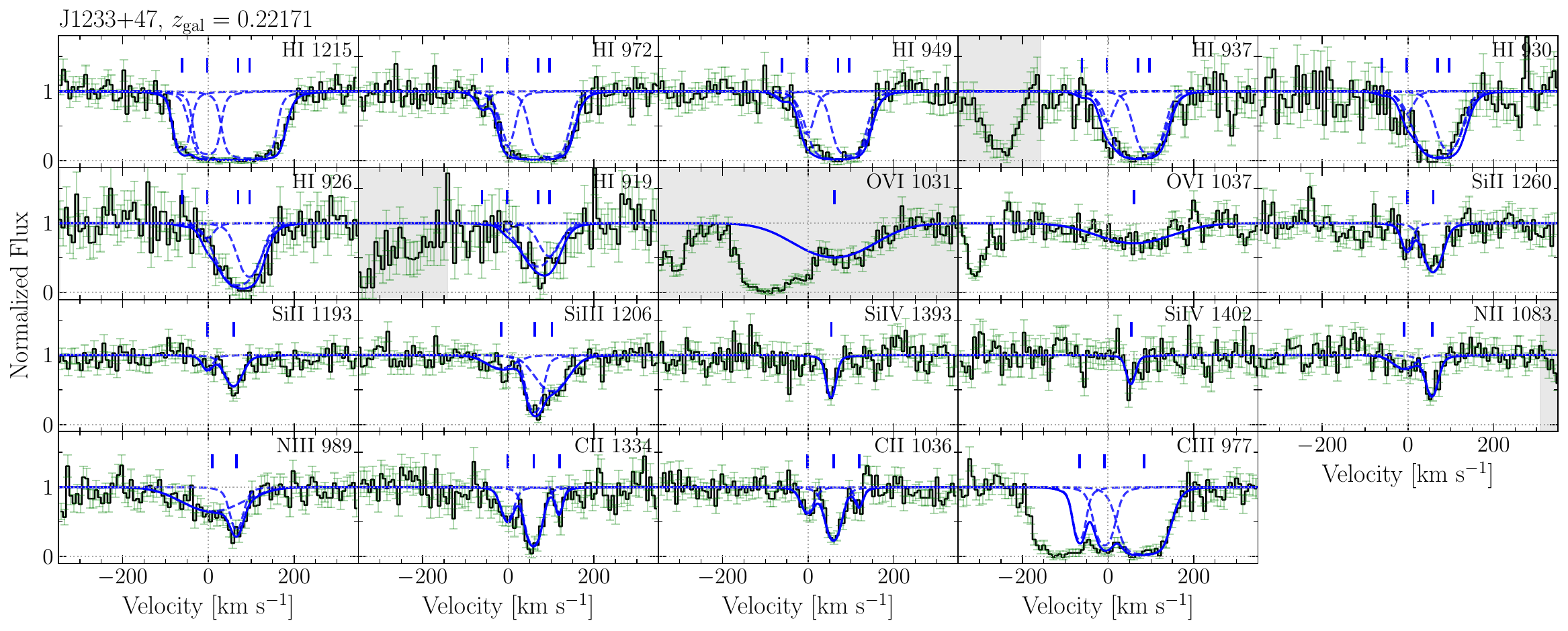}
    \caption{\textit{(Continued)}
            }
\end{figure*}

\begin{figure*}%[h!]
    \centering
    \figurenum{11}
    \includegraphics[width=1.0\linewidth]{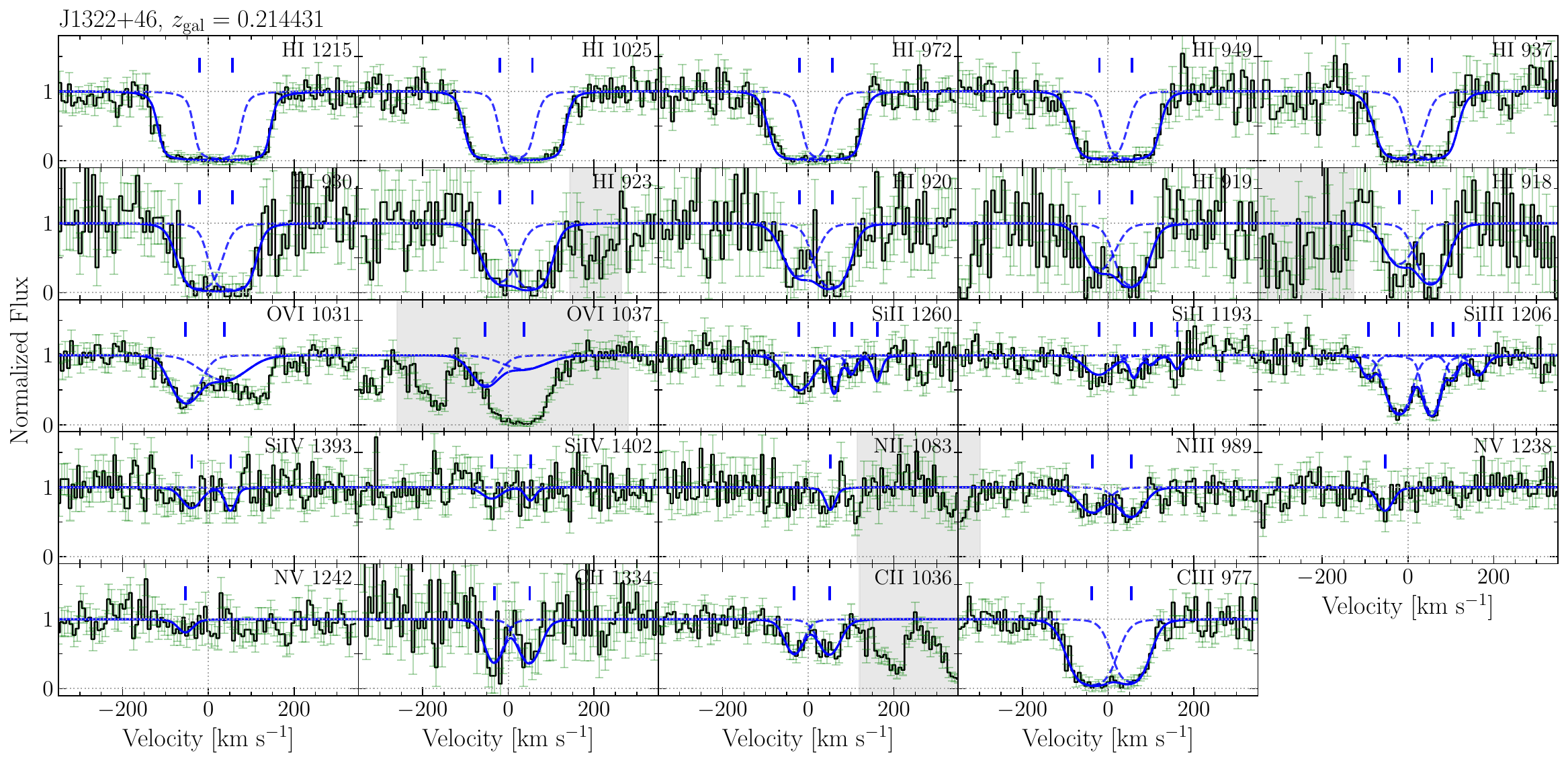}
    \includegraphics[width=1.0\linewidth]{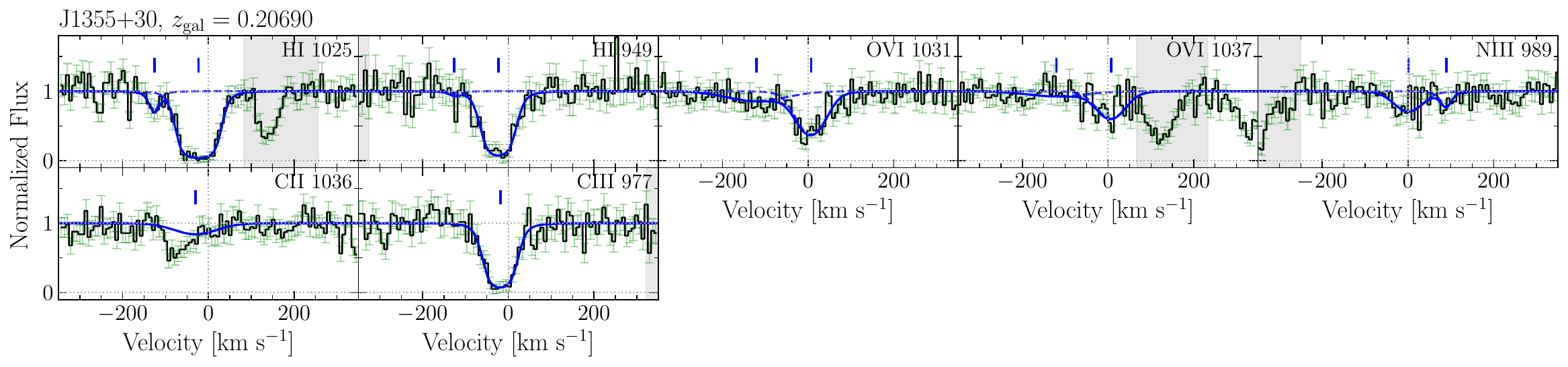}
    \includegraphics[width=1.0\linewidth]{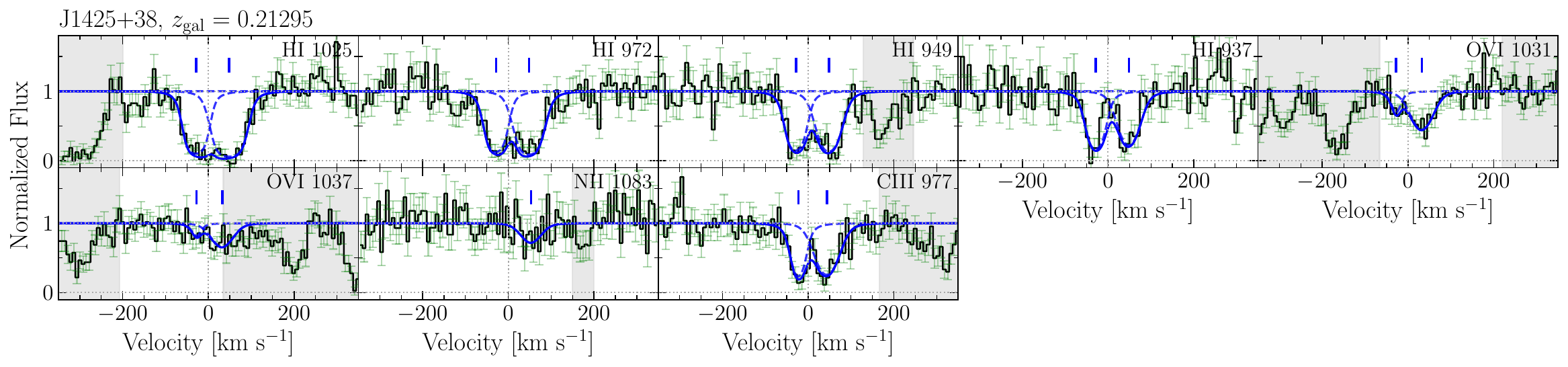}
    \includegraphics[width=1.0\linewidth]{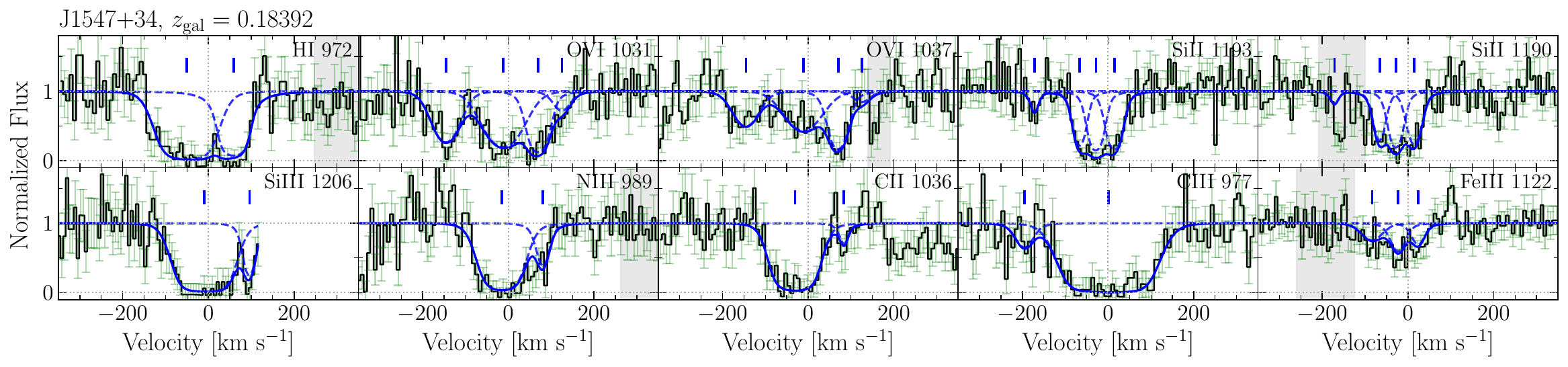}
    \caption{\textit{(Continued)}
            }
\end{figure*}

\begin{figure*}%[h!]
    \centering
    \figurenum{11}
    \includegraphics[width=1.0\linewidth]{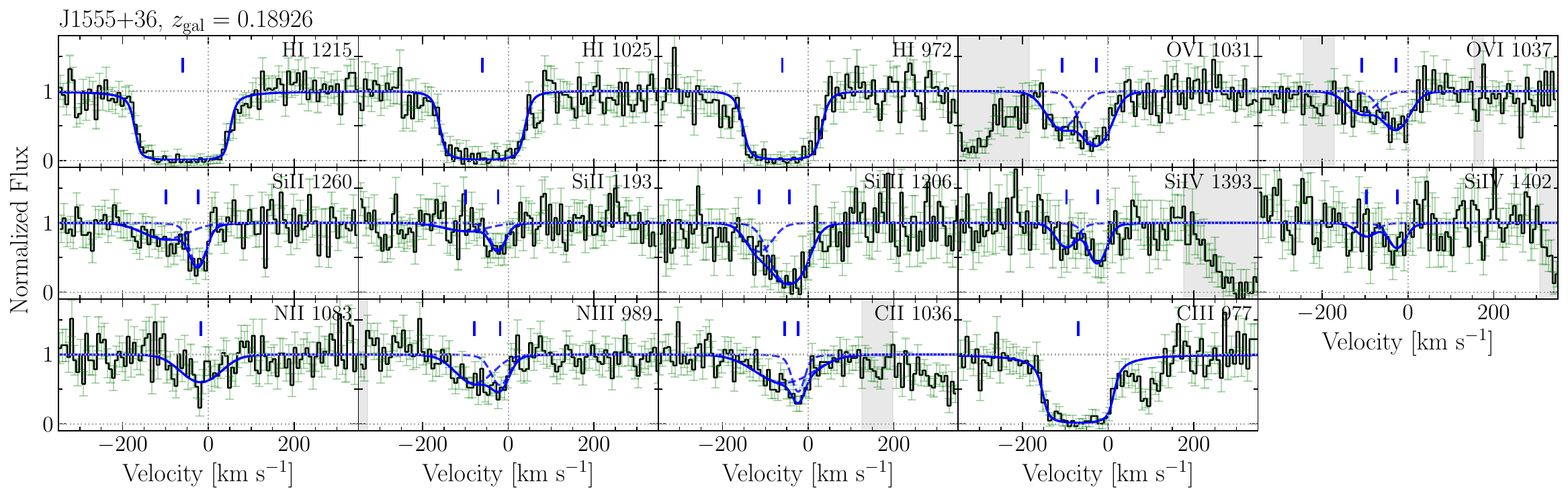}
    \includegraphics[width=1.0\linewidth]{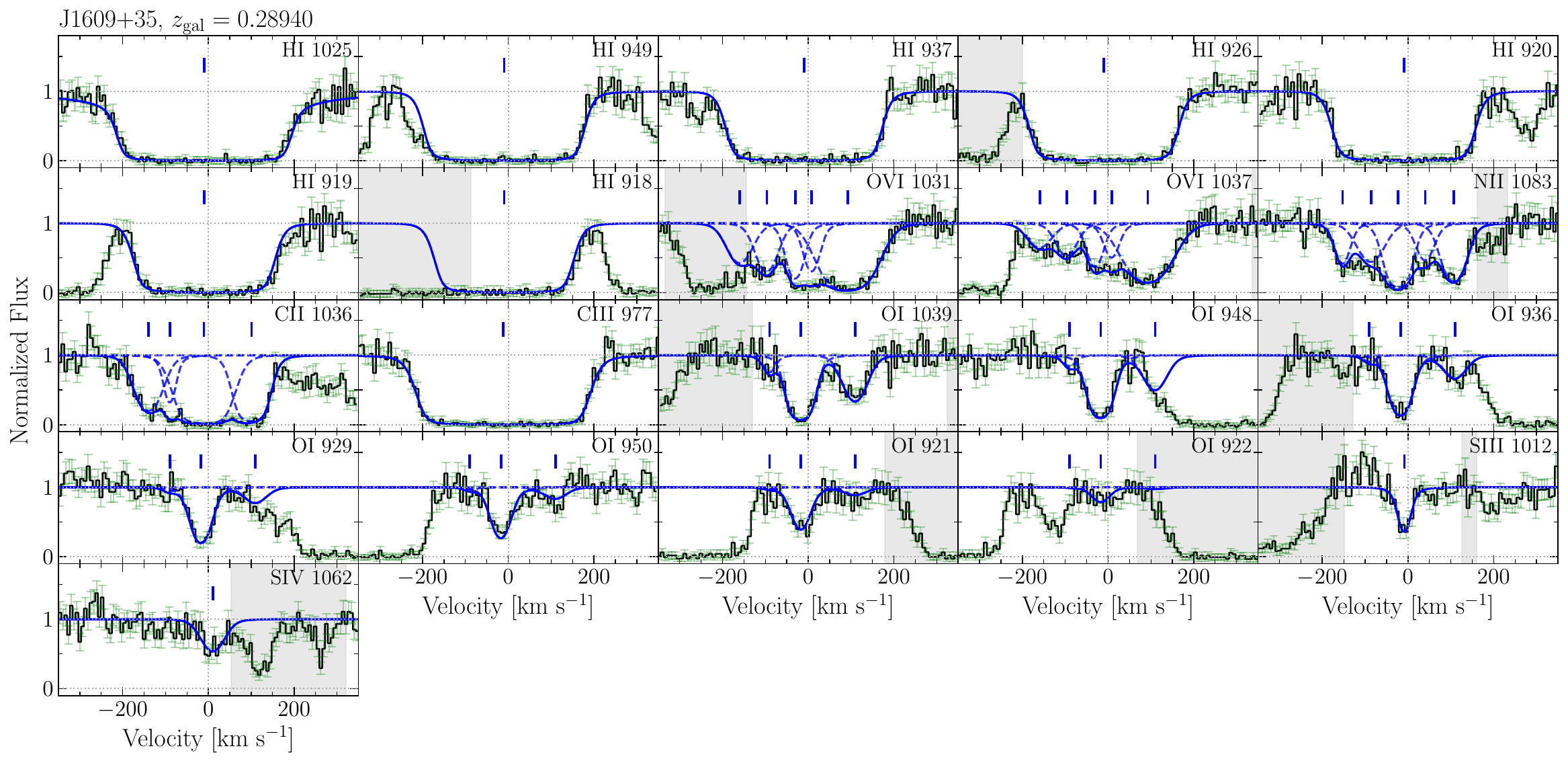}
    \includegraphics[width=1.0\linewidth]{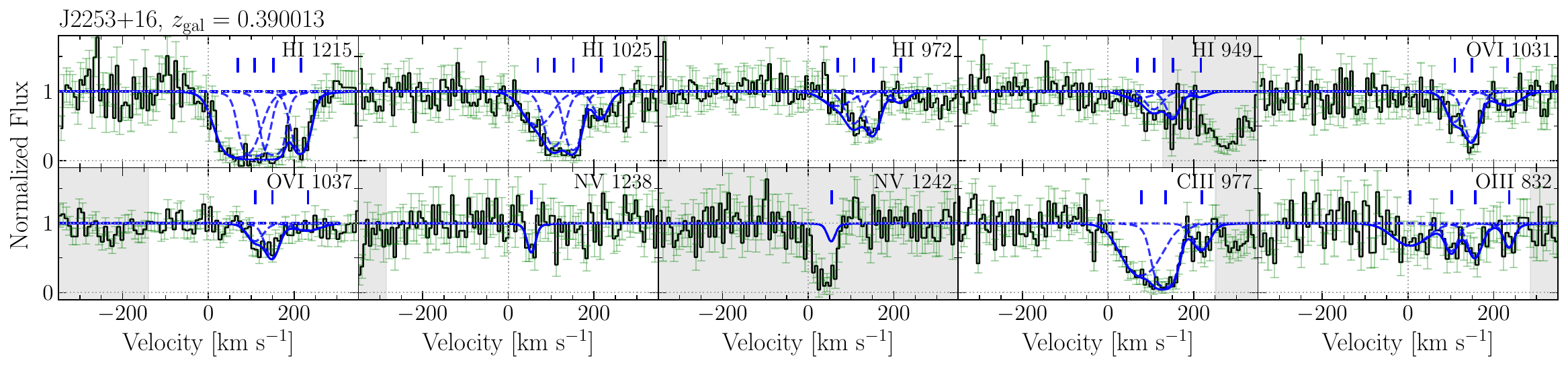}
    \caption{\textit{(Continued)}
            }
\end{figure*}

\end{document}

%% file: setdef.tex
% LAST UPDATE: Nov 12, 2015
% set short-hand definitions for convenience
%
%
\def\etal{{\rm et al.}}
\def\etali{{\it et al.\thinspace}}
\def\etns{{\rm et\thinspace al.}}   % et al.
\def\etaln{et al.\thinspace}

\def\EAGLE{\texttt{EAGLE}}
\def\AREPO{\texttt{AREPO}}
\def\ENZO{\texttt{ENZO}}
\def\GADGET{\texttt{GADGET-3}}
\def\OWLS{\texttt{OWLS}}
\def\FIRE{\texttt{FIRE}}
\def\FIREtwo{\texttt{FIRE-2}}
\def\illustris{\texttt{Illustris}}
\def\illustrisTNG{\texttt{IllustrisTNG}}

\def\zgal{$z_\mathrm{gal}$}

\def\arcsec{$^{\prime\prime}$}

%%%%%%%%%%%%%%%%%%%%
% Ions related
%%%%%%%%%%%%%%%%%%%%
%\def\mgIIdb{Mg II $\lambda\lambda$2796, 2803}
%\def\mgIIdbl{Mg II $\lambda$2796}
%\def\mgIIdbu{Mg II $\lambda$2796}
\def\mgIIdb{\ion{Mg}{2} $\lambda\lambda$2796, 2803}
\def\mgIIdbl{\ion{Mg}{2} $\lambda$2796}
\def\mgIIdbu{\ion{Mg}{2} $\lambda$2803}
\def\oI{[\ion{O}{1}] $\lambda$6300}
\def\oVIdb{\ion{O}{6} $\lambda\lambda$1031, 1037}
\def\oVIA{\ion{O}{6} $\lambda$1031}
\def\oVIB{\ion{O}{6} $\lambda$1037}
\def\siIIIlam{\ion{Si}{3} $\lambda$1206}
\def\siIIdb{\ion{Si}{2} $\lambda\lambda$1190, 1193}
\def\siIIlam{\ion{Si}{2} $\lambda$1260}
\def\sIIlam{[\ion{S}{2}] $\lambda\lambda$6716, 6731}
\def\oIIIlam{[\ion{O}{3}] $\lambda$5007}
\def\nIIlam{[\ion{N}{2}] $\lambda\lambda$6548, 6583}

\def\hI{\mbox {\ion{H}{1}}}
\def\mgI{\mbox {\ion{Mg}{1}}}
\def\mgII{\mbox {\ion{Mg}{2}}}
\def\nII{\mbox [{\ion{N}{2}}]}
\def\oVI{\mbox {\ion{O}{6}}}
\def\oVII{\mbox {\ion{O}{7}}}
\def\oVIII{\mbox {\ion{O}{8}}}
\def\cIV{\mbox {\ion{C}{4}}}
\def\halpha{$\mathrm{H}\alpha$}
\def\hbeta{$\mathrm{H}\beta$}
\def\heII{\mbox {\ion{He}{2}}}
\def\sII{\mbox {\ion{S}{2}}}
\def\siIII{\mbox {\ion{Si}{3}}}
\def\siII{\mbox {\ion{Si}{2}}}
\def\cII{\mbox {\ion{C}{2}}}
\def\cIII{\mbox {\ion{C}{3}}}

\def\mgplus{$\textrm{Mg}^{+}$}
\def\oVIion{O$^\mathrm{5+}$}
\def\siIIion{Si$^\mathrm{+}$}
\def\siIIIion{Si$^\mathrm{++}$}

%%%%%%%%%%%%%%%%%%%%
% Units/variables related
%%%%%%%%%%%%%%%%%%%%
\def\aapix{\mbox{\AA\ pixel$^{-1}$}}
\def\linesmm{\mbox{lines mm$^{-1}$}}
\def\kms{\mbox{km s$^{-1}$}}
\def\kmsMpc{\mbox{km s$^{-1}$ Mpc$^{-1}$}}
\def\kmstb{km s$^{-1}$}
\def\micron{\mbox{$\mu$m}}
\def\modotyr{\mbox {$\rm M_\odot$~yr$^{-1}$}}
\def\msununit{$\rm M_\odot$}
\def\percmsq{\mbox{cm$^{-2}$}}
\def\percmcube{\mbox{cm$^{-3}$}}

\def\rvir{$r_\mathrm{vir}$}
\def\mvir{$M_\mathrm{vir}$}
\def\mstar{$M_\star$}
\def\brvir{$d/r_\mathrm{vir}$}

\def\logmstar{$\log M_\star$}
\def\logmvir{$\log M_\mathrm{vir}$}
\def\logmstarmsun{$\log (M_\star/\mathrm{M_\odot})$}
\def\logmvirmsun{$\log (M_\mathrm{vir}/\mathrm{M_\odot})$}
\def\logmsun{$\log \mathrm{M_\odot}$}

\def\vinf{$V_\mathrm{\infty}$}

\def\NmgII{$N_\mathrm{MgII}$}

%%%%%%%%%%%%% SPECIAL %%%%%%%%%%%%%%
\def\deltavlos{$|\Delta v_\mathrm{LOS}|$}
\def\fmgiimis{$f_\mathrm{MgII,mis}$}
\def\fmgiimisb{$f_\mathrm{MgII,mis}(b)$}
\def\halovc{$v_\mathrm{c,halo}$}
\def\jstar{$\bm{j_\star}$}
\def\jstarscalar{$j_\star$}
\def\logNmgii{log N(Mg\thinspace II)}
\def\logNovi{log N(O\thinspace VI)}

\def\velwovi{$\langle v_\mathrm{O\ VI} \rangle_w$}
\def\absvelwovi{$|\langle v_\mathrm{O\ VI} \rangle_w|$}

\def \dlow {\mbox{$400 {\rm ~l~mm}^{-1}$}}
\def \dhigh {\mbox{$600 {\rm ~l~mm}^{-1}$}}
% PAPER SPECIFIC
% EVAN S.
\newcommand{\be}{\begin{equation}} \newcommand{\ba}{\begin{eqnarray}}
\newcommand{\ee}{\end{equation}} \newcommand{\ea}{\end{eqnarray}}
\def\-{{\em{---}}}
\def \mA {\mbox{${\rm m \AA} $} }
\def \rr {\mbox{${\rm RR}$} }
\def \rarb {\mbox{${\rm R_AR_B}$} }
\def \rara {\mbox{${\rm R_AR_A}$} }
\def \dd {\mbox{${\rm DD}$} }
\def \dada {\mbox{${\rm D_AD_A}$} }
\def \dadb {\mbox{${\rm D_AD_B}$} }
\def \dr {\mbox{${\rm DR}$} }
\def \darb {\mbox{${\rm D_AR_B}$} }
\def \dara {\mbox{${\rm D_AR_A}$} }
\def \dbra {\mbox{${\rm D_BR_A}$} }
\def \hMpc      {h^{-1}{\rm\ Mpc}}
\def \hkpc      {h^{-1}{\rm\ kpc}}
\def \h         {\hbox{$\, h$} }
\def \hinv      {\hbox{$\, h^{-1}$} }
\def \hinvseven    {\hbox{$\, h_{70}^{-1}$} }
\def\ewr{\mbox {EW$_r$}}
\def\ewo{\mbox {EW$_o$}}
\def\H7{\mbox {$h_{0.7}$}}
\def\naI{\mbox {\ion{Na}{1}}}
\def\feI{\mbox {\ion{Fe}{1}}}
\def\znII{\mbox {\ion{Zn}{2}}}
\def\crII{\mbox {\ion{Cr}{2}}}
\def\alI{\mbox {\sc Al~I~}}
\def\alII{\mbox {\sc Al~II~}}
\def\alIII{\mbox {\sc Al~III~}}
\def\mnII{\mbox {\ion{Mn}{2}}}
\def\niII{\mbox {\ion{Ni}{2}}}
\def\feII{\mbox {\ion{Fe}{2}}}
\def\feIII{\mbox {\ion{Fe}{3}}}
\def\sV{\mbox {\ion{S}{5}}}
\def\siIV{\mbox {\ion{Si}{4}}}
\def\siI{\mbox {\ion{Si}{1}}}
\def\llambda{\mbox {$\lambda$}}
\def\hlen{\mbox {$h_{0.7}^{-1}$}}
\def\lstarlya{\mbox {$L^*_{Ly\alpha}$}}
\def\IZw18{I~Zw~18}
\def\m82{M82}
\def\Ab{Abell~}
\def\gi{\mbox {\rm g-i}}
\def\ug{\mbox {\rm u-g}}
\def\br{\mbox {\rm b-r}}
\def\eqn{equation}
\def\vesc{\mbox {$v_{\rm esc}$}}
% HE PAPER
\def\heha{\mbox {He~I~$\lambda 5876$ / H$\alpha$}}
\def\xhe{\mbox {$\chi({\rm He}) / \chi({\rm H})$} }
\def\he{\mbox {\rm He}}
\def\hii{\mbox {${\rm H}^+$}}
\def\h{\mbox {\rm H}}
\def\mab{\mbox {$\rm m_{AB}$}}
\def\ssp{\baselineskip=13pt plus 1pt minus 1pt}
\def\tsp{\baselineskip=5pt plus 1pt minus 1pt}
%
% ASTRO SYMBOLS (revised to work in/out mathmode).
%
\def\deg{\mbox {$^{\circ}$}}
\def\msun{\mbox {${\rm ~M_\odot}$}}
\def\zsun{\mbox {${\rm ~Z_{\odot}}$}}
\def\lsun{\mbox {${~\rm L_\odot}$}}
\def\msunyr{\mbox {$~{\rm M_\odot}$~yr$^{-1}$}}
\def\angs{\mbox {~\AA}}
\def\lya{\mbox {Ly$\alpha$}}
\def\lyb{\mbox {Ly$\beta$}}
\def\lygamma{\mbox {Ly$\gamma$}}
\def\lydelta{\mbox {Ly$\delta$}} % HI 937
\def\Ha{\mbox {H$\alpha$}}
\def\Hb{\mbox {H$\beta$}}
\def\Hg{\mbox {H$\gamma$}}
\def\tion{\mbox {$T_{\rm ion}$~}}
\def\ch{\mbox {$\bigtriangleup$}}
\def\grad{\mbox {$\bigtriangledown$}}
\def\lstar{\mbox {$L^*$}}
\def\line{\mbox {~$\lambda$}}
\def\lines{\mbox {~$\lambda\lambda$~}}
\def\h0{\mbox {~H$_0$}}
\def\q0{\mbox {~q$_0$}}
%
% **** LINE RATIOS ****
%
\def\auroral{[OIII]~$\lambda4363$~}
\def\auroral{[OIII]~$\lambda4363$~}
\def\ohsun{\mbox {(O/H)$_{\odot}$~}}

\def\O1ha{[OI]$\lambda6300$~/~H$\alpha$~}
\def\Ru{[OII]$\lambda\lambda3727$~/~[OIII]$\lambda5007$~}
\def\s2ha{[SII]$\lambda\lambda6717,31$~/~H$\alpha$~}
\def\2z2{HeII~$\lambda4686$~}
\def\z7{[NII]~$\lambda6583$ }
\def\N2{[NII]~$\lambda6583$~/~H$\alpha$~}
\def\16z2{[SII]~$\lambda\lambda6717, 6731$ }
\def\HgI{HgI~$\lambda4358$~}
\def\Sdensity{[SII]~$\lambda6717 / \lambda6731$}
\def\Temp{[OIII]~$\lambda\lambda4959 + 5007 ~{\rm to}~ \lambda4363$~}

%% file: table_obsgal.tex
% Galaxy Spec
\begin{deluxetable*}{lcccccc}
\label{tb:obs_gal}  
\tablecaption{Galaxy Spectroscopy}
\tabletypesize{\footnotesize}
\tablewidth{0pt}
\tablehead{
\colhead{Galaxy} &
\colhead{Galaxy RA} &
\colhead{Galaxy DEC} &
\colhead{Instrument} &
\colhead{Observing Date} &
\colhead{Exposure Time} &
\colhead{Reference}
\\
\colhead{} &
\colhead{(J2000)} &
\colhead{(J2000)} &
\colhead{} &
\colhead{} & 
\colhead{(s)} &
\colhead{}
}
%\decimalcolnumbers
\startdata
J040150$-$054047  & 04:01:50.48 & $-$05:40:47.07  & MMT/Red Channel  & 2021 Jan 9   & 1800 &  This work\\
J080702+360141    & 08:07:02.27 &   +36:01:41.15  & MMT/Blue Channel & 2020 Feb 28  & 2700 &  This work\\
J084356+261855    & 08:43:56.12 &   +26:18:55.28  & MMT/Blue Channel & 2020 Feb 28  & 1800 &  This work\\
J085215+171137    & 08:52:15.36 &   +17:11:37.07  & Keck/LRIS        & 2014 May 2   & 2640 &  \citet{Ho2017}\\
J091954+291345    & 09:19:54.11 &   +29:13:45.34  & Keck/LRIS        & 2015 Mar 22  & 1760 &  \citet{Ho2017}\\
J094330+053118    & 09:43:30.67 &   +05:31:18.11  & Keck/ESI         & 2016 Jan 15  & 4500 &  \citet{Kacprzak2019}$^{\dagger}$\\
J103643+565119    & 10:36:43.44 &   +56:51:19.00  & Keck/LRIS        & 2015 Mar 22  & 1760 &  \citet{Ho2017}\\
J104117+061018    & 10:41:17.80 &   +06:10:18.88  & Keck/ESI         & 2014 Apr 25  & 3300 &  \citet{Kacprzak2019}$^{\dagger}$\\
J113911$-$135108  & 11:39:11.52 & $-$13:51:08.69  & Keck/ESI         & 2016 Jan 15  & 1650 &  \citet{Kacprzak2019}$^{\dagger}$\\
J113909$-$135053  & 11:39:09.80 & $-$13:50:53.08  & Keck/ESI         & 2016 Jun 6   & 1800 &  \citet{Kacprzak2019}$^{\dagger}$\\
J123338+475757    & 12:33:38.87 &   +47:57:57.62  & MMT/Red Channel  & 2021 Jan 9   & 2400 &  This work\\
J132222+464546    & 13:22:22.46 &   +46:45:46.18  & Keck/ESI         & 2016 Jun 6   & 1500 &  \citet{Kacprzak2019}$^{\dagger}$\\ 
J135521+303320    & 13:55:21.20 &   +30:33:20.41  & Keck/LRIS        & 2014 May 3   & 3520 &  \citet{Ho2017}\\
J142459+382113    & 14:24:59.82 &   +38:21:13.37  & APO/DIS          & 2016 Apr 2   & 5100 &  \citet{Ho2017}\\
J154741+343350    & 15:47:41.46 &   +34:33:50.86  & Keck/LRIS        & 2014 May 3   & 1760 &  \citet{Ho2017}\\
J155505+362848    & 15:55:05.27 &   +36:28:48.40  & APO/DIS          & 2016 Jul 4   & 3200 &  \citet{Martin2019}\\
J160951+353838    & 16:09:51.62 &   +35:38:38.58  & Keck/LRIS        & 2014 May 2   & 1760 &  \citet{Ho2017}\\
J225400+160925    & 22:54:00.58 &   +16:09:25.83  & Keck/ESI         & 2016 Jun 6   & 1200 &  \citet{Kacprzak2019}$^{\dagger}$\\ 
\enddata
\tablenotetext{\dagger}{\small
       The ESI data were re-analyzed for galaxy rotation curve measurements.
       }
\end{deluxetable*}

%% file: table_galprop_new.tex
% Target basic info
\begin{deluxetable*}{llcccccccccc}
\label{tb:galprops}  
\tablecaption{Target Information and Galaxy Properties}
%\tablecolumns{8}
%\tabletypesize{\normalsize}
%\tabletypesize{\scriptsize}
\tabletypesize{\footnotesize}
\tablewidth{0pt}
\tablehead{
\colhead{Quasar} &
\colhead{Galaxy} &
\colhead{\zgal} &
\colhead{\logmstar} &
\colhead{\logmvir} & 
\colhead{\rvir} &
\colhead{\vinf} &
\colhead{$R_\mathrm{RC}$} &
\colhead{$i$} &
\colhead{$\alpha$} &
\colhead{$d$} &
\colhead{\brvir}
\\
\colhead{} &
\colhead{} &
\colhead{} &
\colhead{(\logmsun)} &
\colhead{(\logmsun)} &
\colhead{(kpc)} &
\colhead{(\kmstb)} &
\colhead{(kpc)} &
\colhead{(\deg)} &
\colhead{(\deg)} &
\colhead{(kpc)} &
\colhead{}
}
\decimalcolnumbers
\startdata
%QSO             GAL                z_gal                           logMs                   logMvir                 rvir   vc    rv     i                   alpha                   b    b_rvir  
J040148$-$054056 & J040150$-$054047 & 0.21966                     &   9.9\tablenotemark{a}  & 11.6                  & 168 & 140 & 4.8 & 58\tablenotemark{b} & 24\tablenotemark{b} &  89  & 0.53\\
J080704+360353   & J080702+360141   & 0.088061                    &  10.9\tablenotemark{c}  & 12.7                  & 426 & 250 & 1.4 & 71\tablenotemark{b} & 32\tablenotemark{b} &  231 & 0.54\\
J084349+261910   & J084356+261855   & 0.11284                     &  10.5\tablenotemark{c}  & 12.1                  & 261 & 190 & 1.5 & 55\tablenotemark{b} & 42\tablenotemark{b} &  184 & 0.71\\
J085215+171143   & J085215+171137   & 0.16921\tablenotemark{d}    &   9.8\tablenotemark{d}  & 11.5                  & 162 & 200 & 2.7 & 77\tablenotemark{d} & 23\tablenotemark{d} &  20  & 0.12\\
J091954+291408   & J091954+291345   & 0.23288\tablenotemark{d}    &  10.5\tablenotemark{d}  & 12.2                  & 264 & 250 & 3.4 & 73\tablenotemark{d} & 15\tablenotemark{d} &  88  & 0.34\\
J094331+053131   & J094330+053118   & 0.353052\tablenotemark{e}   &   9.4\tablenotemark{a}  & 10.3                  & 128 & 105 & 1.5 & 44\tablenotemark{e} & 8\tablenotemark{e}  &  99  & 0.77\\
J103640+565125   & J103643+565119   & 0.13629\tablenotemark{d}    &   9.9\tablenotemark{d}  & 11.6                  & 174 & 210 & 1.7 & 58\tablenotemark{d} & 21\tablenotemark{d} &  58  & 0.33\\
J104116+061016   & J104117+061018   & 0.442173\tablenotemark{e}   &  10.3                   & 12.0\tablenotemark{e} & 203 & 320 & 1.2 & 50\tablenotemark{e} & 4\tablenotemark{e}  &  57  & 0.28\\
J113910$-$135043 & J113911$-$135108 & 0.204194\tablenotemark{e}   &  10.1                   & 11.7\tablenotemark{e} & 184 & 175 & 2.4 & 82\tablenotemark{e} & 6\tablenotemark{e}  &  96  & 0.52\\
J113910$-$135043 & J113909$-$135053 & 0.319255\tablenotemark{e}   &  10.2                   & 11.9\tablenotemark{e} & 197 & 205 & 2.9 & 83\tablenotemark{e} & 39\tablenotemark{e} &  77  & 0.39\\
J123335+475800   & J123338+475757   & 0.22171                     &  10.6\tablenotemark{a}  & 12.2                  & 267 & 225 & 1.5 & 74\tablenotemark{b} & 5\tablenotemark{b}  &  141 & 0.53\\
J132222+464546   & J132222+464546   & 0.214431\tablenotemark{e}   &  10.6\tablenotemark{a}  & 12.2                  & 277 & 180 & 1.8 & 58\tablenotemark{e} & 14\tablenotemark{e} &  40  & 0.15\\
J135522+303324   & J135521+303320   & 0.20690\tablenotemark{d}    &   9.8\tablenotemark{d}  & 11.5                  & 157 & 155 & 2.8 & 72\tablenotemark{d} & 2\tablenotemark{d}  &  78  & 0.50\\
J142501+382100   & J142459+382113   & 0.21295\tablenotemark{d}    &  10.2\tablenotemark{d}  & 11.7                  & 190 & 190 & 2.2 & 61\tablenotemark{d} & 7\tablenotemark{d}  &  83  & 0.44\\
J154741+343357   & J154741+343350   & 0.18392\tablenotemark{d}    &  10.0\tablenotemark{d}  & 11.6                  & 173 & 175 & 1.6 & 59\tablenotemark{d} & 4\tablenotemark{d}  &  26  & 0.15\\
J155504+362847   & J155505+362848   & 0.18926\tablenotemark{d}    &  10.1\tablenotemark{d}  & 11.7                  & 188 & 255 & 3.6 & 55\tablenotemark{d} & 45\tablenotemark{d} &  35  & 0.19\\
J160951+353843   & J160951+353838   & 0.28940\tablenotemark{d}    &   9.9\tablenotemark{d}  & 11.6                  & 164 & 45  & 2.7 & 65\tablenotemark{d} & 25\tablenotemark{d} &  25  & 0.15\\
J225357+160853   & J225400+160925   & 0.390013\tablenotemark{e}   &  10.5                   & 12.2\tablenotemark{e} & 238 & 175 & 3.3 & 76\tablenotemark{e} & 24\tablenotemark{e} &  284 & 1.19\\
\enddata
\tablecomments{
    (1)  Name of the quasar.  
    (2)  Name of the galaxy.
    (3)  Galaxy systemic redshift. 
         The typical uncertainty is 25 \kms.
    (4)  Galaxy stellar mass, rescaled to Chabrier IMF when necessary following \citet{MadauDickinson2014}.
         The typical uncertainty is 0.2 dex.
    (5)  Halo mass, derived using the stellar mass--halo mass relation in \citet{Behroozi2010} except for those adopted from \citet{Kacprzak2019}.  
         The typical uncertainty of halo mass is 0.2 dex; this does not include the intrinsic scatter in stellar mass at a single halo mass 
         from the mean stellar mass--halo mass relation.
    (6)  Halo virial radius, defined by the overdensity in \citet{BryanNorman1998} with respect to the critical density evaluated at the galaxy systemic redshift.  
         The virial radius has a typical uncertainty of 15 kpc.
    (7)  Asymptotic rotation speed of the galaxy disk.
         The typical uncertainty is 15 \kms.
    (8)  Turnover radius from the arctangent rotation curve model.
         The turnover radius has a typical uncertainty of 0.3 kpc.
    (9)  Inclination angle of galactic disk.
         The typical uncertainty is 3\deg.
    (10) Azimuthal angle, the angle between the quasar sightline and the galaxy major axis.  
         The typical uncertainty of the azimuthal angle is 2\deg, which is dominated by the uncertainty of the position angle of the galaxy major axis.
    (11) Sightline impact parameter, the projected distance between the centers of the galaxy and quasar. 
         The typical uncertainty of impact parameter is less than 1 kpc, which comes from the $\lesssim0\farcs1$ uncertainty in the angular separation between
         the galaxy and the quasar.
    (12) Sightline impact parameter normalized by the halo virial radius.  
         The typical uncertainty is 0.07, dominated by the uncertainty in the virial radius.
}
\tablenotetext{a}{\small
        Adopted from \citet{Werk2012}.  Galaxy stellar masses are rescaled from Salpeter to Chabrier IMF.
        }
\tablenotetext{b}{\small
        Derived from SDSS DR17 \texttt{PhotoObj} catalog for $r$-band images.
        Inclination angle of the galactic disk is calculated from the disk axis ratio using the Hubble formula \citep{Hubble1926} with $q_0=0.2$.
        Azimuthal angle of the quasar sightline is calculated from the galaxy major-axis position angle in the catalog.
        }
\tablenotetext{c}{\small
        Adopted from \citet{Heckman2017}.  Galaxy stellar masses are rescaled from Kroupa to Chabrier IMF.
        }
\tablenotetext{d}{\small
        Adopted from \citet{Martin2019}.
        }
\tablenotetext{e}{\small
        Adopted from \citet{Kacprzak2019}.  With their halo masses, the stellar masses are re-derived using the stellar mass--halo mass relation from \citet{Behroozi2010}.
        }
\end{deluxetable*}

%% file: table_obsqso.tex
% QSO Observation
\begin{deluxetable*}{lcccccc}
\label{tb:obs_qso}  
\tablecaption{New and Archival Quasar Spectroscopy with \textit{HST/COS}}
\tabletypesize{\footnotesize}
\tablewidth{0pt}
\tablehead{
\colhead{Quasar} &
\colhead{Quasar RA} &
\colhead{Quasar DEC} &
\colhead{$z_\mathrm{qso}$} &
\colhead{Grating} &
\colhead{Exposure Time} &
\colhead{Program ID}
\\
\colhead{} &
\colhead{(J2000)} &
\colhead{(J2000)} &
\colhead{} &
\colhead{} &
\colhead{(s)} &
\colhead{}
}
%\decimalcolnumbers
\startdata
J040148$-$054056 & 04:01:48.98 & $-$05:40:56.5 & 0.570 & G130M/G160M  & 5377/5912   & 11598\\
J080704+360353   & 08:07:04.88 &   +36:03:53.5 & 0.468 & G130M/G160M  & 5378/8612   & 13862\\
J084349+261910   & 08:43:49.75 &   +26:19:10.7 & 0.257 & G130M        & 2194        & 13862\\
J085215+171143   & 08:52:15.34 &   +17:11:43.8 & 0.562 & G130M        & 10810       & 15866\\
J091954+291408   & 09:19:54.28 &   +29:14:08.3 & 0.722 & G130M        & 4933        & 15866\\
J094331+053131   & 09:43:31.61 &   +05:31:31.4 & 0.564 & G130M/G160M  & 3662/3945   & 11598\\
J103640+565125   & 10:36:40.74 &   +56:51:25.9 & 1.264 & G130M        & 8641        & 15866\\
J104116+061016   & 10:41:17.16 &   +06:10:16.8 & 1.270 & G160M        & 14183       & 12252\\
J113910$-$135043 & 11:39:10.70 & $-$13:50:43.6 & 0.557 & G130M        & 7669        & 12275\\
J123335+475800   & 12:33:35.07 &   +47:58:00.4 & 0.382 & G130M/G160M  & 3885/4218   & 11598\\
                 &             &               &       & G130M        & 2044        & 13033\\
J132222+464546   & 13:22:22.67 &   +46:45:35.2 & 0.374 & G130M/G160M  & 3902/4249   & 11598\\
                 &             &               &       & G130M        & 2044        & 13033\\
J135522+303324   & 13:55:22.89 &   +30:33:24.7 & 0.382 & G130M        & 8092        & 15866\\
J142501+382100   & 14:25:01.46 &   +38:21:00.5 & 1.145 & G130M        & 8122        & 15866\\
J154741+343357   & 15:47:41.88 &   +34:33:57.3 & 0.928 & G130M        & 5197        & 15866\\
J155504+362847   & 15:55:04.39 &   +36:28:47.9 & 0.714 & G130M/G160M  & 5151/6030   & 11598\\
J160951+353843   & 16:09:51.81 &   +35:38:43.7 & 0.828 & G130M        & 10639       & 15866\\
J225357+160853   & 22:53:57.74 &   +16:08:53.5 & 0.859 & G130M/G160M  & 4473/4514   & 13398\\
\enddata
\end{deluxetable*}

%% file: table_voigt_o6_new.tex
% Voigt Profile for All
\begin{deluxetable*}{llccccc}
\label{tb:voigt_o6}  
\tablecaption{Results from Voigt Profile Fitting of \ion{O}{6} Absorption}
\tabletypesize{\scriptsize}
\tablewidth{0pt}
\tablehead{
\colhead{Quasar} &
\colhead{Galaxy} &
\colhead{$z_\mathrm{gal}$} &
\colhead{$v$} &
\colhead{$b$} &
\colhead{$\log N_\mathrm{OVI}$} & 
%\colhead{$\langle v_\mathrm{O\ VI} \rangle_w$}
\colhead{\velwovi}
\\
\colhead{} &
\colhead{} &
\colhead{} &
\colhead{(\kmstb)} &
\colhead{(\kmstb)} &
\colhead{($\log \mathrm{cm}^{-2}$)} &
\colhead{(\kmstb)}
}
\decimalcolnumbers
\startdata
%QSO               GAL                z_gal         v                        b                      logN      
J040148$-$054056 & J040150$-$054047 & 0.21966    &  $   -40.6  \pm 4.2 $  &  $  26.2  \pm 6.2 $  &  $ 14.26 \pm 0.06 $ & $-6.6^{+3.7}_{-4.3}$    \\
                 &                  &            &  $    23.6  \pm 3.0 $  &  $  23.1  \pm 4.3 $  &  $ 14.31 \pm 0.06 $ &                         \\
J080704+360353   & J080702+360141   & 0.088061   &  $  -115.4  \pm 7.7 $  &  $  46.4 \pm 11.4 $  &  $ 14.43 \pm 0.08 $ & $6.1^{+30.8}_{-29.7}$   \\
                 &                  &            &  $   107.7 \pm 10.5 $  &  $  12.0 \pm 17.5 $  &  $ 14.12 \pm 0.23 $ &                         \\
                 &                  &            &  $   175.2 \pm 12.1 $  &  $  12.9 \pm 21.6 $  &  $ 13.99 \pm 0.29 $ &                         \\
J084349+261910   & J084356+261855   & 0.11284    &  $  -188.3 \pm 15.1 $  &  $  30.6 \pm 22.7 $  &  $ 13.70 \pm 0.19 $ & $32.5^{+17.9}_{-21.5}$  \\
                 &                  &            &  $    80.5  \pm 8.9 $  &  $  70.3 \pm 12.7 $  &  $ 14.37 \pm 0.05 $ &                         \\
J085215+171143   & J085215+171137   & 0.16921    &  $   -39.5  \pm 9.2 $  &  $  64.9 \pm 15.5 $  &  $ 14.64 \pm 0.07 $ & $-39.0^{+9.1}_{-8.9}$   \\
J094331+053131   & J094330+053118   & 0.353052   &  $    49.6  \pm 1.8 $  &  $  20.1  \pm 2.6 $  &  $ 14.64 \pm 0.08 $ & $93.0^{+13.5}_{-10.7}$  \\
                 &                  &            &  $   161.1  \pm 4.4 $  &  $  11.0 \pm 10.0 $  &  $ 13.32 \pm 0.31 $ &                         \\
                 &                  &            &  $   397.9  \pm 4.4 $  &  $  13.6  \pm 7.0 $  &  $ 13.75 \pm 0.12 $ &                         \\
J103640+565125   & J103643+565119   & 0.13629    &  ...                   &  ...                 & $\leq 13.2$$^{\dagger}$ & ...                 \\
J104116+061016   & J104117+061018   & 0.442173   &  $  -182.7  \pm 5.0 $  &  $  15.8  \pm 7.7 $  &  $ 13.82 \pm 0.16 $ & $-111.7^{+7.8}_{-7.8}$  \\
                 &                  &            &  $  -123.9  \pm 4.8 $  &  $  30.9 \pm 10.7 $  &  $ 14.28 \pm 0.09 $ &                         \\
                 &                  &            &  $   -49.9  \pm 7.5 $  &  $  30.2  \pm 9.8 $  &  $ 14.06 \pm 0.12 $ &                         \\
J113910$-$135043 & J113911$-$135108 & 0.204194   &  $  -227.0  \pm 8.7 $  &  $  10.5 \pm 19.8 $  &  $ 13.26 \pm 0.32 $ & $5.1^{+20.7}_{-22.2}$   \\
                 &                  &            &  $   -13.2 \pm 14.9 $  &  $  36.7 \pm 18.6 $  &  $ 14.14 \pm 0.20 $ &                         \\
                 &                  &            &  $    61.0 \pm 11.9 $  &  $  40.7 \pm 29.7 $  &  $ 14.12 \pm 0.29 $ &                         \\
J113910$-$135043 & J113909$-$135053 & 0.319255   &  $   -83.9  \pm 4.3 $  &  $  59.8  \pm 6.9 $  &  $ 14.11 \pm 0.04 $ & $-10.1^{+8.4}_{-7.0}$   \\
                 &                  &            &  $    37.7  \pm 2.0 $  &  $  29.4  \pm 3.4 $  &  $ 14.01 \pm 0.04 $ &                         \\
                 &                  &            &  $   159.8  \pm 6.6 $  &  $  14.3 \pm 10.8 $  &  $ 13.45 \pm 0.16 $ &                         \\
J123335+475800   & J123338+475757   & 0.22171    &  $    60.9 \pm 11.0 $  &  $ 112.2 \pm 15.3 $  &  $ 14.59 \pm 0.05 $ & $60.5^{+11.0}_{-10.6}$  \\
J132222+464546   & J132222+464546   & 0.214431   &  $   -54.1  \pm 6.2 $  &  $  36.8  \pm 7.1 $  &  $ 14.39 \pm 0.10 $ & $-20.3^{+14.0}_{-11.6}$ \\
                 &                  &            &  $    37.3 \pm 22.0 $  &  $  64.8 \pm 22.4 $  &  $ 14.19 \pm 0.17 $ &                         \\
J135522+303324   & J135521+303320   & 0.20690    &  $  -120.5 \pm 28.6 $  &  $  90.8 \pm 44.6 $  &  $ 13.86 \pm 0.21 $ & $-23.6^{+11.8}_{-16.2}$ \\
                 &                  &            &  $     7.0  \pm 4.0 $  &  $  38.7  \pm 5.6 $  &  $ 14.36 \pm 0.05 $ &                         \\
J142501+382100   & J142459+382113   & 0.21295    &  $   -27.7  \pm 5.9 $  &  $   7.3 \pm 13.0 $  &  $ 13.62 \pm 0.34 $ & $19.2^{+7.6}_{-9.9}$    \\
                 &                  &            &  $    32.4  \pm 4.7 $  &  $  27.9  \pm 8.2 $  &  $ 14.18 \pm 0.07 $ &                         \\
J154741+343357   & J154741+343350   & 0.18392    &  $  -145.3  \pm 4.8 $  &  $  33.3  \pm 7.5 $  &  $ 14.47 \pm 0.07 $ & $15.9^{+16.4}_{-18.5}$  \\
                 &                  &            &  $   -11.4  \pm 9.4 $  &  $  49.5 \pm 13.6 $  &  $ 14.68 \pm 0.09 $ &                         \\
                 &                  &            &  $    70.1  \pm 5.1 $  &  $  18.3  \pm 9.7 $  &  $ 14.99 \pm 0.23 $ &                         \\
                 &                  &            &  $   125.3 \pm 16.0 $  &  $  20.8 \pm 25.9 $  &  $ 13.71 \pm 0.38 $ &                         \\
J155504+362847   & J155505+362848   & 0.18926    &  $   -27.5  \pm 6.8 $  &  $  35.8  \pm 7.4 $  &  $ 14.53 \pm 0.09 $ & $-54.5^{+9.1}_{-11.0}$  \\
                 &                  &            &  $  -107.9 \pm 13.9 $  &  $  41.2 \pm 16.6 $  &  $ 14.26 \pm 0.16 $ &                         \\
J160951+353843   & J160951+353838   & 0.28940    &  $  -159.5 \pm 17.4 $  &  $  32.6 \pm 28.9 $  &  $ 14.24 \pm 0.29 $ & $23.2^{+13.4}_{-13.9}$  \\
                 &                  &            &  $   -96.5 \pm 10.0 $  &  $  26.5 \pm 14.1 $  &  $ 14.36 \pm 0.21 $ &                         \\
                 &                  &            &  $   -30.0  \pm 6.7 $  &  $  18.4  \pm 1.7 $  &  $ 14.50 \pm 0.19 $ &                         \\
                 &                  &            &  $     8.4  \pm 9.1 $  &  $  17.6 \pm 16.1 $  &  $ 14.29 \pm 0.29 $ &                         \\
                 &                  &            &  $    92.8  \pm 5.6 $  &  $  57.8  \pm 5.9 $  &  $ 15.10 \pm 0.04 $ &                         \\
J225357+160853   & J225400+160925   & 0.390013   &  $   109.6 \pm 16.0 $  &  $  21.3 \pm 16.4 $  &  $ 13.91 \pm 0.34 $ & $151.7^{+11.5}_{-12.8}$ \\
                 &                  &            &  $   149.5  \pm 6.5 $  &  $  17.6  \pm 7.8 $  &  $ 14.24 \pm 0.17 $ &                         \\
                 &                  &            &  $   232.7 \pm 14.0 $  &  $  38.0 \pm 23.8 $  &  $ 13.67 \pm 0.19 $ &                         \\
\hline
\multicolumn{7}{c}{Excluded from \oVI\ kinematics analysis due to group environment}\\
\hline
J091954+291408   & J091954+291345   & 0.23288    &  $  -209.4  \pm 4.9 $  &  $  16.2  \pm 8.9 $  &  $ 14.21 \pm 0.19 $ & $-52.7^{+30.3}_{-28.3}$ \\
                 &                  &            &  $  -110.9  \pm 5.1 $  &  $  12.0  \pm 9.0 $  &  $ 14.04 \pm 0.19 $ &                         \\
                 &                  &            &  $   -38.4 \pm 11.7 $  &  $  28.8 \pm 20.7 $  &  $ 13.85 \pm 0.17 $ &                         \\
                 &                  &            &  $    46.9 \pm 11.4 $  &  $   7.3 \pm 22.0 $  &  $ 13.61 \pm 0.23 $ &                         \\
                 &                  &            &  $   100.9 \pm 16.3 $  &  $  16.1 \pm 37.8 $  &  $ 13.58 \pm 0.41 $ &                         \\
                 &                  &            &  $   162.9  \pm 9.7 $  &  $  15.2 \pm 16.7 $  &  $ 13.93 \pm 0.16 $ &                         \\
\enddata
\tablenotetext{\dagger}{\small
        The 2$\sigma$ upper limit measured from \oVI\ 1031.
        %within the velocity window of three times the resolution element centered at the galaxy systemic velocity.
       }
\tablecomments{
    (1) Name of the quasar.  
    (2) Name of the galaxy.
    (3) Galaxy systemic redshift.
    (4) Velocity centroid of individual component.
    (5) Line width, i.e., Doppler parameter, of the velocity component.
    (6) Logarithm of the column density of the velocity component.
    (7) Column-density-weighted velocity.
}
\end{deluxetable*}

%% file: table_o6classification.tex
% OVI VELOCITY COMPONENTS WITH MATCHING LOW IONS
\begin{deluxetable}{llcc}
\label{tb:o6class}  
\tablecaption{Classification of \oVI\ Velocity Components}
%Based on Rotation Kinematics and Relationship with Low Ions}
\tabletypesize{\footnotesize}
\tablewidth{0pt}
\tablehead{
\colhead{\oVI\ Category} &
\colhead{Kinematics Class} &
\colhead{$n_\mathrm{comp, O\ VI}$} &
\colhead{Ratio}
}
\decimalcolnumbers
\startdata
%\hline
\multicolumn{4}{c}{All Sightlines (exclude J0919$+$29)}\\
\hline
All                  & Any               & 41    & ... \\
                     & Corotate          & 21    & 0.51\\
                     & Systemic Velocity & 5     & 0.12\\
                     & Counter-rotate    & 15    & 0.37\\
\hline
\multicolumn{4}{c}{Sightlines with \siII\ and \siIII\ Spectral Coverage
($n_\mathrm{comp, O\ VI} = 29$)$^{\dagger}$}\\
\hline
All                  & Any               & 29    & ...  \\
                     & Corotate          & 16    & 0.55 \\
                     & Systemic Velocity & 3     & 0.10 \\
                     & Counter-rotate    & 10    & 0.34 \\
\hline
With \siII\ match    & Any               & 10    & ...  \\
                     & Corotate          & 9     & 0.90 \\
                     & Systemic Velocity & 1     & 0.10 \\
                     & Counter-rotate    & 0     & 0    \\
No \siII\ match      & Any               & 19    & ...  \\
                     & Corotate          & 7     & 0.37 \\
                     & Systemic Velocity & 2     & 0.11 \\
                     & Counter-rotate    & 10    & 0.53 \\
\hline
%
% \hline
% \multicolumn{4}{c}{Sightlines with \siIII\ Spectral Coverage
% ($n_\mathrm{comp, O\ VI} = 29$)}\\
% \hline
With \siIII\ match   & Any               & 14    & ... \\
                     & Corotate          & 10    & 0.71\\
                     & Systemic Velocity & 2     & 0.14\\
                     & Counter-rotate    & 2     & 0.14\\
No \siIII\ match     & Any               & 15    & ... \\
                     & Corotate          & 6     & 0.40\\
                     & Systemic Velocity & 1     & 0.07\\
                     & Counter-rotate    & 8     & 0.53\\
\enddata
\tablenotetext{\dagger}{\small
       The sightlines either have spectral coverage for both
       \siII\ and \siIII\ simultaneously or neither.
       }
\tablecomments{
    (1) Category of \oVI\ velocity components.
        % with or without low-ion match.
        % ``all'' means regardless of the comparison with low ions.
    (2) Classification of \oVI\ velocity components based on their 
        Doppler sign relative to the disk rotation
        at the quasar side of the galaxy disk.
    (3) Number of \oVI\ velocity components.  
    (4) The ratio of the number of \oVI\ velocity components between
        the kinematics class and the total within the same \oVI\
        category.
}
\end{deluxetable}

%% file: table_voigt_all.tex
% Voigt Profile for All
\startlongtable
\begin{deluxetable*}{llccccc}
\label{tb:voigt_all}  
\tablecaption{Results for Voigt Profile Fitting}
\tabletypesize{\scriptsize}
\tablewidth{0pt}
\tablehead{
\colhead{Quasar} &
\colhead{Galaxy} &
\colhead{$z_{gal}$} &
\colhead{ion} &
\colhead{$v$} &
\colhead{$b$} &
\colhead{$\log N$}
\\
\colhead{} &
\colhead{} &
\colhead{} &
\colhead{} &
\colhead{(\kmstb)} &
\colhead{(\kmstb)} &
\colhead{($\log \mathrm{cm}^{-2}$)}
}
\startdata
%QSO               GAL                z_gal        ion             v                        b                      logN      
J040148$-$054056 & J040150$-$054047 & 0.21966    & \ion{H}{1}   &  $   -47.2  \pm 3.2 $  &  $  25.8  \pm 4.0 $  &  $ 15.51 \pm 0.08 $ \\
J040148$-$054056 & J040150$-$054047 & 0.21966    & \ion{H}{1}   &  $   -22.4  \pm 3.7 $  &  $  50.7  \pm 2.1 $  &  $ 15.16 \pm 0.11 $ \\
J040148$-$054056 & J040150$-$054047 & 0.21966    & \ion{H}{1}   &  $  -621.6  \pm 3.4 $  &  $  17.1  \pm 4.9 $  &  $ 13.57 \pm 0.07 $ \\
J040148$-$054056 & J040150$-$054047 & 0.21966    & \ion{H}{1}   &  $  -559.2  \pm 1.5 $  &  $  21.2  \pm 1.4 $  &  $ 14.61 \pm 0.04 $ \\
J040148$-$054056 & J040150$-$054047 & 0.21966    & \ion{O}{6}   &  $   -40.6  \pm 4.2 $  &  $  26.2  \pm 6.2 $  &  $ 14.26 \pm 0.06 $ \\
J040148$-$054056 & J040150$-$054047 & 0.21966    & \ion{O}{6}   &  $    23.6  \pm 3.0 $  &  $  23.1  \pm 4.3 $  &  $ 14.31 \pm 0.06 $ \\
J040148$-$054056 & J040150$-$054047 & 0.21966    & \ion{C}{3}   &  $   -52.4  \pm 4.9 $  &  $  12.2  \pm 7.7 $  &  $ 13.80 \pm 0.33 $ \\
J040148$-$054056 & J040150$-$054047 & 0.21966    & \ion{C}{3}   &  $     0.5  \pm 4.9 $  &  $  22.4  \pm 8.2 $  &  $ 14.10 \pm 0.28 $ \\
J040148$-$054056 & J040150$-$054047 & 0.21966    & \ion{C}{3}   &  $  -556.5 \pm 11.8 $  &  $  29.2 \pm 18.6 $  &  $ 12.92 \pm 0.18 $ \\
J040148$-$054056 & J040150$-$054047 & 0.21966    & \ion{N}{3}   &  $   -30.0  \pm 8.1 $  &  $   9.6 \pm 17.6 $  &  $ 13.72 \pm 0.36 $ \\
J040148$-$054056 & J040150$-$054047 & 0.21966    & \ion{N}{3}   &  $    23.1 \pm 11.8 $  &  $  74.9 \pm 11.2 $  &  $ 14.55 \pm 0.09 $ \\
J040148$-$054056 & J040150$-$054047 & 0.21966    & \ion{Si}{3}  &  $   -26.1  \pm 6.1 $  &  $  32.5  \pm 9.4 $  &  $ 12.64 \pm 0.08 $ \\
J040148$-$054056 & J040150$-$054047 & 0.21966    & \ion{Si}{3}  &  $  -606.1 \pm 17.7 $  &  $  46.0 \pm 27.1 $  &  $ 12.35 \pm 0.18 $ \\
J040148$-$054056 & J040150$-$054047 & 0.21966    & \ion{Si}{4}  &  $   -43.5  \pm 6.1 $  &  $  15.5  \pm 9.3 $  &  $ 12.94 \pm 0.14 $ \\
J040148$-$054056 & J040150$-$054047 & 0.21966    & \ion{S}{4}   &  $   -19.4  \pm 3.9 $  &  $  29.3  \pm 5.6 $  &  $ 14.77 \pm 0.07 $ \\
J080704+360353   & J080702+360141   & 0.088061   & \ion{H}{1}   &  $  -720.2  \pm 3.0 $  &  $  31.9  \pm 4.9 $  &  $ 14.30 \pm 0.12 $ \\
J080704+360353   & J080702+360141   & 0.088061   & \ion{H}{1}   &  $  -634.3 \pm 12.9 $  &  $   7.3 \pm 37.7 $  &  $ 12.99 \pm 0.26 $ \\
J080704+360353   & J080702+360141   & 0.088061   & \ion{H}{1}   &  $  -584.9  \pm 7.2 $  &  $  10.8 \pm 16.7 $  &  $ 13.39 \pm 0.23 $ \\
J080704+360353   & J080702+360141   & 0.088061   & \ion{H}{1}   &  $  -214.9  \pm 3.3 $  &  $  33.0  \pm 3.7 $  &  $ 14.80 \pm 0.15 $ \\
J080704+360353   & J080702+360141   & 0.088061   & \ion{H}{1}   &  $  -106.9  \pm 3.3 $  &  $  23.0  \pm 4.0 $  &  $ 14.80 \pm 0.24 $ \\
J080704+360353   & J080702+360141   & 0.088061   & \ion{H}{1}   &  $   -40.0  \pm 4.7 $  &  $   5.0  \pm 2.5 $  &  $ 14.04 \pm 0.74 $ \\
J080704+360353   & J080702+360141   & 0.088061   & \ion{H}{1}   &  $   105.3  \pm 6.9 $  &  $   7.5 \pm 22.6 $  &  $ 13.20 \pm 0.65 $ \\
J080704+360353   & J080702+360141   & 0.088061   & \ion{H}{1}   &  $   174.1  \pm 4.7 $  &  $  17.9  \pm 8.7 $  &  $ 13.50 \pm 0.10 $ \\
J080704+360353   & J080702+360141   & 0.088061   & \ion{O}{6}   &  $  -115.4  \pm 7.7 $  &  $  46.4 \pm 11.4 $  &  $ 14.43 \pm 0.08 $ \\
J080704+360353   & J080702+360141   & 0.088061   & \ion{O}{6}   &  $   107.7 \pm 10.5 $  &  $  12.0 \pm 17.5 $  &  $ 14.12 \pm 0.23 $ \\
J080704+360353   & J080702+360141   & 0.088061   & \ion{O}{6}   &  $   175.2 \pm 12.1 $  &  $  12.9 \pm 21.6 $  &  $ 13.99 \pm 0.29 $ \\
J080704+360353   & J080702+360141   & 0.088061   & \ion{C}{4}   &  $  -201.1  \pm 9.4 $  &  $  24.9 \pm 14.9 $  &  $ 13.45 \pm 0.21 $ \\
J080704+360353   & J080702+360141   & 0.088061   & \ion{C}{4}   &  $  -105.5  \pm 6.3 $  &  $  39.6  \pm 9.3 $  &  $ 14.07 \pm 0.08 $ \\
J080704+360353   & J080702+360141   & 0.088061   & \ion{C}{4}   &  $   145.5  \pm 7.7 $  &  $  66.7 \pm 10.5 $  &  $ 14.21 \pm 0.06 $ \\
J080704+360353   & J080702+360141   & 0.088061   & \ion{N}{5}   &  $   146.9 \pm 12.1 $  &  $  63.8 \pm 18.3 $  &  $ 13.86 \pm 0.11 $ \\
J084349+261910   & J084356+261855   & 0.11284    & \ion{H}{1}   &  $  -175.1 \pm 10.8 $  &  $  11.4 \pm 28.5 $  &  $ 12.90 \pm 0.23 $ \\
J084349+261910   & J084356+261855   & 0.11284    & \ion{H}{1}   &  $    92.7  \pm 2.4 $  &  $  45.4  \pm 3.3 $  &  $ 14.50 \pm 0.06 $ \\
J084349+261910   & J084356+261855   & 0.11284    & \ion{H}{1}   &  $   254.0  \pm 9.4 $  &  $  42.6 \pm 12.5 $  &  $ 13.42 \pm 0.11 $ \\
J084349+261910   & J084356+261855   & 0.11284    & \ion{O}{6}   &  $  -188.3 \pm 15.1 $  &  $  30.6 \pm 22.7 $  &  $ 13.70 \pm 0.19 $ \\
J084349+261910   & J084356+261855   & 0.11284    & \ion{O}{6}   &  $    80.5  \pm 8.9 $  &  $  70.3 \pm 12.7 $  &  $ 14.37 \pm 0.05 $ \\
J084349+261910   & J084356+261855   & 0.11284    & \ion{Si}{3}  &  $  -156.8  \pm 9.7 $  &  $  21.5 \pm 20.3 $  &  $ 12.40 \pm 0.18 $ \\
J084349+261910   & J084356+261855   & 0.11284    & \ion{Si}{3}  &  $    87.6  \pm 8.4 $  &  $  10.3 \pm 29.0 $  &  $ 12.44 \pm 0.65 $ \\
J084349+261910   & J084356+261855   & 0.11284    & \ion{Si}{2}  &  $  -152.5  \pm 7.8 $  &  $  13.0 \pm 19.1 $  &  $ 12.86 \pm 0.16 $ \\
J085215+171143   & J085215+171137   & 0.16921    & \ion{H}{1}   &  $  -807.7  \pm 3.3 $  &  $  36.2  \pm 4.8 $  &  $ 13.74 \pm 0.04 $ \\
J085215+171143   & J085215+171137   & 0.16921    & \ion{H}{1}   &  $  -600.5  \pm 8.5 $  &  $  34.4 \pm 13.0 $  &  $ 13.23 \pm 0.11 $ \\
J085215+171143   & J085215+171137   & 0.16921    & \ion{H}{1}   &  $   -57.7  \pm 1.5 $  &  $  41.5  \pm 0.6 $  &  $ 18.72 \pm 0.05 $ \\
J085215+171143   & J085215+171137   & 0.16921    & \ion{H}{1}   &  $    91.0 \pm 42.8 $  &  $  69.1 \pm 11.0 $  &  $ 13.74 \pm 0.37 $ \\
J085215+171143   & J085215+171137   & 0.16921    & \ion{C}{2}   &  $  -899.7  \pm 4.6 $  &  $  18.9  \pm 8.4 $  &  $ 13.77 \pm 0.09 $ \\
J085215+171143   & J085215+171137   & 0.16921    & \ion{C}{2}   &  $  -832.5  \pm 2.3 $  &  $  23.2  \pm 4.1 $  &  $ 14.24 \pm 0.04 $ \\
J085215+171143   & J085215+171137   & 0.16921    & \ion{O}{6}   &  $  -882.8  \pm 4.1 $  &  $   4.1  \pm 9.7 $  &  $ 13.54 \pm 0.23 $ \\
J085215+171143   & J085215+171137   & 0.16921    & \ion{O}{6}   &  $  -800.8  \pm 2.6 $  &  $   6.8  \pm 5.2 $  &  $ 13.94 \pm 0.21 $ \\
J085215+171143   & J085215+171137   & 0.16921    & \ion{O}{6}   &  $  -607.9  \pm 8.5 $  &  $  45.6 \pm 13.2 $  &  $ 13.85 \pm 0.09 $ \\
J085215+171143   & J085215+171137   & 0.16921    & \ion{O}{6}   &  $   -39.5  \pm 9.2 $  &  $  64.9 \pm 15.5 $  &  $ 14.64 \pm 0.07 $ \\
J085215+171143   & J085215+171137   & 0.16921    & \ion{C}{3}   &  $  -808.7 \pm 17.7 $  &  $  40.9 \pm 31.1 $  &  $ 12.92 \pm 0.20 $ \\
J085215+171143   & J085215+171137   & 0.16921    & \ion{C}{3}   &  $  -246.7 \pm 10.3 $  &  $   8.5 \pm 21.9 $  &  $ 12.68 \pm 0.23 $ \\
J085215+171143   & J085215+171137   & 0.16921    & \ion{C}{3}   &  $   -70.8  \pm 6.2 $  &  $  59.5 \pm 10.8 $  &  $ 15.11 \pm 0.34 $ \\
J085215+171143   & J085215+171137   & 0.16921    & \ion{C}{3}   &  $    50.5  \pm 6.9 $  &  $  11.1 \pm 13.1 $  &  $ 13.55 \pm 0.48 $ \\
J085215+171143   & J085215+171137   & 0.16921    & \ion{N}{2}   &  $  -802.3 \pm 15.4 $  &  $  39.8 \pm 25.5 $  &  $ 13.58 \pm 0.18 $ \\
J085215+171143   & J085215+171137   & 0.16921    & \ion{N}{2}   &  $  -224.6  \pm 8.5 $  &  $   4.9 \pm 27.5 $  &  $ 13.25 \pm 0.54 $ \\
J085215+171143   & J085215+171137   & 0.16921    & \ion{N}{2}   &  $  -133.3  \pm 7.7 $  &  $  16.6 \pm 11.9 $  &  $ 13.77 \pm 0.21 $ \\
J085215+171143   & J085215+171137   & 0.16921    & \ion{N}{2}   &  $   -47.7  \pm 9.7 $  &  $  43.9 \pm 13.4 $  &  $ 14.83 \pm 0.10 $ \\
J085215+171143   & J085215+171137   & 0.16921    & \ion{N}{2}   &  $     6.9  \pm 5.1 $  &  $  11.8 \pm 11.8 $  &  $ 14.51 \pm 0.55 $ \\
J085215+171143   & J085215+171137   & 0.16921    & \ion{N}{2}   &  $    70.3  \pm 7.7 $  &  $  19.5 \pm 13.6 $  &  $ 13.65 \pm 0.15 $ \\
J085215+171143   & J085215+171137   & 0.16921    & \ion{N}{3}   &  $  -121.0 \pm 13.3 $  &  $  34.5 \pm 12.8 $  &  $ 14.48 \pm 0.19 $ \\
J085215+171143   & J085215+171137   & 0.16921    & \ion{N}{3}   &  $   -42.1 \pm 12.8 $  &  $  34.1 \pm 22.7 $  &  $ 14.97 \pm 0.18 $ \\
J085215+171143   & J085215+171137   & 0.16921    & \ion{N}{3}   &  $    15.1 \pm 10.0 $  &  $  16.0 \pm 12.4 $  &  $ 14.90 \pm 0.97 $ \\
J085215+171143   & J085215+171137   & 0.16921    & \ion{O}{1}   &  $  -835.9  \pm 3.6 $  &  $   9.6  \pm 7.8 $  &  $ 14.44 \pm 0.23 $ \\
J085215+171143   & J085215+171137   & 0.16921    & \ion{O}{1}   &  $  -645.9  \pm 6.9 $  &  $  24.5 \pm 12.6 $  &  $ 14.25 \pm 0.10 $ \\
J085215+171143   & J085215+171137   & 0.16921    & \ion{O}{1}   &  $  -220.5  \pm 9.2 $  &  $  26.6 \pm 17.1 $  &  $ 14.13 \pm 0.12 $ \\
J085215+171143   & J085215+171137   & 0.16921    & \ion{O}{1}   &  $   -62.6  \pm 4.1 $  &  $  34.6  \pm 6.3 $  &  $ 14.88 \pm 0.05 $ \\
J085215+171143   & J085215+171137   & 0.16921    & \ion{O}{1}   &  $     7.2  \pm 4.9 $  &  $   6.2 \pm 11.1 $  &  $ 14.30 \pm 0.74 $ \\
J085215+171143   & J085215+171137   & 0.16921    & \ion{O}{1}   &  $    79.7  \pm 6.7 $  &  $   2.6  \pm 2.3 $  &  $ 14.40 \pm 0.68 $ \\
J085215+171143   & J085215+171137   & 0.16921    & \ion{S}{3}   &  $  -221.5  \pm 4.4 $  &  $  17.4  \pm 7.4 $  &  $ 14.25 \pm 0.08 $ \\
J085215+171143   & J085215+171137   & 0.16921    & \ion{S}{3}   &  $   -68.2  \pm 6.7 $  &  $  38.6 \pm 10.3 $  &  $ 14.49 \pm 0.08 $ \\
J085215+171143   & J085215+171137   & 0.16921    & \ion{S}{3}   &  $    -2.8  \pm 3.6 $  &  $   9.2  \pm 7.4 $  &  $ 14.41 \pm 0.17 $ \\
J085215+171143   & J085215+171137   & 0.16921    & \ion{Si}{2}  &  $  -264.9  \pm 7.9 $  &  $  15.8 \pm 14.6 $  &  $ 12.76 \pm 0.17 $ \\
J085215+171143   & J085215+171137   & 0.16921    & \ion{Si}{2}  &  $   -64.1 \pm 12.3 $  &  $  50.4  \pm 9.2 $  &  $ 14.16 \pm 0.12 $ \\
J085215+171143   & J085215+171137   & 0.16921    & \ion{Si}{2}  &  $    -5.1  \pm 5.1 $  &  $  22.7  \pm 6.1 $  &  $ 14.00 \pm 0.16 $ \\
J085215+171143   & J085215+171137   & 0.16921    & \ion{Si}{3}  &  $  -188.2 \pm 36.4 $  &  $  40.4 \pm 27.3 $  &  $ 13.09 \pm 0.50 $ \\
J085215+171143   & J085215+171137   & 0.16921    & \ion{Si}{3}  &  $   -76.7  \pm 9.0 $  &  $  52.2 \pm 37.1 $  &  $ 14.17 \pm 0.34 $ \\
J085215+171143   & J085215+171137   & 0.16921    & \ion{Si}{3}  &  $    17.7  \pm 8.2 $  &  $  18.1 \pm 14.0 $  &  $ 13.77 \pm 0.72 $ \\
J085215+171143   & J085215+171137   & 0.16921    & \ion{Si}{3}  &  $    76.9  \pm 7.4 $  &  $  15.6 \pm 13.7 $  &  $ 12.55 \pm 0.18 $ \\
J085215+171143   & J085215+171137   & 0.16921    & \ion{Fe}{2}  &  $   -58.2  \pm 5.9 $  &  $  31.8  \pm 9.8 $  &  $ 14.15 \pm 0.08 $ \\
J085215+171143   & J085215+171137   & 0.16921    & \ion{Fe}{2}  &  $     5.6  \pm 3.8 $  &  $   8.0  \pm 9.2 $  &  $ 13.97 \pm 0.25 $ \\
J085215+171143   & J085215+171137   & 0.16921    & \ion{Fe}{3}  &  $   -44.4 \pm 12.1 $  &  $  47.2 \pm 15.5 $  &  $ 14.70 \pm 0.12 $ \\
J085215+171143   & J085215+171137   & 0.16921    & \ion{Fe}{3}  &  $    12.8  \pm 6.9 $  &  $  12.1 \pm 17.9 $  &  $ 14.08 \pm 0.45 $ \\
J091954+291408   & J091954+291345   & 0.23288    & \ion{H}{1}   &  $  -283.5  \pm 5.6 $  &  $   2.5  \pm 1.7 $  &  $ 14.57 \pm 0.59 $ \\
J091954+291408   & J091954+291345   & 0.23288    & \ion{H}{1}   &  $  -150.0  \pm 3.2 $  &  $  31.6  \pm 3.1 $  &  $ 19.14 \pm 0.59 $ \\
J091954+291408   & J091954+291345   & 0.23288    & \ion{H}{1}   &  $   -19.0  \pm 3.6 $  &  $  22.6  \pm 3.9 $  &  $ 16.76 \pm 0.21 $ \\
J091954+291408   & J091954+291345   & 0.23288    & \ion{H}{1}   &  $   149.1  \pm 1.7 $  &  $  37.6  \pm 3.9 $  &  $ 18.82 \pm 0.60 $ \\
J091954+291408   & J091954+291345   & 0.23288    & \ion{C}{2}   &  $  -157.6  \pm 8.3 $  &  $  37.9 \pm 12.7 $  &  $ 14.21 \pm 0.10 $ \\
J091954+291408   & J091954+291345   & 0.23288    & \ion{C}{2}   &  $   122.8 \pm 46.7 $  &  $  51.1 \pm 32.8 $  &  $ 14.64 \pm 0.47 $ \\
J091954+291408   & J091954+291345   & 0.23288    & \ion{C}{2}   &  $   173.6 \pm 17.3 $  &  $  19.0 \pm 26.0 $  &  $ 14.73 \pm 0.73 $ \\
J091954+291408   & J091954+291345   & 0.23288    & \ion{O}{6}   &  $  -209.4  \pm 4.9 $  &  $  16.2  \pm 8.9 $  &  $ 14.21 \pm 0.19 $ \\
J091954+291408   & J091954+291345   & 0.23288    & \ion{O}{6}   &  $  -110.9  \pm 5.1 $  &  $  12.0  \pm 9.0 $  &  $ 14.04 \pm 0.19 $ \\
J091954+291408   & J091954+291345   & 0.23288    & \ion{O}{6}   &  $   -38.4 \pm 11.7 $  &  $  28.8 \pm 20.7 $  &  $ 13.85 \pm 0.17 $ \\
J091954+291408   & J091954+291345   & 0.23288    & \ion{O}{6}   &  $    46.9 \pm 11.4 $  &  $   7.3 \pm 22.0 $  &  $ 13.61 \pm 0.23 $ \\
J091954+291408   & J091954+291345   & 0.23288    & \ion{O}{6}   &  $   100.9 \pm 16.3 $  &  $  16.1 \pm 37.8 $  &  $ 13.58 \pm 0.41 $ \\
J091954+291408   & J091954+291345   & 0.23288    & \ion{O}{6}   &  $   162.9  \pm 9.7 $  &  $  15.2 \pm 16.7 $  &  $ 13.93 \pm 0.16 $ \\
J091954+291408   & J091954+291345   & 0.23288    & \ion{C}{3}   &  $  -167.3 \pm 27.2 $  &  $  53.4 \pm 21.4 $  &  $ 14.00 \pm 0.24 $ \\
J091954+291408   & J091954+291345   & 0.23288    & \ion{C}{3}   &  $  -105.5 \pm 10.7 $  &  $  19.1 \pm 19.5 $  &  $ 14.01 \pm 0.74 $ \\
J091954+291408   & J091954+291345   & 0.23288    & \ion{C}{3}   &  $   -18.0  \pm 5.8 $  &  $  29.3 \pm 11.1 $  &  $ 13.61 \pm 0.09 $ \\
J091954+291408   & J091954+291345   & 0.23288    & \ion{C}{3}   &  $    45.0  \pm 8.8 $  &  $   8.4 \pm 18.7 $  &  $ 13.26 \pm 0.32 $ \\
J091954+291408   & J091954+291345   & 0.23288    & \ion{C}{3}   &  $   151.7  \pm 6.8 $  &  $  50.4 \pm 15.0 $  &  $ 14.73 \pm 0.45 $ \\
J091954+291408   & J091954+291345   & 0.23288    & \ion{N}{2}   &  $    94.6  \pm 5.8 $  &  $  13.4 \pm 11.2 $  &  $ 14.08 \pm 0.15 $ \\
J091954+291408   & J091954+291345   & 0.23288    & \ion{N}{2}   &  $   168.5  \pm 5.1 $  &  $  27.2  \pm 9.2 $  &  $ 14.33 \pm 0.09 $ \\
J094331+053131   & J094330+053118   & 0.353052   & \ion{H}{1}   &  $    36.3  \pm 2.9 $  &  $  43.7  \pm 2.9 $  &  $ 14.78 \pm 0.04 $ \\
J094331+053131   & J094330+053118   & 0.353052   & \ion{H}{1}   &  $   245.5  \pm 9.5 $  &  $  43.9  \pm 6.2 $  &  $ 15.16 \pm 0.09 $ \\
J094331+053131   & J094330+053118   & 0.353052   & \ion{H}{1}   &  $   334.3  \pm 1.8 $  &  $  26.9  \pm 2.5 $  &  $ 16.12 \pm 0.04 $ \\
J094331+053131   & J094330+053118   & 0.353052   & \ion{H}{1}   &  $   413.0  \pm 2.2 $  &  $  25.7  \pm 2.9 $  &  $ 15.88 \pm 0.04 $ \\
J094331+053131   & J094330+053118   & 0.353052   & \ion{H}{1}   &  $   483.2 \pm 19.9 $  &  $  74.7 \pm 13.4 $  &  $ 14.98 \pm 0.13 $ \\
J094331+053131   & J094330+053118   & 0.353052   & \ion{H}{1}   &  $   672.7  \pm 6.6 $  &  $  17.2 \pm 10.5 $  &  $ 13.42 \pm 0.16 $ \\
J094331+053131   & J094330+053118   & 0.353052   & \ion{H}{1}   &  $   816.0  \pm 4.2 $  &  $   6.1  \pm 5.0 $  &  $ 13.51 \pm 0.38 $ \\
J094331+053131   & J094330+053118   & 0.353052   & \ion{H}{1}   &  $   940.6  \pm 6.0 $  &  $  19.9 \pm 10.9 $  &  $ 13.06 \pm 0.31 $ \\
J094331+053131   & J094330+053118   & 0.353052   & \ion{O}{6}   &  $    49.6  \pm 1.8 $  &  $  20.1  \pm 2.6 $  &  $ 14.64 \pm 0.08 $ \\
J094331+053131   & J094330+053118   & 0.353052   & \ion{O}{6}   &  $   161.1  \pm 4.4 $  &  $  11.0 \pm 10.0 $  &  $ 13.32 \pm 0.31 $ \\
J094331+053131   & J094330+053118   & 0.353052   & \ion{O}{6}   &  $   397.9  \pm 4.4 $  &  $  13.6  \pm 7.0 $  &  $ 13.75 \pm 0.12 $ \\
J094331+053131   & J094330+053118   & 0.353052   & \ion{C}{3}   &  $    37.4  \pm 2.2 $  &  $   9.1  \pm 4.5 $  &  $ 13.36 \pm 0.25 $ \\
J094331+053131   & J094330+053118   & 0.353052   & \ion{C}{3}   &  $   195.0  \pm 1.8 $  &  $  24.1  \pm 6.9 $  &  $ 14.72 \pm 0.67 $ \\
J094331+053131   & J094330+053118   & 0.353052   & \ion{C}{3}   &  $   274.1  \pm 7.3 $  &  $  10.8 \pm 13.2 $  &  $ 12.70 \pm 0.22 $ \\
J094331+053131   & J094330+053118   & 0.353052   & \ion{C}{3}   &  $   329.5  \pm 2.7 $  &  $  15.0  \pm 4.1 $  &  $ 13.42 \pm 0.10 $ \\
J094331+053131   & J094330+053118   & 0.353052   & \ion{C}{3}   &  $   407.9  \pm 2.0 $  &  $  11.6  \pm 4.2 $  &  $ 13.55 \pm 0.24 $ \\
J094331+053131   & J094330+053118   & 0.353052   & \ion{C}{3}   &  $   531.1  \pm 4.9 $  &  $  19.9  \pm 7.5 $  &  $ 13.12 \pm 0.10 $ \\
J094331+053131   & J094330+053118   & 0.353052   & \ion{C}{3}   &  $   667.4  \pm 5.3 $  &  $  12.5  \pm 8.6 $  &  $ 12.88 \pm 0.15 $ \\
J094331+053131   & J094330+053118   & 0.353052   & \ion{C}{3}   &  $   808.5  \pm 6.0 $  &  $   4.0 \pm 18.7 $  &  $ 12.58 \pm 0.61 $ \\
J094331+053131   & J094330+053118   & 0.353052   & \ion{C}{3}   &  $   923.5  \pm 5.1 $  &  $   2.6 \pm 11.7 $  &  $ 12.65 \pm 0.31 $ \\
J094331+053131   & J094330+053118   & 0.353052   & \ion{N}{5}   &  $    43.2  \pm 3.8 $  &  $  13.0  \pm 6.0 $  &  $ 13.90 \pm 0.16 $ \\
J094331+053131   & J094330+053118   & 0.353052   & \ion{N}{5}   &  $   129.6  \pm 4.7 $  &  $  14.3  \pm 7.9 $  &  $ 13.37 \pm 0.20 $ \\
J094331+053131   & J094330+053118   & 0.353052   & \ion{N}{5}   &  $   383.8  \pm 4.7 $  &  $   9.6  \pm 8.2 $  &  $ 13.71 \pm 0.18 $ \\
J094331+053131   & J094330+053118   & 0.353052   & \ion{Si}{3}  &  $   411.2  \pm 5.5 $  &  $  11.3  \pm 9.6 $  &  $ 12.51 \pm 0.24 $ \\
J103640+565125   & J103643+565119   & 0.13629    & \ion{H}{1}   &  $   -38.5  \pm 2.9 $  &  $  17.6  \pm 3.4 $  &  $ 14.35 \pm 0.11 $ \\
J103640+565125   & J103643+565119   & 0.13629    & \ion{H}{1}   &  $    18.2  \pm 5.3 $  &  $  19.7  \pm 8.1 $  &  $ 13.57 \pm 0.10 $ \\
J103640+565125   & J103643+565119   & 0.13629    & \ion{Si}{3}  &  $   -56.7 \pm 14.0 $  &  $  29.7 \pm 23.0 $  &  $ 12.64 \pm 0.20 $ \\
J103640+565125   & J103643+565119   & 0.13629    & \ion{Si}{3}  &  $    29.0  \pm 4.7 $  &  $  16.6  \pm 7.9 $  &  $ 12.86 \pm 0.14 $ \\
J104116+061016   & J104117+061018   & 0.442173   & \ion{H}{1}   &  $  -160.5  \pm 5.8 $  &  $  38.3  \pm 4.0 $  &  $ 16.66 \pm 0.29 $ \\
J104116+061016   & J104117+061018   & 0.442173   & \ion{H}{1}   &  $   -67.8  \pm 9.6 $  &  $  30.1  \pm 4.8 $  &  $ 15.21 \pm 0.17 $ \\
J104116+061016   & J104117+061018   & 0.442173   & \ion{H}{1}   &  $    63.4 \pm 16.2 $  &  $  54.0 \pm 31.9 $  &  $ 13.27 \pm 0.18 $ \\
J104116+061016   & J104117+061018   & 0.442173   & \ion{H}{1}   &  $   173.6  \pm 4.2 $  &  $   9.5  \pm 9.2 $  &  $ 13.33 \pm 0.28 $ \\
J104116+061016   & J104117+061018   & 0.442173   & \ion{H}{1}   &  $   199.6 \pm 10.4 $  &  $  40.9  \pm 8.4 $  &  $ 13.70 \pm 0.13 $ \\
J104116+061016   & J104117+061018   & 0.442173   & \ion{C}{2}   &  $  -207.7  \pm 4.8 $  &  $  17.7  \pm 5.4 $  &  $ 14.11 \pm 0.12 $ \\
J104116+061016   & J104117+061018   & 0.442173   & \ion{C}{2}   &  $  -152.8  \pm 2.5 $  &  $  21.5  \pm 4.6 $  &  $ 14.75 \pm 0.13 $ \\
J104116+061016   & J104117+061018   & 0.442173   & \ion{C}{2}   &  $   -80.4  \pm 2.5 $  &  $  19.2  \pm 4.2 $  &  $ 14.18 \pm 0.06 $ \\
J104116+061016   & J104117+061018   & 0.442173   & \ion{C}{2}   &  $    57.2  \pm 4.4 $  &  $  21.9  \pm 6.5 $  &  $ 13.91 \pm 0.08 $ \\
J104116+061016   & J104117+061018   & 0.442173   & \ion{O}{6}   &  $  -182.7  \pm 5.0 $  &  $  15.8  \pm 7.7 $  &  $ 13.82 \pm 0.16 $ \\
J104116+061016   & J104117+061018   & 0.442173   & \ion{O}{6}   &  $  -123.9  \pm 4.8 $  &  $  30.9 \pm 10.7 $  &  $ 14.28 \pm 0.09 $ \\
J104116+061016   & J104117+061018   & 0.442173   & \ion{O}{6}   &  $   -49.9  \pm 7.5 $  &  $  30.2  \pm 9.8 $  &  $ 14.06 \pm 0.12 $ \\
J104116+061016   & J104117+061018   & 0.442173   & \ion{C}{3}   &  $  -135.3  \pm 1.2 $  &  $  45.6  \pm 3.7 $  &  $ 14.96 \pm 0.17 $ \\
J104116+061016   & J104117+061018   & 0.442173   & \ion{C}{3}   &  $    93.5  \pm 1.5 $  &  $  30.4  \pm 2.2 $  &  $ 13.82 \pm 0.03 $ \\
J104116+061016   & J104117+061018   & 0.442173   & \ion{N}{2}   &  $  -151.3  \pm 4.4 $  &  $  23.1  \pm 6.3 $  &  $ 14.43 \pm 0.11 $ \\
J104116+061016   & J104117+061018   & 0.442173   & \ion{N}{3}   &  $  -140.3  \pm 2.1 $  &  $  41.1  \pm 3.5 $  &  $ 14.58 \pm 0.02 $ \\
J104116+061016   & J104117+061018   & 0.442173   & \ion{N}{3}   &  $   -63.6  \pm 2.1 $  &  $   7.2  \pm 4.3 $  &  $ 13.93 \pm 0.14 $ \\
J104116+061016   & J104117+061018   & 0.442173   & \ion{Si}{2}  &  $  -187.3 \pm 22.0 $  &  $  19.1 \pm 25.3 $  &  $ 12.87 \pm 0.47 $ \\
J104116+061016   & J104117+061018   & 0.442173   & \ion{Si}{2}  &  $  -149.9  \pm 3.1 $  &  $  11.7  \pm 4.6 $  &  $ 13.72 \pm 0.22 $ \\
J104116+061016   & J104117+061018   & 0.442173   & \ion{Si}{2}  &  $   -64.6  \pm 8.5 $  &  $  32.7 \pm 13.7 $  &  $ 13.02 \pm 0.12 $ \\
J104116+061016   & J104117+061018   & 0.442173   & \ion{Si}{2}  &  $    11.8  \pm 5.2 $  &  $  13.6  \pm 8.7 $  &  $ 12.84 \pm 0.15 $ \\
J104116+061016   & J104117+061018   & 0.442173   & \ion{Si}{2}  &  $   209.7  \pm 5.2 $  &  $  18.0  \pm 8.2 $  &  $ 12.90 \pm 0.12 $ \\
J104116+061016   & J104117+061018   & 0.442173   & \ion{Si}{3}  &  $  -199.8  \pm 8.3 $  &  $  12.2  \pm 8.7 $  &  $ 12.70 \pm 0.28 $ \\
J104116+061016   & J104117+061018   & 0.442173   & \ion{Si}{3}  &  $  -155.5  \pm 5.0 $  &  $  18.5 \pm 13.2 $  &  $ 13.65 \pm 0.34 $ \\
J104116+061016   & J104117+061018   & 0.442173   & \ion{Si}{3}  &  $   -98.9 \pm 14.1 $  &  $  50.6 \pm 11.6 $  &  $ 13.54 \pm 0.14 $ \\
J104116+061016   & J104117+061018   & 0.442173   & \ion{Si}{3}  &  $    57.4  \pm 7.5 $  &  $  43.4 \pm 11.3 $  &  $ 12.76 \pm 0.08 $ \\
J113910$-$135043 & J113911$-$135108 & 0.204194   & \ion{H}{1}   &  $  -414.3  \pm 3.5 $  &  $   7.7  \pm 7.0 $  &  $ 13.04 \pm 0.17 $ \\
J113910$-$135043 & J113911$-$135108 & 0.204194   & \ion{H}{1}   &  $  -319.7  \pm 7.0 $  &  $  18.3 \pm 11.3 $  &  $ 13.04 \pm 0.16 $ \\
J113910$-$135043 & J113911$-$135108 & 0.204194   & \ion{H}{1}   &  $  -271.9  \pm 2.5 $  &  $   3.0  \pm 1.4 $  &  $ 13.94 \pm 0.29 $ \\
J113910$-$135043 & J113911$-$135108 & 0.204194   & \ion{H}{1}   &  $  -248.0  \pm 3.0 $  &  $   7.1  \pm 5.3 $  &  $ 13.50 \pm 0.30 $ \\
J113910$-$135043 & J113911$-$135108 & 0.204194   & \ion{H}{1}   &  $   -20.7  \pm 4.5 $  &  $  23.9  \pm 3.1 $  &  $ 16.27 \pm 0.30 $ \\
J113910$-$135043 & J113911$-$135108 & 0.204194   & \ion{H}{1}   &  $    41.1 \pm 13.9 $  &  $  25.6 \pm 16.1 $  &  $ 15.11 \pm 0.27 $ \\
J113910$-$135043 & J113911$-$135108 & 0.204194   & \ion{H}{1}   &  $    95.4  \pm 8.2 $  &  $  22.8 \pm 14.1 $  &  $ 14.25 \pm 0.46 $ \\
J113910$-$135043 & J113911$-$135108 & 0.204194   & \ion{H}{1}   &  $   151.1  \pm 2.0 $  &  $  14.0  \pm 3.5 $  &  $ 16.15 \pm 0.30 $ \\
J113910$-$135043 & J113911$-$135108 & 0.204194   & \ion{H}{1}   &  $   193.7  \pm 5.5 $  &  $  17.5  \pm 3.6 $  &  $ 14.58 \pm 0.16 $ \\
J113910$-$135043 & J113911$-$135108 & 0.204194   & \ion{C}{2}   &  $   -18.2 \pm 10.7 $  &  $  27.3 \pm 16.9 $  &  $ 13.76 \pm 0.17 $ \\
J113910$-$135043 & J113911$-$135108 & 0.204194   & \ion{C}{2}   &  $    55.3 \pm 17.7 $  &  $  23.7 \pm 33.0 $  &  $ 13.19 \pm 0.49 $ \\
J113910$-$135043 & J113911$-$135108 & 0.204194   & \ion{O}{6}   &  $  -227.0  \pm 8.7 $  &  $  10.5 \pm 19.8 $  &  $ 13.26 \pm 0.32 $ \\
J113910$-$135043 & J113911$-$135108 & 0.204194   & \ion{O}{6}   &  $   -13.2 \pm 14.9 $  &  $  36.7 \pm 18.6 $  &  $ 14.14 \pm 0.20 $ \\
J113910$-$135043 & J113911$-$135108 & 0.204194   & \ion{O}{6}   &  $    61.0 \pm 11.9 $  &  $  40.7 \pm 29.7 $  &  $ 14.12 \pm 0.29 $ \\
J113910$-$135043 & J113911$-$135108 & 0.204194   & \ion{C}{3}   &  $  -391.4  \pm 4.2 $  &  $   5.0 \pm 10.5 $  &  $ 12.80 \pm 0.34 $ \\
J113910$-$135043 & J113911$-$135108 & 0.204194   & \ion{C}{3}   &  $    -8.5  \pm 6.5 $  &  $  31.9  \pm 6.2 $  &  $ 14.32 \pm 0.16 $ \\
J113910$-$135043 & J113911$-$135108 & 0.204194   & \ion{C}{3}   &  $    45.8  \pm 6.0 $  &  $  10.9 \pm 13.8 $  &  $ 13.58 \pm 0.31 $ \\
J113910$-$135043 & J113911$-$135108 & 0.204194   & \ion{C}{3}   &  $    80.7  \pm 6.0 $  &  $  15.2  \pm 6.9 $  &  $ 13.46 \pm 0.13 $ \\
J113910$-$135043 & J113911$-$135108 & 0.204194   & \ion{C}{3}   &  $   156.1  \pm 2.7 $  &  $   5.8  \pm 6.8 $  &  $ 13.09 \pm 0.47 $ \\
J113910$-$135043 & J113911$-$135108 & 0.204194   & \ion{N}{3}   &  $    -9.7  \pm 7.0 $  &  $  31.7 \pm 10.2 $  &  $ 13.98 \pm 0.10 $ \\
J113910$-$135043 & J113911$-$135108 & 0.204194   & \ion{N}{3}   &  $    73.9 \pm 13.2 $  &  $  35.8 \pm 19.5 $  &  $ 13.76 \pm 0.17 $ \\
J113910$-$135043 & J113911$-$135108 & 0.204194   & \ion{Si}{3}  &  $  -435.2  \pm 9.0 $  &  $  35.1 \pm 13.2 $  &  $ 12.71 \pm 0.11 $ \\
J113910$-$135043 & J113911$-$135108 & 0.204194   & \ion{Si}{3}  &  $  -233.0 \pm 16.7 $  &  $  67.9 \pm 24.0 $  &  $ 12.72 \pm 0.14 $ \\
J113910$-$135043 & J113911$-$135108 & 0.204194   & \ion{Si}{3}  &  $    -5.7  \pm 2.7 $  &  $  27.5  \pm 4.5 $  &  $ 13.46 \pm 0.07 $ \\
J113910$-$135043 & J113911$-$135108 & 0.204194   & \ion{Si}{3}  &  $    46.3  \pm 2.7 $  &  $   6.6  \pm 6.8 $  &  $ 12.89 \pm 0.49 $ \\
J113910$-$135043 & J113911$-$135108 & 0.204194   & \ion{Si}{3}  &  $    83.7  \pm 3.0 $  &  $  13.4  \pm 4.9 $  &  $ 12.85 \pm 0.10 $ \\
J113910$-$135043 & J113911$-$135108 & 0.204194   & \ion{Si}{3}  &  $   158.3  \pm 3.2 $  &  $  10.1  \pm 5.4 $  &  $ 12.63 \pm 0.12 $ \\
J113910$-$135043 & J113909$-$135053 & 0.319255   & \ion{H}{1}   &  $  -602.0  \pm 0.7 $  &  $  16.7  \pm 1.0 $  &  $ 15.38 \pm 0.10 $ \\
J113910$-$135043 & J113909$-$135053 & 0.319255   & \ion{H}{1}   &  $  -101.4  \pm 0.9 $  &  $  30.8  \pm 1.2 $  &  $ 14.91 \pm 0.02 $ \\
J113910$-$135043 & J113909$-$135053 & 0.319255   & \ion{H}{1}   &  $    48.4  \pm 0.5 $  &  $  23.4  \pm 0.3 $  &  $ 16.14 \pm 0.02 $ \\
J113910$-$135043 & J113909$-$135053 & 0.319255   & \ion{O}{6}   &  $   -83.9  \pm 4.3 $  &  $  59.8  \pm 6.9 $  &  $ 14.11 \pm 0.04 $ \\
J113910$-$135043 & J113909$-$135053 & 0.319255   & \ion{O}{6}   &  $    37.7  \pm 2.0 $  &  $  29.4  \pm 3.4 $  &  $ 14.01 \pm 0.04 $ \\
J113910$-$135043 & J113909$-$135053 & 0.319255   & \ion{O}{6}   &  $   159.8  \pm 6.6 $  &  $  14.3 \pm 10.8 $  &  $ 13.45 \pm 0.16 $ \\
J113910$-$135043 & J113909$-$135053 & 0.319255   & \ion{C}{3}   &  $  -601.3  \pm 2.5 $  &  $  11.8  \pm 4.5 $  &  $ 13.16 \pm 0.10 $ \\
J113910$-$135043 & J113909$-$135053 & 0.319255   & \ion{C}{3}   &  $  -113.8  \pm 1.8 $  &  $  19.1  \pm 2.8 $  &  $ 13.53 \pm 0.04 $ \\
J113910$-$135043 & J113909$-$135053 & 0.319255   & \ion{C}{3}   &  $   -73.2  \pm 2.5 $  &  $   9.0  \pm 5.1 $  &  $ 12.94 \pm 0.11 $ \\
J113910$-$135043 & J113909$-$135053 & 0.319255   & \ion{C}{3}   &  $    17.0 \pm 14.3 $  &  $  33.1  \pm 9.2 $  &  $ 13.06 \pm 0.24 $ \\
J113910$-$135043 & J113909$-$135053 & 0.319255   & \ion{C}{3}   &  $    43.4  \pm 1.8 $  &  $  14.2  \pm 4.0 $  &  $ 13.58 \pm 0.11 $ \\
J123335+475800   & J123338+475757   & 0.22171    & \ion{H}{1}   &  $   -61.1  \pm 4.7 $  &  $  14.6  \pm 3.9 $  &  $ 14.14 \pm 0.12 $ \\
J123335+475800   & J123338+475757   & 0.22171    & \ion{H}{1}   &  $    -2.7 \pm 13.5 $  &  $  15.1 \pm 12.8 $  &  $ 15.17 \pm 0.60 $ \\
J123335+475800   & J123338+475757   & 0.22171    & \ion{H}{1}   &  $    69.7 \pm 14.5 $  &  $  43.7  \pm 8.0 $  &  $ 16.50 \pm 0.30 $ \\
J123335+475800   & J123338+475757   & 0.22171    & \ion{H}{1}   &  $    96.2 \pm 24.0 $  &  $  25.7 \pm 17.4 $  &  $ 16.16 \pm 0.78 $ \\
J123335+475800   & J123338+475757   & 0.22171    & \ion{C}{2}   &  $    -1.2  \pm 2.7 $  &  $  16.0  \pm 4.5 $  &  $ 13.79 \pm 0.07 $ \\
J123335+475800   & J123338+475757   & 0.22171    & \ion{C}{2}   &  $    59.4  \pm 1.2 $  &  $  17.0  \pm 2.2 $  &  $ 14.42 \pm 0.07 $ \\
J123335+475800   & J123338+475757   & 0.22171    & \ion{C}{2}   &  $   119.5  \pm 2.7 $  &  $   7.2  \pm 6.2 $  &  $ 13.49 \pm 0.13 $ \\
J123335+475800   & J123338+475757   & 0.22171    & \ion{O}{6}   &  $    60.9 \pm 11.0 $  &  $ 112.2 \pm 15.3 $  &  $ 14.59 \pm 0.05 $ \\
J123335+475800   & J123338+475757   & 0.22171    & \ion{C}{3}   &  $   -66.7  \pm 1.5 $  &  $   9.5  \pm 4.4 $  &  $ 14.02 \pm 0.61 $ \\
J123335+475800   & J123338+475757   & 0.22171    & \ion{C}{3}   &  $    -7.9  \pm 3.2 $  &  $  19.5  \pm 5.2 $  &  $ 13.88 \pm 0.13 $ \\
J123335+475800   & J123338+475757   & 0.22171    & \ion{C}{3}   &  $    84.4  \pm 3.4 $  &  $  37.7  \pm 7.2 $  &  $ 14.64 \pm 0.28 $ \\
J123335+475800   & J123338+475757   & 0.22171    & \ion{N}{2}   &  $    -9.3 \pm 10.1 $  &  $  31.3 \pm 16.7 $  &  $ 13.65 \pm 0.16 $ \\
J123335+475800   & J123338+475757   & 0.22171    & \ion{N}{2}   &  $    56.2  \pm 2.2 $  &  $  14.1  \pm 3.7 $  &  $ 14.09 \pm 0.07 $ \\
J123335+475800   & J123338+475757   & 0.22171    & \ion{N}{3}   &  $     9.1 \pm 18.9 $  &  $  89.8 \pm 18.3 $  &  $ 14.35 \pm 0.11 $ \\
J123335+475800   & J123338+475757   & 0.22171    & \ion{N}{3}   &  $    66.0  \pm 3.4 $  &  $  14.1  \pm 6.6 $  &  $ 14.18 \pm 0.13 $ \\
J123335+475800   & J123338+475757   & 0.22171    & \ion{Si}{2}  &  $    -1.7  \pm 3.4 $  &  $  12.4  \pm 6.2 $  &  $ 12.67 \pm 0.10 $ \\
J123335+475800   & J123338+475757   & 0.22171    & \ion{Si}{2}  &  $    59.4  \pm 2.5 $  &  $  22.8  \pm 3.5 $  &  $ 13.25 \pm 0.05 $ \\
J123335+475800   & J123338+475757   & 0.22171    & \ion{Si}{3}  &  $   -16.9 \pm 15.9 $  &  $  42.0 \pm 22.6 $  &  $ 12.54 \pm 0.21 $ \\
J123335+475800   & J123338+475757   & 0.22171    & \ion{Si}{3}  &  $    62.1  \pm 2.7 $  &  $  15.0  \pm 8.2 $  &  $ 13.42 \pm 0.24 $ \\
J123335+475800   & J123338+475757   & 0.22171    & \ion{Si}{3}  &  $   101.6 \pm 28.0 $  &  $  43.9 \pm 21.2 $  &  $ 13.11 \pm 0.31 $ \\
J123335+475800   & J123338+475757   & 0.22171    & \ion{Si}{4}  &  $    54.0  \pm 2.0 $  &  $  10.3  \pm 3.1 $  &  $ 13.28 \pm 0.10 $ \\
J132222+464546   & J132222+464546   & 0.214431   & \ion{H}{1}   &  $   -20.2  \pm 6.9 $  &  $  36.9  \pm 2.6 $  &  $ 16.52 \pm 0.12 $ \\
J132222+464546   & J132222+464546   & 0.214431   & \ion{H}{1}   &  $    56.0  \pm 5.9 $  &  $  31.8  \pm 2.2 $  &  $ 16.91 \pm 0.14 $ \\
J132222+464546   & J132222+464546   & 0.214431   & \ion{C}{2}   &  $   -32.3  \pm 3.5 $  &  $  23.4  \pm 5.2 $  &  $ 14.08 \pm 0.06 $ \\
J132222+464546   & J132222+464546   & 0.214431   & \ion{C}{2}   &  $    49.9  \pm 3.5 $  &  $  28.1  \pm 5.5 $  &  $ 14.14 \pm 0.06 $ \\
J132222+464546   & J132222+464546   & 0.214431   & \ion{O}{6}   &  $   -54.1  \pm 6.2 $  &  $  36.8  \pm 7.1 $  &  $ 14.39 \pm 0.10 $ \\
J132222+464546   & J132222+464546   & 0.214431   & \ion{O}{6}   &  $    37.3 \pm 22.0 $  &  $  64.8 \pm 22.4 $  &  $ 14.19 \pm 0.17 $ \\
J132222+464546   & J132222+464546   & 0.214431   & \ion{C}{3}   &  $   -38.0  \pm 7.4 $  &  $  44.5  \pm 7.3 $  &  $ 14.26 \pm 0.09 $ \\
J132222+464546   & J132222+464546   & 0.214431   & \ion{C}{3}   &  $    53.8  \pm 7.7 $  &  $  36.1  \pm 7.5 $  &  $ 14.07 \pm 0.10 $ \\
J132222+464546   & J132222+464546   & 0.214431   & \ion{N}{2}   &  $    52.1  \pm 5.4 $  &  $  11.5  \pm 9.6 $  &  $ 13.64 \pm 0.13 $ \\
J132222+464546   & J132222+464546   & 0.214431   & \ion{N}{3}   &  $   -37.0 \pm 11.1 $  &  $  36.6 \pm 17.4 $  &  $ 14.04 \pm 0.12 $ \\
J132222+464546   & J132222+464546   & 0.214431   & \ion{N}{3}   &  $    54.6  \pm 7.9 $  &  $  33.1 \pm 11.9 $  &  $ 14.10 \pm 0.12 $ \\
J132222+464546   & J132222+464546   & 0.214431   & \ion{N}{5}   &  $   -53.1  \pm 4.7 $  &  $  20.2  \pm 7.6 $  &  $ 13.60 \pm 0.12 $ \\
J132222+464546   & J132222+464546   & 0.214431   & \ion{Si}{2}  &  $   -21.5  \pm 4.4 $  &  $  37.1  \pm 6.5 $  &  $ 13.14 \pm 0.06 $ \\
J132222+464546   & J132222+464546   & 0.214431   & \ion{Si}{2}  &  $    61.7  \pm 3.7 $  &  $   9.4  \pm 6.3 $  &  $ 12.86 \pm 0.11 $ \\
J132222+464546   & J132222+464546   & 0.214431   & \ion{Si}{2}  &  $   101.5  \pm 6.9 $  &  $  10.4 \pm 12.6 $  &  $ 12.38 \pm 0.20 $ \\
J132222+464546   & J132222+464546   & 0.214431   & \ion{Si}{2}  &  $   161.7  \pm 4.0 $  &  $   8.4  \pm 7.2 $  &  $ 12.59 \pm 0.16 $ \\
J132222+464546   & J132222+464546   & 0.214431   & \ion{Si}{3}  &  $   -92.8  \pm 6.2 $  &  $  15.6 \pm 10.7 $  &  $ 12.46 \pm 0.15 $ \\
J132222+464546   & J132222+464546   & 0.214431   & \ion{Si}{3}  &  $   -21.0  \pm 2.2 $  &  $  27.3  \pm 4.2 $  &  $ 13.42 \pm 0.06 $ \\
J132222+464546   & J132222+464546   & 0.214431   & \ion{Si}{3}  &  $    56.3  \pm 2.5 $  &  $  16.9  \pm 4.8 $  &  $ 13.41 \pm 0.14 $ \\
J132222+464546   & J132222+464546   & 0.214431   & \ion{Si}{3}  &  $   105.2 \pm 10.1 $  &  $  12.5 \pm 17.6 $  &  $ 12.40 \pm 0.27 $ \\
J132222+464546   & J132222+464546   & 0.214431   & \ion{Si}{3}  &  $   166.1  \pm 5.9 $  &  $  18.1 \pm 10.0 $  &  $ 12.46 \pm 0.14 $ \\
J132222+464546   & J132222+464546   & 0.214431   & \ion{Si}{4}  &  $   -38.5  \pm 6.4 $  &  $  26.9  \pm 9.4 $  &  $ 13.04 \pm 0.11 $ \\
J132222+464546   & J132222+464546   & 0.214431   & \ion{Si}{4}  &  $    52.1  \pm 7.2 $  &  $  13.2 \pm 11.4 $  &  $ 12.89 \pm 0.17 $ \\
J135522+303324   & J135521+303320   & 0.20690    & \ion{H}{1}   &  $  -126.2  \pm 5.2 $  &  $   8.0 \pm 11.7 $  &  $ 13.80 \pm 0.17 $ \\
J135522+303324   & J135521+303320   & 0.20690    & \ion{H}{1}   &  $   -23.1  \pm 1.2 $  &  $  25.1  \pm 2.3 $  &  $ 16.21 \pm 0.27 $ \\
J135522+303324   & J135521+303320   & 0.20690    & \ion{C}{2}   &  $   -29.8 \pm 16.6 $  &  $  64.4 \pm 25.4 $  &  $ 13.82 \pm 0.12 $ \\
J135522+303324   & J135521+303320   & 0.20690    & \ion{O}{6}   &  $  -120.5 \pm 28.6 $  &  $  90.8 \pm 44.6 $  &  $ 13.86 \pm 0.21 $ \\
J135522+303324   & J135521+303320   & 0.20690    & \ion{O}{6}   &  $     7.0  \pm 4.0 $  &  $  38.7  \pm 5.6 $  &  $ 14.36 \pm 0.05 $ \\
J135522+303324   & J135521+303320   & 0.20690    & \ion{C}{3}   &  $   -18.6  \pm 2.2 $  &  $  24.8  \pm 7.3 $  &  $ 14.42 \pm 0.52 $ \\
J135522+303324   & J135521+303320   & 0.20690    & \ion{N}{3}   &  $     1.0  \pm 7.5 $  &  $  39.4 \pm 12.4 $  &  $ 13.96 \pm 0.10 $ \\
J135522+303324   & J135521+303320   & 0.20690    & \ion{N}{3}   &  $    89.2  \pm 6.7 $  &  $  10.3 \pm 19.3 $  &  $ 13.51 \pm 0.49 $ \\
J142501+382100   & J142459+382113   & 0.21295    & \ion{H}{1}   &  $  -518.5  \pm 1.7 $  &  $  28.9  \pm 2.0 $  &  $ 15.35 \pm 0.05 $ \\
J142501+382100   & J142459+382113   & 0.21295    & \ion{H}{1}   &  $   -28.4  \pm 1.7 $  &  $  13.1  \pm 2.8 $  &  $ 16.53 \pm 0.64 $ \\
J142501+382100   & J142459+382113   & 0.21295    & \ion{H}{1}   &  $    48.7  \pm 1.7 $  &  $  23.1  \pm 2.0 $  &  $ 15.78 \pm 0.11 $ \\
J142501+382100   & J142459+382113   & 0.21295    & \ion{O}{6}   &  $   -27.7  \pm 5.9 $  &  $   7.3 \pm 13.0 $  &  $ 13.62 \pm 0.34 $ \\
J142501+382100   & J142459+382113   & 0.21295    & \ion{O}{6}   &  $    32.4  \pm 4.7 $  &  $  27.9  \pm 8.2 $  &  $ 14.18 \pm 0.07 $ \\
J142501+382100   & J142459+382113   & 0.21295    & \ion{C}{3}   &  $   -22.7  \pm 3.2 $  &  $  11.1  \pm 9.6 $  &  $ 14.12 \pm 0.84 $ \\
J142501+382100   & J142459+382113   & 0.21295    & \ion{C}{3}   &  $    43.7  \pm 4.4 $  &  $  27.8  \pm 6.9 $  &  $ 13.72 \pm 0.07 $ \\
J142501+382100   & J142459+382113   & 0.21295    & \ion{N}{2}   &  $    53.1 \pm 10.6 $  &  $  26.4 \pm 19.5 $  &  $ 13.81 \pm 0.21 $ \\
J154741+343357   & J154741+343350   & 0.18392    & \ion{H}{1}   &  $   -50.6 \pm 15.4 $  &  $  46.7 \pm 21.4 $  &  $ 16.11 \pm 0.66 $ \\
J154741+343357   & J154741+343350   & 0.18392    & \ion{H}{1}   &  $    59.8 \pm 12.4 $  &  $  11.4  \pm 4.2 $  &  $ 19.34 \pm 0.88 $ \\
J154741+343357   & J154741+343350   & 0.18392    & \ion{O}{6}   &  $  -145.3  \pm 4.8 $  &  $  33.3  \pm 7.5 $  &  $ 14.47 \pm 0.07 $ \\
J154741+343357   & J154741+343350   & 0.18392    & \ion{O}{6}   &  $   -11.4  \pm 9.4 $  &  $  49.5 \pm 13.6 $  &  $ 14.68 \pm 0.09 $ \\
J154741+343357   & J154741+343350   & 0.18392    & \ion{O}{6}   &  $    70.1  \pm 5.1 $  &  $  18.3  \pm 9.7 $  &  $ 14.99 \pm 0.23 $ \\
J154741+343357   & J154741+343350   & 0.18392    & \ion{O}{6}   &  $   125.3 \pm 16.0 $  &  $  20.8 \pm 25.9 $  &  $ 13.71 \pm 0.38 $ \\
J154741+343357   & J154741+343350   & 0.18392    & \ion{C}{2}   &  $   -30.4  \pm 3.0 $  &  $  35.5 \pm 10.8 $  &  $ 15.54 \pm 0.67 $ \\
J154741+343357   & J154741+343350   & 0.18392    & \ion{C}{2}   &  $    83.6  \pm 7.6 $  &  $   4.5 \pm 19.3 $  &  $ 13.76 \pm 0.53 $ \\
J154741+343357   & J154741+343350   & 0.18392    & \ion{C}{3}   &  $  -195.0 \pm 14.2 $  &  $  21.3 \pm 22.5 $  &  $ 13.11 \pm 0.23 $ \\
J154741+343357   & J154741+343350   & 0.18392    & \ion{C}{3}   &  $     0.5  \pm 5.1 $  &  $  67.3 \pm 16.6 $  &  $ 15.08 \pm 0.58 $ \\
J154741+343357   & J154741+343350   & 0.18392    & \ion{N}{3}   &  $   -15.7  \pm 7.1 $  &  $  37.4 \pm 20.7 $  &  $ 15.27 \pm 0.66 $ \\
J154741+343357   & J154741+343350   & 0.18392    & \ion{N}{3}   &  $    81.0 \pm 10.4 $  &  $  11.5 \pm 19.8 $  &  $ 14.28 \pm 0.35 $ \\
J154741+343357   & J154741+343350   & 0.18392    & \ion{Si}{2}  &  $  -171.7  \pm 7.8 $  &  $   5.7 \pm 16.8 $  &  $ 12.87 \pm 0.25 $ \\
J154741+343357   & J154741+343350   & 0.18392    & \ion{Si}{2}  &  $   -66.1  \pm 9.6 $  &  $  11.1  \pm 8.3 $  &  $ 13.92 \pm 0.39 $ \\
J154741+343357   & J154741+343350   & 0.18392    & \ion{Si}{2}  &  $   -28.1  \pm 7.8 $  &  $  13.4 \pm 19.3 $  &  $ 14.20 \pm 0.65 $ \\
J154741+343357   & J154741+343350   & 0.18392    & \ion{Si}{2}  &  $    14.9  \pm 7.8 $  &  $  14.8  \pm 7.4 $  &  $ 14.00 \pm 0.23 $ \\
J154741+343357   & J154741+343350   & 0.18392    & \ion{Si}{3}  &  $    -9.9  \pm 5.1 $  &  $  46.3 \pm 13.2 $  &  $ 14.31 \pm 0.48 $ \\
J154741+343357   & J154741+343350   & 0.18392    & \ion{Si}{3}  &  $    96.0  \pm 4.8 $  &  $  11.6 \pm 12.6 $  &  $ 13.26 \pm 0.53 $ \\
J154741+343357   & J154741+343350   & 0.18392    & \ion{Fe}{3}  &  $   -83.6 \pm 13.4 $  &  $  27.8 \pm 26.4 $  &  $ 14.05 \pm 0.23 $ \\
J154741+343357   & J154741+343350   & 0.18392    & \ion{Fe}{3}  &  $   -22.8  \pm 9.1 $  &  $  17.4 \pm 19.5 $  &  $ 14.21 \pm 0.23 $ \\
J154741+343357   & J154741+343350   & 0.18392    & \ion{Fe}{3}  &  $    24.1 \pm 11.1 $  &  $  14.3 \pm 18.8 $  &  $ 14.01 \pm 0.26 $ \\
J155504+362847   & J155505+362848   & 0.18926    & \ion{H}{1}   &  $   -60.2  \pm 1.0 $  &  $  36.1  \pm 2.2 $  &  $ 17.60 \pm 0.35 $ \\
J155504+362847   & J155505+362848   & 0.18926    & \ion{C}{2}   &  $   -22.9  \pm 5.8 $  &  $  14.3 \pm 11.2 $  &  $ 14.04 \pm 0.22 $ \\
J155504+362847   & J155505+362848   & 0.18926    & \ion{C}{2}   &  $   -55.5 \pm 12.6 $  &  $  82.7 \pm 16.0 $  &  $ 14.40 \pm 0.08 $ \\
J155504+362847   & J155505+362848   & 0.18926    & \ion{O}{6}   &  $   -27.5  \pm 6.8 $  &  $  35.8  \pm 7.4 $  &  $ 14.53 \pm 0.09 $ \\
J155504+362847   & J155505+362848   & 0.18926    & \ion{O}{6}   &  $  -107.9 \pm 13.9 $  &  $  41.2 \pm 16.6 $  &  $ 14.26 \pm 0.16 $ \\
J155504+362847   & J155505+362848   & 0.18926    & \ion{C}{3}   &  $   -69.1  \pm 2.5 $  &  $  27.4  \pm 2.1 $  &  $ 17.15 \pm 0.32 $ \\
J155504+362847   & J155505+362848   & 0.18926    & \ion{N}{2}   &  $   -17.4  \pm 9.8 $  &  $  57.5 \pm 13.8 $  &  $ 14.25 \pm 0.08 $ \\
J155504+362847   & J155505+362848   & 0.18926    & \ion{N}{3}   &  $   -18.9  \pm 9.1 $  &  $  22.7 \pm 15.7 $  &  $ 14.01 \pm 0.44 $ \\
J155504+362847   & J155505+362848   & 0.18926    & \ion{N}{3}   &  $   -78.9 \pm 29.7 $  &  $  52.4 \pm 30.0 $  &  $ 14.24 \pm 0.27 $ \\
J155504+362847   & J155505+362848   & 0.18926    & \ion{Si}{2}  &  $   -23.9  \pm 4.0 $  &  $  19.9  \pm 6.8 $  &  $ 13.11 \pm 0.12 $ \\
J155504+362847   & J155505+362848   & 0.18926    & \ion{Si}{2}  &  $   -99.3 \pm 24.7 $  &  $  68.0 \pm 36.5 $  &  $ 12.94 \pm 0.19 $ \\
J155504+362847   & J155505+362848   & 0.18926    & \ion{Si}{3}  &  $   -43.6  \pm 9.6 $  &  $  41.5  \pm 8.5 $  &  $ 13.57 \pm 0.10 $ \\
J155504+362847   & J155505+362848   & 0.18926    & \ion{Si}{3}  &  $  -114.4 \pm 26.7 $  &  $  36.8 \pm 24.4 $  &  $ 12.89 \pm 0.47 $ \\
J155504+362847   & J155505+362848   & 0.18926    & \ion{Si}{4}  &  $   -24.7  \pm 6.3 $  &  $  23.9  \pm 9.5 $  &  $ 13.40 \pm 0.11 $ \\
J155504+362847   & J155505+362848   & 0.18926    & \ion{Si}{4}  &  $   -97.3  \pm 8.8 $  &  $  28.4 \pm 13.4 $  &  $ 13.14 \pm 0.18 $ \\
J160951+353843   & J160951+353838   & 0.28940    & \ion{H}{1}   &  $    -9.5  \pm 0.7 $  &  $  57.0  \pm 0.8 $  &  $ 20.05 \pm 0.09 $ \\
J160951+353843   & J160951+353838   & 0.28940    & \ion{O}{1}   &  $   -90.2  \pm 8.6 $  &  $  13.3 \pm 14.8 $  &  $ 14.68 \pm 0.16 $ \\
J160951+353843   & J160951+353838   & 0.28940    & \ion{O}{1}   &  $   -17.0  \pm 1.2 $  &  $  21.8  \pm 1.4 $  &  $ 16.36 \pm 0.04 $ \\
J160951+353843   & J160951+353838   & 0.28940    & \ion{O}{1}   &  $   110.0  \pm 6.0 $  &  $  29.3  \pm 3.9 $  &  $ 15.49 \pm 0.04 $ \\
J160951+353843   & J160951+353838   & 0.28940    & \ion{C}{2}   &  $  -139.5  \pm 7.2 $  &  $  32.2  \pm 5.1 $  &  $ 14.60 \pm 0.09 $ \\
J160951+353843   & J160951+353838   & 0.28940    & \ion{C}{2}   &  $   -89.7 \pm 10.7 $  &  $  10.3  \pm 4.6 $  &  $ 14.40 \pm 0.67 $ \\
J160951+353843   & J160951+353838   & 0.28940    & \ion{C}{2}   &  $   -10.7  \pm 9.5 $  &  $  45.4  \pm 5.8 $  &  $ 15.43 \pm 0.89 $ \\
J160951+353843   & J160951+353838   & 0.28940    & \ion{C}{2}   &  $   100.9 \pm 13.5 $  &  $  27.9  \pm 2.2 $  &  $ 15.26 \pm 0.54 $ \\
J160951+353843   & J160951+353838   & 0.28940    & \ion{O}{6}   &  $  -159.5 \pm 17.4 $  &  $  32.6 \pm 28.9 $  &  $ 14.24 \pm 0.29 $ \\
J160951+353843   & J160951+353838   & 0.28940    & \ion{O}{6}   &  $   -96.5 \pm 10.0 $  &  $  26.5 \pm 14.1 $  &  $ 14.36 \pm 0.21 $ \\
J160951+353843   & J160951+353838   & 0.28940    & \ion{O}{6}   &  $   -30.0  \pm 6.7 $  &  $  18.4  \pm 1.7 $  &  $ 14.50 \pm 0.19 $ \\
J160951+353843   & J160951+353838   & 0.28940    & \ion{O}{6}   &  $     8.4  \pm 9.1 $  &  $  17.6 \pm 16.1 $  &  $ 14.29 \pm 0.29 $ \\
J160951+353843   & J160951+353838   & 0.28940    & \ion{O}{6}   &  $    92.8  \pm 5.6 $  &  $  57.8  \pm 5.9 $  &  $ 15.10 \pm 0.04 $ \\
J160951+353843   & J160951+353838   & 0.28940    & \ion{C}{3}   &  $   -11.6  \pm 1.2 $  &  $  73.2  \pm 5.7 $  &  $ 17.05 \pm 0.48 $ \\
J160951+353843   & J160951+353838   & 0.28940    & \ion{N}{2}   &  $  -152.5  \pm 4.4 $  &  $  17.4  \pm 6.0 $  &  $ 14.15 \pm 0.12 $ \\
J160951+353843   & J160951+353838   & 0.28940    & \ion{N}{2}   &  $   -86.5 \pm 22.8 $  &  $  35.3 \pm 34.5 $  &  $ 14.29 \pm 0.42 $ \\
J160951+353843   & J160951+353838   & 0.28940    & \ion{N}{2}   &  $   -22.8  \pm 5.3 $  &  $  25.5 \pm 13.7 $  &  $ 14.95 \pm 0.21 $ \\
J160951+353843   & J160951+353838   & 0.28940    & \ion{N}{2}   &  $    40.2 \pm 10.0 $  &  $  22.2 \pm 17.9 $  &  $ 14.13 \pm 0.29 $ \\
J160951+353843   & J160951+353838   & 0.28940    & \ion{N}{2}   &  $   107.4  \pm 3.3 $  &  $  27.7  \pm 4.6 $  &  $ 14.64 \pm 0.05 $ \\
J160951+353843   & J160951+353838   & 0.28940    & \ion{S}{3}   &  $    -8.4  \pm 3.0 $  &  $  12.3  \pm 5.7 $  &  $ 14.71 \pm 0.15 $ \\
J160951+353843   & J160951+353838   & 0.28940    & \ion{S}{4}   &  $    10.2  \pm 4.7 $  &  $  30.7  \pm 7.7 $  &  $ 14.49 \pm 0.07 $ \\
J225357+160853   & J225400+160925   & 0.390013   & \ion{H}{1}   &  $ -1239.5  \pm 3.0 $  &  $  35.9  \pm 0.0 $  &  $ 14.08 \pm 0.06 $ \\
J225357+160853   & J225400+160925   & 0.390013   & \ion{H}{1}   &  $ -1136.0  \pm 3.7 $  &  $   3.9  \pm 2.9 $  &  $ 13.33 \pm 0.43 $ \\
J225357+160853   & J225400+160925   & 0.390013   & \ion{H}{1}   &  $  -771.7  \pm 5.6 $  &  $  11.5 \pm 10.8 $  &  $ 13.43 \pm 0.41 $ \\
J225357+160853   & J225400+160925   & 0.390013   & \ion{H}{1}   &  $  -711.3  \pm 7.8 $  &  $  56.0  \pm 9.3 $  &  $ 13.97 \pm 0.06 $ \\
J225357+160853   & J225400+160925   & 0.390013   & \ion{H}{1}   &  $  -607.3  \pm 8.6 $  &  $  10.7 \pm 17.0 $  &  $ 12.72 \pm 0.28 $ \\
J225357+160853   & J225400+160925   & 0.390013   & \ion{H}{1}   &  $    68.8 \pm 31.5 $  &  $  40.6 \pm 18.0 $  &  $ 14.42 \pm 0.47 $ \\
J225357+160853   & J225400+160925   & 0.390013   & \ion{H}{1}   &  $   107.6  \pm 6.7 $  &  $  22.3 \pm 13.7 $  &  $ 14.66 \pm 0.31 $ \\
J225357+160853   & J225400+160925   & 0.390013   & \ion{H}{1}   &  $   151.8  \pm 5.0 $  &  $  16.9  \pm 4.4 $  &  $ 14.78 \pm 0.11 $ \\
J225357+160853   & J225400+160925   & 0.390013   & \ion{H}{1}   &  $   216.5  \pm 2.8 $  &  $  19.7  \pm 3.6 $  &  $ 14.03 \pm 0.09 $ \\
J225357+160853   & J225400+160925   & 0.390013   & \ion{O}{6}   &  $ -1322.5 \pm 11.6 $  &  $   6.6 \pm 21.4 $  &  $ 13.69 \pm 0.25 $ \\
J225357+160853   & J225400+160925   & 0.390013   & \ion{O}{6}   &  $ -1277.7  \pm 5.0 $  &  $   5.5 \pm 14.6 $  &  $ 13.92 \pm 0.72 $ \\
J225357+160853   & J225400+160925   & 0.390013   & \ion{O}{6}   &  $  -717.8  \pm 2.6 $  &  $   4.2  \pm 3.8 $  &  $ 14.02 \pm 0.42 $ \\
J225357+160853   & J225400+160925   & 0.390013   & \ion{O}{6}   &  $  -603.9  \pm 7.1 $  &  $  21.7 \pm 11.2 $  &  $ 13.62 \pm 0.14 $ \\
J225357+160853   & J225400+160925   & 0.390013   & \ion{O}{6}   &  $   109.6 \pm 16.0 $  &  $  21.3 \pm 16.4 $  &  $ 13.91 \pm 0.34 $ \\
J225357+160853   & J225400+160925   & 0.390013   & \ion{O}{6}   &  $   149.5  \pm 6.5 $  &  $  17.6  \pm 7.8 $  &  $ 14.24 \pm 0.17 $ \\
J225357+160853   & J225400+160925   & 0.390013   & \ion{O}{6}   &  $   232.7 \pm 14.0 $  &  $  38.0 \pm 23.8 $  &  $ 13.67 \pm 0.19 $ \\
J225357+160853   & J225400+160925   & 0.390013   & \ion{C}{3}   &  $    77.4 \pm 16.0 $  &  $  49.5 \pm 15.0 $  &  $ 13.85 \pm 0.19 $ \\
J225357+160853   & J225400+160925   & 0.390013   & \ion{C}{3}   &  $   133.7  \pm 6.0 $  &  $  21.2  \pm 9.2 $  &  $ 14.26 \pm 0.34 $ \\
J225357+160853   & J225400+160925   & 0.390013   & \ion{C}{3}   &  $   218.7  \pm 6.3 $  &  $  21.5 \pm 11.3 $  &  $ 13.09 \pm 0.13 $ \\
J225357+160853   & J225400+160925   & 0.390013   & \ion{N}{3}   &  $   241.3  \pm 7.8 $  &  $   9.7 \pm 13.8 $  &  $ 13.61 \pm 0.19 $ \\
J225357+160853   & J225400+160925   & 0.390013   & \ion{N}{5}   &  $    54.4  \pm 4.7 $  &  $   8.2 \pm 10.8 $  &  $ 13.56 \pm 0.27 $ \\
J225357+160853   & J225400+160925   & 0.390013   & \ion{O}{3}   &  $     5.2 \pm 14.9 $  &  $  48.3 \pm 23.8 $  &  $ 14.21 \pm 0.15 $ \\
J225357+160853   & J225400+160925   & 0.390013   & \ion{O}{3}   &  $   101.8  \pm 8.4 $  &  $  13.5 \pm 17.3 $  &  $ 14.01 \pm 0.21 $ \\
J225357+160853   & J225400+160925   & 0.390013   & \ion{O}{3}   &  $   156.6  \pm 7.3 $  &  $  16.0 \pm 12.8 $  &  $ 14.16 \pm 0.16 $ \\
J225357+160853   & J225400+160925   & 0.390013   & \ion{O}{3}   &  $   236.4  \pm 8.4 $  &  $   9.6  \pm 0.2 $  &  $ 13.88 \pm 1.42 $ \\
\enddata
\end{deluxetable*}

%% file: ms_o6major_x2.bbl
\begin{thebibliography}{}
\expandafter\ifx\csname natexlab\endcsname\relax\def\natexlab#1{#1}\fi
\providecommand{\url}[1]{\href{#1}{#1}}
\providecommand{\dodoi}[1]{doi:~\href{http://doi.org/#1}{\nolinkurl{#1}}}
\providecommand{\doeprint}[1]{\href{http://ascl.net/#1}{\nolinkurl{http://ascl.net/#1}}}
\providecommand{\doarXiv}[1]{\href{https://arxiv.org/abs/#1}{\nolinkurl{https://arxiv.org/abs/#1}}}

\bibitem[{{Behroozi} {et~al.}(2010){Behroozi}, {Conroy}, \&
  {Wechsler}}]{Behroozi2010}
{Behroozi}, P.~S., {Conroy}, C., \& {Wechsler}, R.~H. 2010, \apj, 717, 379

\bibitem[{{Blanton} {et~al.}(2017){Blanton}, {Bershady}, {Abolfathi},
  {Albareti}, {Allende Prieto}, {Almeida}, {Alonso-Garc{\'\i}a}, {Anders},
  {Anderson}, {Andrews}, {Aquino-Ort{\'\i}z}, {Arag{\'o}n-Salamanca},
  {Argudo-Fern{\'a}ndez}, {Armengaud}, {Aubourg}, {Avila-Reese}, {Badenes},
  {Bailey}, {Barger}, {Barrera-Ballesteros}, {Bartosz}, {Bates}, {Baumgarten},
  {Bautista}, {Beaton}, {Beers}, {Belfiore}, {Bender}, {Berlind}, {Bernardi},
  {Beutler}, {Bird}, {Bizyaev}, {Blanc}, {Blomqvist}, {Bolton}, {Boquien},
  {Borissova}, {van den Bosch}, {Bovy}, {Brandt}, {Brinkmann}, {Brownstein},
  {Bundy}, {Burgasser}, {Burtin}, {Busca}, {Cappellari}, {Delgado Carigi},
  {Carlberg}, {Carnero Rosell}, {Carrera}, {Chanover}, {Cherinka}, {Cheung},
  {G{\'o}mez Maqueo Chew}, {Chiappini}, {Choi}, {Chojnowski}, {Chuang},
  {Chung}, {Cirolini}, {Clerc}, {Cohen}, {Comparat}, {da Costa}, {Cousinou},
  {Covey}, {Crane}, {Croft}, {Cruz-Gonzalez}, {Garrido Cuadra}, {Cunha},
  {Damke}, {Darling}, {Davies}, {Dawson}, {de la Macorra}, {Dell'Agli}, {De
  Lee}, {Delubac}, {Di Mille}, {Diamond-Stanic}, {Cano-D{\'\i}az}, {Donor},
  {Downes}, {Drory}, {du Mas des Bourboux}, {Duckworth}, {Dwelly}, {Dyer},
  {Ebelke}, {Eigenbrot}, {Eisenstein}, {Emsellem}, {Eracleous}, {Escoffier},
  {Evans}, {Fan}, {Fern{\'a}ndez-Alvar}, {Fernandez-Trincado}, {Feuillet},
  {Finoguenov}, {Fleming}, {Font-Ribera}, {Fredrickson}, {Freischlad},
  {Frinchaboy}, {Fuentes}, {Galbany}, {Garcia-Dias},
  {Garc{\'\i}a-Hern{\'a}ndez}, {Gaulme}, {Geisler}, {Gelfand},
  {Gil-Mar{\'\i}n}, {Gillespie}, {Goddard}, {Gonzalez-Perez}, {Grabowski},
  {Green}, {Grier}, {Gunn}, {Guo}, {Guy}, {Hagen}, {Hahn}, {Hall}, {Harding},
  {Hasselquist}, {Hawley}, {Hearty}, {Gonzalez Hern{\'a}ndez}, {Ho}, {Hogg},
  {Holley-Bockelmann}, {Holtzman}, {Holzer}, {Huehnerhoff}, {Hutchinson},
  {Hwang}, {Ibarra-Medel}, {da Silva Ilha}, {Ivans}, {Ivory}, {Jackson},
  {Jensen}, {Johnson}, {Jones}, {J{\"o}nsson}, {Jullo}, {Kamble}, {Kinemuchi},
  {Kirkby}, {Kitaura}, {Klaene}, {Knapp}, {Kneib}, {Kollmeier}, {Lacerna},
  {Lane}, {Lang}, {Law}, {Lazarz}, {Lee}, {Le Goff}, {Liang}, {Li}, {Li},
  {Lian}, {Lima}, {Lin}, {Lin}, {Bertran de Lis}, {Liu}, {de Icaza Lizaola},
  {Long}, {Lucatello}, {Lundgren}, {MacDonald}, {Deconto Machado}, {MacLeod},
  {Mahadevan}, {Geimba Maia}, {Maiolino}, {Majewski}, {Malanushenko},
  {Malanushenko}, {Manchado}, {Mao}, {Maraston}, {Marques-Chaves}, {Masseron},
  {Masters}, {McBride}, {McDermid}, {McGrath}, {McGreer}, {Medina Pe{\~n}a},
  {Melendez}, {Merloni}, {Merrifield}, {Meszaros}, {Meza}, {Minchev},
  {Minniti}, {Miyaji}, {More}, {Mulchaey}, {M{\"u}ller-S{\'a}nchez}, {Muna},
  {Munoz}, {Myers}, {Nair}, {Nandra}, {Correa do Nascimento}, {Negrete},
  {Ness}, {Newman}, {Nichol}, {Nidever}, {Nitschelm}, {Ntelis}, {O'Connell},
  {Oelkers}, {Oravetz}, {Oravetz}, {Pace}, {Padilla}, {Palanque-Delabrouille},
  {Alonso Palicio}, {Pan}, {Parejko}, {Parikh}, {P{\^a}ris}, {Park}, {Patten},
  {Peirani}, {Pellejero-Ibanez}, {Penny}, {Percival}, {Perez-Fournon},
  {Petitjean}, {Pieri}, {Pinsonneault}, {Pisani}, {Poleski}, {Prada},
  {Prakash}, {Queiroz}, {Raddick}, {Raichoor}, {Barboza Rembold}, {Richstein},
  {Riffel}, {Riffel}, {Rix}, {Robin}, {Rockosi}, {Rodr{\'\i}guez-Torres},
  {Roman-Lopes}, {Rom{\'a}n-Z{\'u}{\~n}iga}, {Rosado}, {Ross}, {Rossi}, {Ruan},
  {Ruggeri}, {Rykoff}, {Salazar-Albornoz}, {Salvato}, {S{\'a}nchez}, {Aguado},
  {S{\'a}nchez-Gallego}, {Santana}, {Santiago}, {Sayres}, {Schiavon}, {da Silva
  Schimoia}, {Schlafly}, {Schlegel}, {Schneider}, {Schultheis}, {Schuster},
  {Schwope}, {Seo}, {Shao}, {Shen}, {Shetrone}, {Shull}, {Simon}, {Skinner},
  {Skrutskie}, {Slosar}, {Smith}, {Sobeck}, {Sobreira}, {Somers}, {Souto},
  {Stark}, {Stassun}, {Stauffer}, {Steinmetz}, {Storchi-Bergmann},
  {Streblyanska}, {Stringfellow}, {Su{\'a}rez}, {Sun}, {Suzuki}, {Szigeti},
  {Taghizadeh-Popp}, {Tang}, {Tao}, {Tayar}, {Tembe}, {Teske}, {Thakar},
  {Thomas}, {Thompson}, {Tinker}, {Tissera}, {Tojeiro}, {Hernandez Toledo}, {de
  la Torre}, {Tremonti}, {Troup}, {Valenzuela}, {Martinez Valpuesta},
  {Vargas-Gonz{\'a}lez}, {Vargas-Maga{\~n}a}, {Vazquez}, {Villanova}, {Vivek},
  {Vogt}, {Wake}, {Walterbos}, {Wang}, {Weaver}, {Weijmans}, {Weinberg},
  {Westfall}, {Whelan}, {Wild}, {Wilson}, {Wood-Vasey}, {Wylezalek}, {Xiao},
  {Yan}, {Yang}, {Ybarra}, {Y{\`e}che}, {Zakamska}, {Zamora}, {Zarrouk},
  {Zasowski}, {Zhang}, {Zhao}, {Zheng}, {Zheng}, {Zhou}, {Zhou}, {Zhu},
  {Zoccali}, \& {Zou}}]{Blanton2017}
{Blanton}, M.~R., {Bershady}, M.~A., {Abolfathi}, B., {et~al.} 2017, \aj, 154,
  28

\bibitem[{{Bordoloi} {et~al.}(2011){Bordoloi}, {Lilly}, {Knobel}, {Bolzonella},
  {Kampczyk}, {Carollo}, {Iovino}, {Zucca}, {Contini}, {Kneib}, {Le Fevre},
  {Mainieri}, {Renzini}, {Scodeggio}, {Zamorani}, {Balestra}, {Bardelli},
  {Bongiorno}, {Caputi}, {Cucciati}, {de la Torre}, {de Ravel}, {Garilli},
  {Kova{\v c}}, {Lamareille}, {Le Borgne}, {Le Brun}, {Maier}, {Mignoli},
  {Pello}, {Peng}, {Perez Montero}, {Presotto}, {Scarlata}, {Silverman},
  {Tanaka}, {Tasca}, {Tresse}, {Vergani}, {Barnes}, {Cappi}, {Cimatti},
  {Coppa}, {Diener}, {Franzetti}, {Koekemoer}, {L{\'o}pez-Sanjuan},
  {McCracken}, {Moresco}, {Nair}, {Oesch}, {Pozzetti}, \&
  {Welikala}}]{Bordoloi2011}
{Bordoloi}, R., {Lilly}, S.~J., {Knobel}, C., {et~al.} 2011, \apj, 743, 10

\bibitem[{{Bouch{\'e}} {et~al.}(2012){Bouch{\'e}}, {Hohensee}, {Vargas},
  {Kacprzak}, {Martin}, {Cooke}, \& {Churchill}}]{Bouche2012}
{Bouch{\'e}}, N., {Hohensee}, W., {Vargas}, R., {et~al.} 2012, \mnras, 426, 801

\bibitem[{{Bouch{\'e}} {et~al.}(2013){Bouch{\'e}}, {Murphy}, {Kacprzak},
  {P{\'e}roux}, {Contini}, {Martin}, \& {Dessauges-Zavadsky}}]{Bouche2013}
{Bouch{\'e}}, N., {Murphy}, M.~T., {Kacprzak}, G.~G., {et~al.} 2013, Science,
  341, 50

\bibitem[{{Bouch{\'e}} {et~al.}(2016){Bouch{\'e}}, {Finley}, {Schroetter},
  {Murphy}, {Richter}, {Bacon}, {Contini}, {Richard}, {Wendt}, {Kamann},
  {Epinat}, {Cantalupo}, {Straka}, {Schaye}, {Martin}, {P{\'e}roux},
  {Wisotzki}, {Soto}, {Lilly}, {Carollo}, {Brinchmann}, \&
  {Kollatschny}}]{Bouche2016}
{Bouch{\'e}}, N., {Finley}, H., {Schroetter}, I., {et~al.} 2016, \apj, 820, 121

\bibitem[{{Bryan} \& {Norman}(1998)}]{BryanNorman1998}
{Bryan}, G.~L., \& {Norman}, M.~L. 1998, \apj, 495, 80

\bibitem[{{Chabrier}(2003)}]{Chabrier2003}
{Chabrier}, G. 2003, \pasp, 115, 763

\bibitem[{{Chen} {et~al.}(2014){Chen}, {Gauthier}, {Sharon}, {Johnson}, {Nair},
  \& {Liang}}]{Chen2014}
{Chen}, H.-W., {Gauthier}, J.-R., {Sharon}, K., {et~al.} 2014, \mnras, 438,
  1435

\bibitem[{{Chen} {et~al.}(2010){Chen}, {Helsby}, {Gauthier}, {Shectman},
  {Thompson}, \& {Tinker}}]{Chen2010}
{Chen}, H.-W., {Helsby}, J.~E., {Gauthier}, J.-R., {et~al.} 2010, \apj, 714,
  1521

\bibitem[{{Chen} {et~al.}(2020){Chen}, {Zahedy}, {Boettcher}, {Cooper},
  {Johnson}, {Rudie}, {Chen}, {Walth}, {Cantalupo}, {Cooksey},
  {Faucher-Gigu{\`e}re}, {Greene}, {Lopez}, {Mulchaey}, {Penton}, {Petitjean},
  {Putman}, {Rafelski}, {Rauch}, {Schaye}, {Simcoe}, \& {Weiner}}]{Chen2020}
{Chen}, H.-W., {Zahedy}, F.~S., {Boettcher}, E., {et~al.} 2020, \mnras, 497,
  498

\bibitem[{{Churchill} {et~al.}(2013){Churchill}, {Trujillo-Gomez}, {Nielsen},
  \& {Kacprzak}}]{Churchill2013magiicat}
{Churchill}, C.~W., {Trujillo-Gomez}, S., {Nielsen}, N.~M., \& {Kacprzak},
  G.~G. 2013, \apj, 779, 87

\bibitem[{{Davies} {et~al.}(2020){Davies}, {Crain}, {Oppenheimer}, \&
  {Schaye}}]{Davies2020}
{Davies}, J.~J., {Crain}, R.~A., {Oppenheimer}, B.~D., \& {Schaye}, J. 2020,
  \mnras, 491, 4462

\bibitem[{{DeFelippis} {et~al.}(2020){DeFelippis}, {Genel}, {Bryan}, {Nelson},
  {Pillepich}, \& {Hernquist}}]{DeFelippis2020}
{DeFelippis}, D., {Genel}, S., {Bryan}, G.~L., {et~al.} 2020, \apj, 895, 17

\bibitem[{{Diamond-Stanic} {et~al.}(2016){Diamond-Stanic}, {Coil}, {Moustakas},
  {Tremonti}, {Sell}, {Mendez}, {Hickox}, \& {Rudnick}}]{DiamondStanic2016}
{Diamond-Stanic}, A.~M., {Coil}, A.~L., {Moustakas}, J., {et~al.} 2016, \apj,
  824, 24

\bibitem[{{Dutta} {et~al.}(2025){Dutta}, {Muzahid}, {Schaye}, {Cantalupo},
  {Chen}, \& {Johnson}}]{Dutta2025}
{Dutta}, S., {Muzahid}, S., {Schaye}, J., {et~al.} 2025, \apj, 980, 264

\bibitem[{{Dutta} {et~al.}(2024){Dutta}, {Muzahid}, {Schaye}, {Johnson},
  {Bouch{\'e}}, {Chen}, \& {Cantalupo}}]{Dutta2024}
---. 2024, arXiv e-prints, arXiv:2409.15423

\bibitem[{{Faucher-Gigu{\`e}re} \& {Oh}(2023)}]{FaucherOh2023}
{Faucher-Gigu{\`e}re}, C.-A., \& {Oh}, S.~P. 2023, \araa, 61, 131

\bibitem[{{Fielding} {et~al.}(2017){Fielding}, {Quataert}, {McCourt}, \&
  {Thompson}}]{Fielding2017}
{Fielding}, D., {Quataert}, E., {McCourt}, M., \& {Thompson}, T.~A. 2017,
  \mnras, 466, 3810

\bibitem[{{Geller} {et~al.}(2014){Geller}, {Hwang}, {Fabricant}, {Kurtz},
  {Dell'Antonio}, \& {Zahid}}]{Geller2014}
{Geller}, M.~J., {Hwang}, H.~S., {Fabricant}, D.~G., {et~al.} 2014, \apjs, 213,
  35

\bibitem[{{Hafen} {et~al.}(2022){Hafen}, {Stern}, {Bullock}, {Gurvich}, {Yu},
  {Faucher-Gigu{\`e}re}, {Fielding}, {Angl{\'e}s-Alc{\'a}zar}, {Quataert},
  {Wetzel}, {Starkenburg}, {Boylan-Kolchin}, {Moreno}, {Feldmann}, {El-Badry},
  {Chan}, {Trapp}, {Kere{\v{s}}}, \& {Hopkins}}]{Hafen2022}
{Hafen}, Z., {Stern}, J., {Bullock}, J., {et~al.} 2022, \mnras, 514, 5056

\bibitem[{{Hanuschik}(2003)}]{Hanuschik2003}
{Hanuschik}, R.~W. 2003, \aap, 407, 1157

\bibitem[{{Heckman} {et~al.}(2017){Heckman}, {Borthakur}, {Wild},
  {Schiminovich}, \& {Bordoloi}}]{Heckman2017}
{Heckman}, T., {Borthakur}, S., {Wild}, V., {Schiminovich}, D., \& {Bordoloi},
  R. 2017, \apj, 846, 151

\bibitem[{{Ho} \& {Martin}(2020)}]{HoMartin2020}
{Ho}, S.~H., \& {Martin}, C.~L. 2020, \apj, 888, 14

\bibitem[{{Ho} {et~al.}(2017){Ho}, {Martin}, {Kacprzak}, \&
  {Churchill}}]{Ho2017}
{Ho}, S.~H., {Martin}, C.~L., {Kacprzak}, G.~G., \& {Churchill}, C.~W. 2017,
  \apj, 835, 267

\bibitem[{{Huang} {et~al.}(2021){Huang}, {Chen}, {Shectman}, {Johnson},
  {Zahedy}, {Helsby}, {Gauthier}, \& {Thompson}}]{Huang2021}
{Huang}, Y.-H., {Chen}, H.-W., {Shectman}, S.~A., {et~al.} 2021, \mnras, 502,
  4743

\bibitem[{{Hubble}(1926)}]{Hubble1926}
{Hubble}, E.~P. 1926, \apj, 64, 321

\bibitem[{{Johnson} {et~al.}(2013){Johnson}, {Chen}, \&
  {Mulchaey}}]{Johnson2013}
{Johnson}, S.~D., {Chen}, H.-W., \& {Mulchaey}, J.~S. 2013, \mnras, 434, 1765

\bibitem[{{Johnson} {et~al.}(2015){Johnson}, {Chen}, \&
  {Mulchaey}}]{Johnson2015}
---. 2015, \mnras, 449, 3263

\bibitem[{{Johnson} {et~al.}(2017){Johnson}, {Chen}, {Mulchaey}, {Schaye}, \&
  {Straka}}]{Johnson2017}
{Johnson}, S.~D., {Chen}, H.-W., {Mulchaey}, J.~S., {Schaye}, J., \& {Straka},
  L.~A. 2017, \apjl, 850, L10

\bibitem[{{Kacprzak} {et~al.}(2011){Kacprzak}, {Churchill}, {Barton}, \&
  {Cooke}}]{Kacprzak2011ApJ}
{Kacprzak}, G.~G., {Churchill}, C.~W., {Barton}, E.~J., \& {Cooke}, J. 2011,
  \apj, 733, 105

\bibitem[{{Kacprzak} {et~al.}(2010){Kacprzak}, {Churchill}, {Ceverino},
  {Steidel}, {Klypin}, \& {Murphy}}]{Kacprzak2010}
{Kacprzak}, G.~G., {Churchill}, C.~W., {Ceverino}, D., {et~al.} 2010, \apj,
  711, 533

\bibitem[{{Kacprzak} {et~al.}(2012){Kacprzak}, {Churchill}, \&
  {Nielsen}}]{Kacprzak2012}
{Kacprzak}, G.~G., {Churchill}, C.~W., \& {Nielsen}, N.~M. 2012, \apjl, 760, L7

\bibitem[{{Kacprzak} {et~al.}(2015){Kacprzak}, {Muzahid}, {Churchill},
  {Nielsen}, \& {Charlton}}]{Kacprzak2015}
{Kacprzak}, G.~G., {Muzahid}, S., {Churchill}, C.~W., {Nielsen}, N.~M., \&
  {Charlton}, J.~C. 2015, \apj, 815, 22

\bibitem[{{Kacprzak} {et~al.}(2019){Kacprzak}, {Vander Vliet}, {Nielsen},
  {Muzahid}, {Pointon}, {Churchill}, {Ceverino}, {Arraki}, {Klypin},
  {Charlton}, \& {Lewis}}]{Kacprzak2019}
{Kacprzak}, G.~G., {Vander Vliet}, J.~R., {Nielsen}, N.~M., {et~al.} 2019,
  \apj, 870, 137

\bibitem[{{Kacprzak} {et~al.}(2025){Kacprzak}, {Oppenheimer}, {Nielsen},
  {Fernandez-Figueroa}, {Murphy}, {Allen}, {Barone}, {Sameer}, {Churchill},
  {Burchett}, {Gupta}, {Charlton}, \& {Platukis}}]{Kacprzak2025}
{Kacprzak}, G.~G., {Oppenheimer}, B.~D., {Nielsen}, N.~M., {et~al.} 2025, arXiv
  e-prints, arXiv:2507.11613

\bibitem[{{Kakoly} {et~al.}(2025){Kakoly}, {Stern}, {Faucher-Gigu{\`e}re},
  {Fielding}, {Goldner}, {Sun}, \& {Hummels}}]{Kakoly2025}
{Kakoly}, A., {Stern}, J., {Faucher-Gigu{\`e}re}, C.-A., {et~al.} 2025, arXiv
  e-prints, arXiv:2504.17001

\bibitem[{{Keeney} {et~al.}(2013){Keeney}, {Stocke}, {Rosenberg}, {Danforth},
  {Ryan-Weber}, {Shull}, {Savage}, \& {Green}}]{Keeney2013}
{Keeney}, B.~A., {Stocke}, J.~T., {Rosenberg}, J.~L., {et~al.} 2013, \apj, 765,
  27

\bibitem[{{Krogager}(2018)}]{Krogager2018}
{Krogager}, J.-K. 2018, arXiv e-prints, arXiv:1803.01187

\bibitem[{{Kroupa}(2001)}]{Kroupa2001}
{Kroupa}, P. 2001, \mnras, 322, 231

\bibitem[{{Lehner} {et~al.}(2025){Lehner}, {Howk}, {Collins}, {Sameer},
  {Wakker}, {Augustin}, {Barger}, {Berg}, {Bordoloi}, {Brown}, {Cashman},
  {Faucher-Gigu{\`e}re}, {Fox}, {French}, {Gilbert}, {Guhathakurta}, {O'Meara},
  {O'Shea}, {Peeples}, {Pisano}, {Prochaska}, {Stern}, {Tumlinson}, {Werk}, \&
  {Williams}}]{Lehner2025}
{Lehner}, N., {Howk}, J.~C., {Collins}, L., {et~al.} 2025, arXiv e-prints,
  arXiv:2506.16573

\bibitem[{{Lochhaas} {et~al.}(2020){Lochhaas}, {Bryan}, {Li}, {Li}, \&
  {Fielding}}]{Lochhaas2020}
{Lochhaas}, C., {Bryan}, G.~L., {Li}, Y., {Li}, M., \& {Fielding}, D. 2020,
  \mnras, 493, 1461

\bibitem[{{Lopez} {et~al.}(2020){Lopez}, {Tejos}, {Barrientos}, {Ledoux},
  {Sharon}, {Katsianis}, {Florian}, {Rivera-Thorsen}, {Bayliss}, {Dahle},
  {Fernand ez-Figueroa}, {Gladders}, {Gronke}, {Hamel}, {Pessa}, \&
  {Rigby}}]{Lopez2020}
{Lopez}, S., {Tejos}, N., {Barrientos}, L.~F., {et~al.} 2020, \mnras, 491, 4442

\bibitem[{{Madau} \& {Dickinson}(2014)}]{MadauDickinson2014}
{Madau}, P., \& {Dickinson}, M. 2014, \araa, 52, 415

\bibitem[{{Marasco} {et~al.}(2019){Marasco}, {Fraternali}, {Heald}, {de Blok},
  {Oosterloo}, {Kamphuis}, {J{\'o}zsa}, {Vargas}, {Winkel}, {Walterbos},
  {Dettmar}, \& {Juẗte}}]{Marasco2019}
{Marasco}, A., {Fraternali}, F., {Heald}, G., {et~al.} 2019, \aap, 631, A50

\bibitem[{{Martin} {et~al.}(2019){Martin}, {Ho}, {Kacprzak}, \&
  {Churchill}}]{Martin2019}
{Martin}, C.~L., {Ho}, S.~H., {Kacprzak}, G.~G., \& {Churchill}, C.~W. 2019,
  \apj, 878, 84

\bibitem[{{Mathews} \& {Prochaska}(2017)}]{MathewsProchaska2017}
{Mathews}, W.~G., \& {Prochaska}, J.~X. 2017, \apjl, 846, L24

\bibitem[{{Mishra} {et~al.}(2024){Mishra}, {Johnson}, {Rudie}, {Chen},
  {Schaye}, {Qu}, {Zahedy}, {Boettcher}, {Cantalupo}, {Chen},
  {Faucher-Gigu{\'e}re}, {Greene}, {Li}, {Liu}, {Lopez}, \&
  {Petitjean}}]{Mishra2024}
{Mishra}, N., {Johnson}, S.~D., {Rudie}, G.~C., {et~al.} 2024, \apj, 976, 149

\bibitem[{{Morton}(2003)}]{Morton2003}
{Morton}, D.~C. 2003, \apjs, 149, 205

\bibitem[{{Nateghi} {et~al.}(2024){Nateghi}, {Kacprzak}, {Nielsen}, {Sameer},
  {Murphy}, {Churchill}, \& {Charlton}}]{Nateghi2024multiphase}
{Nateghi}, H., {Kacprzak}, G.~G., {Nielsen}, N.~M., {et~al.} 2024, \mnras, 534,
  930

\bibitem[{{Nelson} {et~al.}(2018){Nelson}, {Kauffmann}, {Pillepich}, {Genel},
  {Springel}, {Pakmor}, {Hernquist}, {Weinberger}, {Torrey}, {Vogelsberger}, \&
  {Marinacci}}]{Nelson2018}
{Nelson}, D., {Kauffmann}, G., {Pillepich}, A., {et~al.} 2018, \mnras, 477, 450

\bibitem[{{Ng} {et~al.}(2019){Ng}, {Nielsen}, {Kacprzak}, {Pointon}, {Muzahid},
  {Churchill}, \& {Charlton}}]{Ng2019}
{Ng}, M., {Nielsen}, N.~M., {Kacprzak}, G.~G., {et~al.} 2019, \apj, 886, 66

\bibitem[{{Nielsen} {et~al.}(2013){Nielsen}, {Churchill}, \&
  {Kacprzak}}]{Nielsen2013_ii}
{Nielsen}, N.~M., {Churchill}, C.~W., \& {Kacprzak}, G.~G. 2013, \apj, 776, 115

\bibitem[{{Nielsen} {et~al.}(2017){Nielsen}, {Kacprzak}, {Muzahid},
  {Churchill}, {Murphy}, \& {Charlton}}]{Nielsen2017}
{Nielsen}, N.~M., {Kacprzak}, G.~G., {Muzahid}, S., {et~al.} 2017, \apj, 834,
  148

\bibitem[{{Nielsen} {et~al.}(2018){Nielsen}, {Kacprzak}, {Pointon},
  {Churchill}, \& {Murphy}}]{Nielsen2018}
{Nielsen}, N.~M., {Kacprzak}, G.~G., {Pointon}, S.~K., {Churchill}, C.~W., \&
  {Murphy}, M.~T. 2018, \apj, 869, 153

\bibitem[{{Oke} {et~al.}(1995){Oke}, {Cohen}, {Carr}, {Cromer}, {Dingizian},
  {Harris}, {Labrecque}, {Lucinio}, {Schaal}, {Epps}, \& {Miller}}]{Oke1995}
{Oke}, J.~B., {Cohen}, J.~G., {Carr}, M., {et~al.} 1995, \pasp, 107, 375

\bibitem[{{Oppenheimer} {et~al.}(2016){Oppenheimer}, {Crain}, {Schaye},
  {Rahmati}, {Richings}, {Trayford}, {Tumlinson}, {Bower}, {Schaller}, \&
  {Theuns}}]{Oppenheimer2016}
{Oppenheimer}, B.~D., {Crain}, R.~A., {Schaye}, J., {et~al.} 2016, \mnras, 460,
  2157

\bibitem[{{Pandya} {et~al.}(2023){Pandya}, {Fielding}, {Bryan}, {Carr},
  {Somerville}, {Stern}, {Faucher-Gigu{\`e}re}, {Hafen},
  {Angl{\'e}s-Alc{\'a}zar}, \& {Forbes}}]{Pandya2023}
{Pandya}, V., {Fielding}, D.~B., {Bryan}, G.~L., {et~al.} 2023, \apj, 956, 118

\bibitem[{{Planck Collaboration} {et~al.}(2016){Planck Collaboration}, {Ade},
  {Aghanim}, {Arnaud}, {Ashdown}, {Aumont}, {Baccigalupi}, {Banday},
  {Barreiro}, {Bartlett}, \& et~al.}]{PlanckXIII2015}
{Planck Collaboration}, {Ade}, P.~A.~R., {Aghanim}, N., {et~al.} 2016, \aap,
  594, A13

\bibitem[{{Pointon} {et~al.}(2017){Pointon}, {Nielsen}, {Kacprzak}, {Muzahid},
  {Churchill}, \& {Charlton}}]{Pointon2017}
{Pointon}, S.~K., {Nielsen}, N.~M., {Kacprzak}, G.~G., {et~al.} 2017, \apj,
  844, 23

\bibitem[{{Prochaska} {et~al.}(2017){Prochaska}, {Tejos}, {Crighton},
  {Jnburchett}, {Tiffanyhsyu}, {Tuo-Ji}, {Marijana777}, {Ktirimba}, {Jhennawi},
  {Cooke}, {O'Meara}, \& {Werk}}]{linetoolsZenodo}
{Prochaska}, J.~X., {Tejos}, N., {Crighton}, N., {et~al.} 2017,
  {linetools/linetools: Third Minor Release}, v0.3,  Zenodo

\bibitem[{{Qu} {et~al.}(2024){Qu}, {Chen}, {Johnson}, {Rudie}, {Zahedy},
  {DePalma}, {Schaye}, {Boettcher}, {Cantalupo}, {Chen}, {Faucher-Gigu{\`e}re},
  {Li}, {Mulchaey}, {Petitjean}, \& {Rafelski}}]{Qu2024}
{Qu}, Z., {Chen}, H.-W., {Johnson}, S.~D., {et~al.} 2024, \apj, 968, 8

\bibitem[{{Rockosi} {et~al.}(2010){Rockosi}, {Stover}, {Kibrick}, {Lockwood},
  {Peck}, {Cowley}, {Bolte}, {Adkins}, {Alcott}, {Allen}, {Brown}, {Cabak},
  {Deich}, {Hilyard}, {Kassis}, {Lanclos}, {Lewis}, {Pfister}, {Phillips},
  {Robinson}, {Saylor}, {Thompson}, {Ward}, {Wei}, \& {Wright}}]{Rockosi2010}
{Rockosi}, C., {Stover}, R., {Kibrick}, R., {et~al.} 2010, in Society of
  Photo-Optical Instrumentation Engineers (SPIE) Conference Series, Vol. 7735,
  Society of Photo-Optical Instrumentation Engineers (SPIE) Conference Series,
  0

\bibitem[{{Salpeter}(1955)}]{Salpeter1955}
{Salpeter}, E.~E. 1955, \apj, 121, 161

\bibitem[{{Sameer} {et~al.}(2024){Sameer}, {Charlton}, {Wakker}, {Kacprzak},
  {Nielsen}, {Churchill}, {Richter}, {Muzahid}, {Ho}, {Nateghi}, {Rosenwasser},
  {Narayanan}, \& {Ganguly}}]{Sameer2024}
{Sameer}, {Charlton}, J.~C., {Wakker}, B.~P., {et~al.} 2024, \mnras, 530, 3827

\bibitem[{{Schroetter} {et~al.}(2019){Schroetter}, {Bouch{\'e}}, {Zabl},
  {Contini}, {Wendt}, {Schaye}, {Mitchell}, {Muzahid}, {Marino}, {Bacon},
  {Lilly}, {Richard}, \& {Wisotzki}}]{Schroetter2019}
{Schroetter}, I., {Bouch{\'e}}, N.~F., {Zabl}, J., {et~al.} 2019, \mnras, 490,
  4368

\bibitem[{{Sheinis} {et~al.}(2002){Sheinis}, {Bolte}, {Epps}, {Kibrick},
  {Miller}, {Radovan}, {Bigelow}, \& {Sutin}}]{Sheinis2002}
{Sheinis}, A.~I., {Bolte}, M., {Epps}, H.~W., {et~al.} 2002, \pasp, 114, 851

\bibitem[{{Steidel} {et~al.}(2002){Steidel}, {Kollmeier}, {Shapley},
  {Churchill}, {Dickinson}, \& {Pettini}}]{Steidel2002}
{Steidel}, C.~C., {Kollmeier}, J.~A., {Shapley}, A.~E., {et~al.} 2002, \apj,
  570, 526

\bibitem[{{Stern} {et~al.}(2018){Stern}, {Faucher-Gigu{\`e}re}, {Hennawi},
  {Hafen}, {Johnson}, \& {Fielding}}]{Stern2018}
{Stern}, J., {Faucher-Gigu{\`e}re}, C.-A., {Hennawi}, J.~F., {et~al.} 2018,
  \apj, 865, 91

\bibitem[{{Stewart} {et~al.}(2013){Stewart}, {Brooks}, {Bullock}, {Maller},
  {Diemand}, {Wadsley}, \& {Moustakas}}]{Stewart2013}
{Stewart}, K.~R., {Brooks}, A.~M., {Bullock}, J.~S., {et~al.} 2013, \apj, 769,
  74

\bibitem[{{Sultan} {et~al.}(2025){Sultan}, {Faucher-Gigu{\`e}re}, {Stern},
  {Rotshtein}, {Byrne}, \& {Wijers}}]{Sultan2025}
{Sultan}, I., {Faucher-Gigu{\`e}re}, C.-A., {Stern}, J., {et~al.} 2025, \mnras,
  540, 1017

\bibitem[{{Suresh} {et~al.}(2017){Suresh}, {Rubin}, {Kannan}, {Werk},
  {Hernquist}, \& {Vogelsberger}}]{Suresh2017}
{Suresh}, J., {Rubin}, K. H.~R., {Kannan}, R., {et~al.} 2017, \mnras, 465, 2966

\bibitem[{{Tchernyshyov} {et~al.}(2022){Tchernyshyov}, {Werk}, {Wilde},
  {Prochaska}, {Tripp}, {Burchett}, {Bordoloi}, {Howk}, {Lehner}, {O'Meara},
  {Tejos}, \& {Tumlinson}}]{Tchernyshyov2022}
{Tchernyshyov}, K., {Werk}, J.~K., {Wilde}, M.~C., {et~al.} 2022, \apj, 927,
  147

\bibitem[{{Tchernyshyov} {et~al.}(2023){Tchernyshyov}, {Werk}, {Wilde},
  {Prochaska}, {Tripp}, {Burchett}, {Bordoloi}, {Howk}, {Lehner}, {O'Meara},
  {Tejos}, \& {Tumlinson}}]{Tchernyshyov2023}
---. 2023, \apj, 949, 41

\bibitem[{{Tejos} {et~al.}(2021){Tejos}, {L{\'o}pez}, {Ledoux},
  {Fern{\'a}ndez-Figueroa}, {Rivas}, {Sharon}, {Johnston}, {Florian}, {D'Ago},
  {Katsianis}, {Barrientos}, {Berg}, {Corro-Guerra}, {Hamel}, {Moya-Sierralta},
  {Poudel}, {Rigby}, \& {Solimano}}]{Tejos2021}
{Tejos}, N., {L{\'o}pez}, S., {Ledoux}, C., {et~al.} 2021, \mnras, 507, 663

\bibitem[{{Tody}(1986)}]{Tody1986}
{Tody}, D. 1986, in Society of Photo-Optical Instrumentation Engineers (SPIE)
  Conference Series, Vol. 627, Instrumentation in astronomy VI, ed. D.~L.
  {Crawford}, 733

\bibitem[{{Tody}(1993)}]{Tody1993}
{Tody}, D. 1993, in Astronomical Society of the Pacific Conference Series,
  Vol.~52, Astronomical Data Analysis Software and Systems II, ed. R.~J.
  {Hanisch}, R.~J.~V. {Brissenden}, \& J.~{Barnes}, 173

\bibitem[{{Tumlinson} {et~al.}(2017){Tumlinson}, {Peeples}, \&
  {Werk}}]{Tumlinson2017}
{Tumlinson}, J., {Peeples}, M.~S., \& {Werk}, J.~K. 2017, \araa, 55, 389

\bibitem[{{Tumlinson} {et~al.}(2011){Tumlinson}, {Thom}, {Werk}, {Prochaska},
  {Tripp}, {Weinberg}, {Peeples}, {O'Meara}, {Oppenheimer}, {Meiring}, {Katz},
  {Dav{\'e}}, {Ford}, \& {Sembach}}]{Tumlinson2011}
{Tumlinson}, J., {Thom}, C., {Werk}, J.~K., {et~al.} 2011, Science, 334, 948

\bibitem[{{Tumlinson} {et~al.}(2013){Tumlinson}, {Thom}, {Werk}, {Prochaska},
  {Tripp}, {Katz}, {Dav{\'e}}, {Oppenheimer}, {Meiring}, {Ford}, {O'Meara},
  {Peeples}, {Sembach}, \& {Weinberg}}]{Tumlinson2013}
---. 2013, \apj, 777, 59

\bibitem[{{van Dokkum}(2001)}]{vanDokkum2001}
{van Dokkum}, P.~G. 2001, \pasp, 113, 1420

\bibitem[{{Wenger} {et~al.}(2000){Wenger}, {Ochsenbein}, {Egret}, {Dubois},
  {Bonnarel}, {Borde}, {Genova}, {Jasniewicz}, {Lalo{\"e}}, {Lesteven}, \&
  {Monier}}]{SIMBAD2020}
{Wenger}, M., {Ochsenbein}, F., {Egret}, D., {et~al.} 2000, \aaps, 143, 9

\bibitem[{{Werk} {et~al.}(2012){Werk}, {Prochaska}, {Thom}, {Tumlinson},
  {Tripp}, {O'Meara}, \& {Meiring}}]{Werk2012}
{Werk}, J.~K., {Prochaska}, J.~X., {Thom}, C., {et~al.} 2012, \apjs, 198, 3

\bibitem[{{Werk} {et~al.}(2013){Werk}, {Prochaska}, {Thom}, {Tumlinson},
  {Tripp}, {O'Meara}, \& {Peeples}}]{Werk2013}
---. 2013, \apjs, 204, 17

\bibitem[{{Werk} {et~al.}(2016){Werk}, {Prochaska}, {Cantalupo}, {Fox},
  {Oppenheimer}, {Tumlinson}, {Tripp}, {Lehner}, \& {McQuinn}}]{Werk2016}
{Werk}, J.~K., {Prochaska}, J.~X., {Cantalupo}, S., {et~al.} 2016, \apj, 833,
  54

\bibitem[{{Wizinowich} {et~al.}(2006){Wizinowich}, {Le Mignant}, {Bouchez},
  {Campbell}, {Chin}, {Contos}, {van Dam}, {Hartman}, {Johansson}, {Lafon},
  {Lewis}, {Stomski}, {Summers}, {Brown}, {Danforth}, {Max}, \&
  {Pennington}}]{Wizinowich2006}
{Wizinowich}, P.~L., {Le Mignant}, D., {Bouchez}, A.~H., {et~al.} 2006, \pasp,
  118, 297

\bibitem[{{Zabl} {et~al.}(2019){Zabl}, {Bouch{\'e}}, {Schroetter}, {Wendt},
  {Finley}, {Schaye}, {Conseil}, {Contini}, {Marino}, {Mitchell}, {Muzahid},
  {Pezzulli}, \& {Wisotzki}}]{Zabl2019}
{Zabl}, J., {Bouch{\'e}}, N.~F., {Schroetter}, I., {et~al.} 2019, \mnras, 485,
  1961

\end{thebibliography}
